\documentclass{aa}
\usepackage{psfig,afterpage,multirow}
\begin{document}
\title{Crystalline silicate dust around
evolved stars\thanks{Based on observations with ISO, an ESA
project with instruments funded by ESA Member States (especially
the PI countries: France, Germany, the Netherlands and the United
Kingdom) and with the participation of ISAS and NASA}$^,$\thanks{Tables 4 to 20
are only available in electronic form at the CDS via anonymous ftp to
cdsarc.u-strasbg.fr (130.79.128.5) or via 
http://cdsweb.u-strasbg.fr/cgi-bin/qcat?J/A+A/} }

\subtitle{I. The sample stars}

\author{F.J. Molster\inst{1,2,\dagger}, L.B.F.M. Waters\inst{1,3},
A.G.G.M. Tielens\inst{4}, M.J. Barlow\inst{5}}
\institute{Astronomical Institute `Anton Pannekoek', University of
Amsterdam, Kruislaan 403, NL-1098 SJ Amsterdam, the Netherlands
\and
School of Materials Science and Engineering,
Georgia Tech, Atlanta, GA 30332-0245, USA
\and
Instituut voor Sterrenkunde, Katholieke Universiteit Leuven, Celestijnenlaan
200B, B-3001 Heverlee, Belgium
\and
SRON Laboratory for Space Research Groningen, P.O. Box 800,
NL-9700 AV Groningen, The Netherlands
\and
Department of Physics and Astronomy, University College London, Gower Street,
London WC1E 6BT, United Kingdom
}
\offprints{F.J. Molster: fmolster@so.estec.esa.nl\\
\mbox{ }$\, \dagger \!\!$ Present address: F.J. Molster, ESA/ESTEC, SCI-SO, Postbus 299, 2200 AG  Noordwijk, The Netherlands}
\date{received date; accepted date}

\authorrunning{F.J. Molster et al.}
\titlerunning{The sample stars}

\abstract{This is the first paper in a series of three where we present the
first comprehensive inventory of solid state emission bands
observed in a sample of 17 oxygen-rich circumstellar dust shells
surrounding evolved stars. The data were taken with the Short and
Long Wavelength Spectrographs on board of the Infrared Space
Observatory~(ISO) and cover the 2.4 to 195 $\mu$m wavelength range.
The spectra show the presence of broad 10 and 18 $\mu$m bands that
can be attributed to amorphous silicates. In addition, at least
49 narrow bands are found whose position and width indicate they
can be attributed to crystalline silicates. Almost all of these
bands were not known before ISO. The incredible richness of the
crystalline silicate spectra observed by ISO allows detailed
studies of the mineralogy of these dust shells, and is a telltale 
about the origin and evolution of the dust.
We have measured the peak positions, widths and strengths of the
individual, continuum subtracted bands. Based on these
measurements, we were able to order the spectra in sequence of
decreasing crystalline silicate band strength. We found that the
strength of the emission bands correlates with the geometry of the
circumstellar shell, as derived from direct imaging or inferred
from the shape of the spectral energy distribution. This naturally
divides the sample into objects that show a disk-like geometry
(strong crystalline silicate bands), and objects whose dust shell
is characteristic of an outflow (weak crystalline silicate bands).
All stars with the 33.6 $\mu$m forsterite band stronger than 20
percent over continuum are disk sources. We define spectral
regions (called complexes) where a concentration of emission bands
is evident, at 10, 18, 23, 28, 33, 40 and 60 $\mu$m. We derive
average shapes for these complexes and compare these to the
individual band shapes of the programme stars. In an Appendix, we
provide detailed comments on the measured band positions and
strengths of individual sources.}
\maketitle
\keywords{Infrared: stars - circumstellar matter - 
Stars: AGB and post-AGB; mass loss - Planetary Nebulae: general - 
Dust, extinction}


\section{Introduction}

Red giants and supergiants are characterized by low surface
temperatures resulting in the presence of many different molecules
in their atmosphere. These objects are also known to have dense
stellar winds, presumably driven by a combination of pulsations
and radiation pressure on the dust which forms in the cooling
outflow. Since dust efficiently absorbes radiation at short
wavelengths, the central star can easily become obscured and most
of the luminosity of the star is re-radiated at mid-IR
wavelengths. At these wavelengths, the most important
ro-vibrational bands of abundant molecules can be found, and
indeed many solid state bands from various dust components have
been found using infrared spectrographs. The Infrared Space
Observatory (ISO, Kessler et al. 1996) has allowed for the first
time a comprehensive inventory of solid state bands in
astrophysical objects with uninterrupted wavelength coverage from
2.4 to 200 $\mu$m and with a spectral resolution which is well
suited for the detection of solid state bands.

We have undertaken detailed studies of the dust emission and
absorption spectra of evolved stars, ranging from Asymptotic Giant
Branch~(AGB) stars to Planetary Nebulae~(PNe) and to (post) Red
Supergiants~(RSG).
Preliminary results of these studies have been published elsewhere
and can be summarized as follows: Many oxygen-rich evolved stars
have a surprisingly rich spectrum of solid state emission bands
between 10 and 100 $\mu$m, dominated by both amorphous and
crystalline silicates (e.g. Waters et al. 1996; Molster et al.
1999a; 1999b). The crystalline silicates were not known to be
present in dust shells around evolved stars before ISO was
launched, and allow for the first time a mineralogical analysis of
the dust composition around these objects. Some objects show a
very high abundance of crystalline silicates (e.g. Molster et al.
2001a), which seems to be related to the geometry of the
circumstellar dust shell (Molster et al 1999a). Surprisingly,
stars which were believed to have a carbon-rich dust chemistry
also showed the presence of crystalline silicate emission,
pointing to a complex chemical composition of the circumstellar
environment, possibly due to rapid changes in the chemistry of the
stellar photosphere (e.g. Waters et al. 1998;
Cohen et al. 1999, Molster et al. 2001a).
These observations demonstrate the use of the
crystalline silicate bands for a better understanding of the
evolution of late type giants and supergiants.

\begin{table*}[b]
\caption[]{The programme stars and logbook of the observations used for this
study. The boldface revolution numbers are those used for the final spectrum.}
{\footnotesize
\begin{tabular}{|l|l|l|l|l|l|}
\hline
Object & Type &Revolution     & AOT & T$_{\rm {int}}$ & Comment\\
       &      &               & (AOT nr)& (sec)       & \\
\hline
IRAS09425-6040 & C-rich AGB star with O-rich dust   & \bf 084   & SWS01 & 1816  & \\
               &                                    & \bf 254   & SWS01 & 6538  & \\
NGC~6537        & Planetary Nebula, hot central star &  470      & SWS01 & 1912  & mispointed\\
               &                                    &  470      & LWS01 & 1318  & \\
               &                                    & \bf 703   & SWS01 & 3454  & \\
               &                                    &\bf 703    & LWS01 & 2230  & \\
NGC~6302        & Planetary Nebula, hot central star & \bf 094   & SWS01 & 6528  & \\
               &                                    & 479       & SWS06 & 8532  & $2.4 - 7.0$ and $12.0 - 27.5 \mu$m\\
               &                                    &   479     & SWS06 & 12165 & $7.0 - 12.0$ and $27.5 - 45.2 \mu$m\\
               &                                    & \bf289-678& LWS01 & 13243 & combination of rev 289, 482,\\
               &                                    &           &       &       & 489, 503, 510, 671, and 678\\
MWC~922         & Peculiar object                    & \bf 153   & SWS01 & 1834  & combined with rev 703\\
               &                                    & \bf 478   & LWS01 & 1316  & \\
               &                                    & \bf 703   & SWS01 & 1912  & combined with rev 153\\
AC Her         & Binary post-AGB star               & 106       & SWS01 & 1834  & \\
               &                                    & \bf 471   & LWS01 & 1910  & \\
               &                                    & \bf 520   & SWS01 & 6538  & \\
HD~45677        & B[e] star, nature unclear          & \bf 711   & SWS01 & 6538  & \\
89~Her         & Binary post-AGB star               & 082       & SWS01 & 1044  & \\
               &                                    & \bf 336   & LWS01 & 1860  & \\
               &                                    & \bf 518   & SWS01 & 6538  & \\
MWC~300         & Evolved star, B supergiant         & \bf 516   & SWS01 & 3454  & \\
Vy~2-2         & Proto-planetary nebula             & \bf 320   & SWS01 & 1140  & \\
               &                                    & \bf 547   & LWS01 & 1318  & $\approx 7$ arcsec mispointed \\
HD~44179        & Red Rectangle; binary post-AGB star& 702       & SWS06 & 1174  & $31.4 - 35.1 \mu$m \\
               &                                    & 702       & SWS06 & 856   & $19.5 - 25 \mu$m\\
               &                                    & \bf 702   & SWS01 & 6538  & \\
               &                                    & \bf 709   & LWS01 & 3428  & \\
               &                                    & 870       & SWS06 & 8406  & $12 - 19.5 \mu$m\\
HD~161796       & Post-AGB star                      & 071       & SWS01 & 1044  & \\
               &                                    & \bf 080   & LWS01 & 1554  & \\
               &                                    & \bf 342   & SWS01 & 1912  & \\
               &                                    & \bf 521   & SWS06 & 1744  & $29.0 - 45.2 \mu$m\\
OH~26.5+0.6     & OH/IR star, high mass loss rate    & \bf 330   & SWS01 & 1912  & \\
               &                                    &   \bf 330 & LWS01 & 1268  & \\
               &                                    &     340   & LWS01 &  822  & \\
Roberts~22     & post-AGB star, A supergiant        &     084   & SWS01 & 1044  & mispointed    \\
               &                                    & \bf 103   & LWS01 & 478   &  \\
               &                                    & \bf 254   & SWS01 & 3454  & mispointed    \\
HD~179821       & post-AGB or post-RSG star          &     113   & SWS01 & 1834  & \\
               &                                    & \bf 520   & SWS01 & 6538  & \\
               &                                    & \bf 319   & LWS01 & 1266  & \\
AFGL~4106       & post-RSG, binary                   &  060      & SWS01 & 1130  & \\
               &                                    &     104   & SWS01 & 1834  & \\
               &                                    & \bf 104   & LWS01 &  476  & \\
               &                                    & \bf 249   & SWS01 & 3454  & \\
NML~Cyg        & Red Supergiant, high mass loss rate& \bf 052   & SWS01 & 6544  & \\
               &                                    &     342   & SWS01 & 1140  & \\
               &                                    & \bf 530   & SWS06 & 1688  & $29.5 - 45.2 \mu$m\\
               &                                    & \bf 555   & LWS01 & 2798  & \\
               &                                    &     741   & SWS01 & 1140  & \\
IRC+10420      & post-RSG, A supergiant             & \bf 128   & SWS01 & 3462  & \\
               &                                    & \bf 316   & SWS06 & 1718  & $29.3-44.7 \mu$m \\
               &                                    & \bf 724   & LWS01 & 3430  & \\
\hline
\end{tabular}}
\label{tab:log}
\end{table*}

\afterpage{\clearpage}

This paper is the first in a series of three where we present
a detailed and comprehensive overview of the solid state emission
bands in oxygen-rich dust shells surrounding evolved stars and
related objects. The purpose of these papers is to quantify as
best as possible the presence and characteristics of the numerous
new emission bands that have been discovered using the ISO data.
In paper~II (Molster et al. 2002a) of this series, we describe the average band
profiles of seven "complexes" that can be recognized in the
combined Short- (de Graauw et al. 1996) and Long- (Clegg et al.
1996) wavelength spectrometers (hereafter referred to as SWS and
LWS respectively) that were on board of ISO. Based on the strength of
the crystalline silicate bands, we divide in paper~II
the sample of 17 stars into two groups. This division is also one which
separates objects with a highly flattened dust distribution
(referred to here as "disk" sources) from those with a more
spherical distribution of dust (the non-disk or 
spherical outflow sources, hereafter referred to as "outflow" sources). 
In the present study, we present the 17 programme stars, we give
an overview of the individual spectra, and we quantify the
position and strength of the bands. In Paper III (Molster et al. 2002b)
we will investigate several trends in the spectrum
and correlate them with other information available about these sources.

This paper is organized as follows: Sect.~2 presents the sample,
the observations and data reduction; Sect.~3 describes the nature
of the individual sources, and the shape of the complexes as
compared to the mean. Sect~4 summarizes the results of this study.
In Appendix~A we present the band strength data and some more
detailed discussion about individual spectral features with
respect to their reliability and blending.

\section{The observations and data reduction}
\label{sec:obser}

\subsection{The sample}

In order to get as broad as possible an overview of the O-rich
dust features around evolved stars, we aimed for a sample covering
the evolution of stars from the Asymptotic Giant Branch~(AGB) to
the Planetary Nebula~(PN) phase, and also including some massive
(post-) Red Supergiants (RSG). Several objects were also included
whose evolutionary status is unclear (e.g. MWC~922, MWC~300 and
HD~45677). We inspected the spectra for the presence of
crystalline silicate bands.
Since we are interested in (sometimes weak) bands on
top of a continuum, we selected bright objects with fluxes higher
than about 20~Jy in the 20 to 45 $\mu$m region. We preferentially
included objects for which both an SWS and an LWS spectrum are
available. We finally arrived at a sample of 17 objects,
listed in Table~\ref{tab:log}.

In several cases more than one observation was available. The boldface revolution
numbers in Table~\ref{tab:log} indicate
which observations were used for the final spectrum. The other observations
were less accurately reduced and only used as a reference in case of
doubt in the main spectrum.

\subsection{Data reduction}

\begin{figure*}[t]
\centerline{\psfig{figure=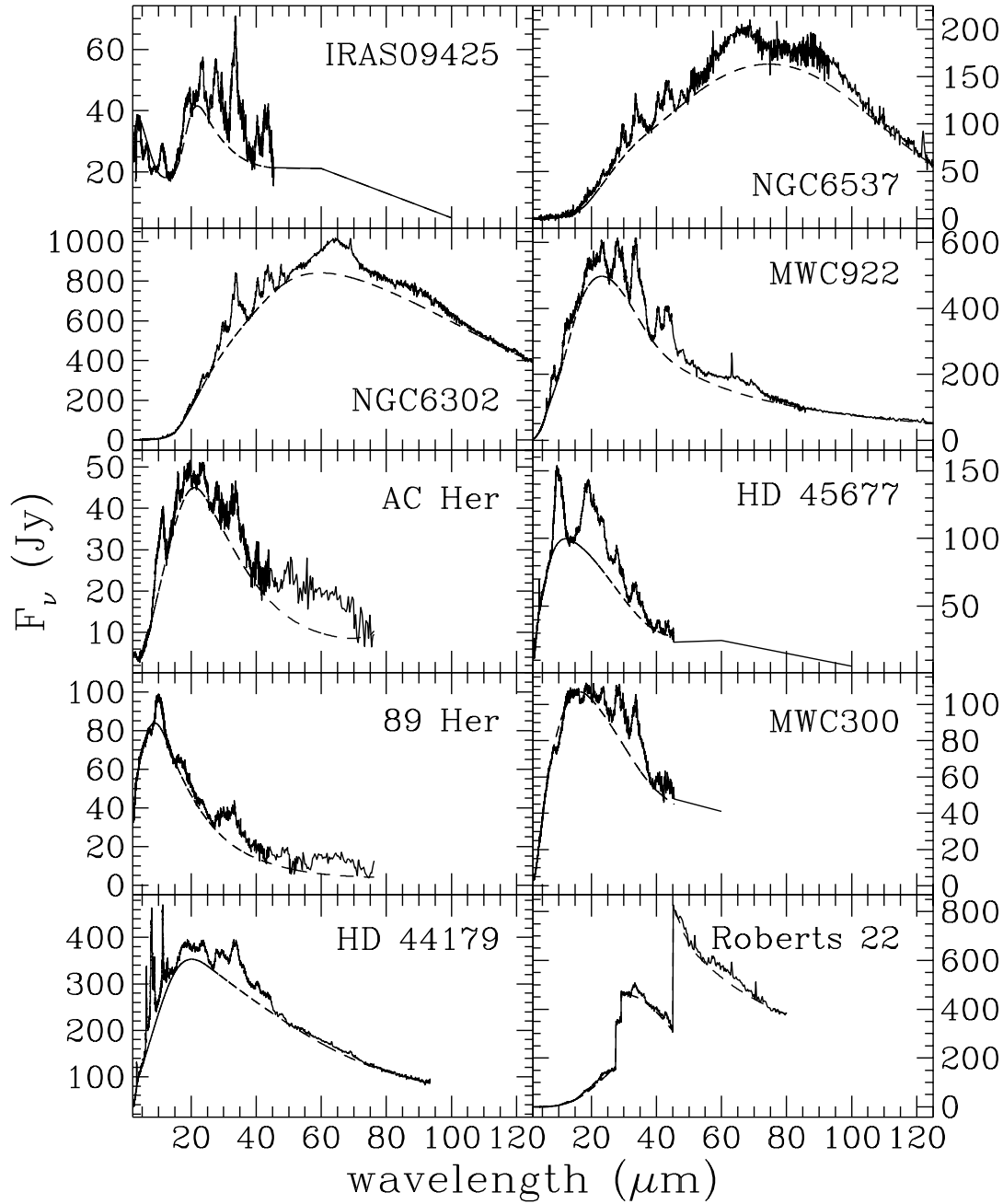,width=143mm}}
\caption[]{\small The spectra of the disk sources (solid line) and
their eye-ball spline-fit continuum (dashed line). The jumps in
the spectrum of Roberts~22 are due to a mispointing of the
satellite. The straight lines after 45$\mu$m in the spectra of
IRAS09425-6040, HD~45677 and MWC~300 are the connections with
the IRAS datapoints.} \label{fig:diskcont}
\end{figure*}

\begin{figure*}[t]
\centerline{\psfig{figure=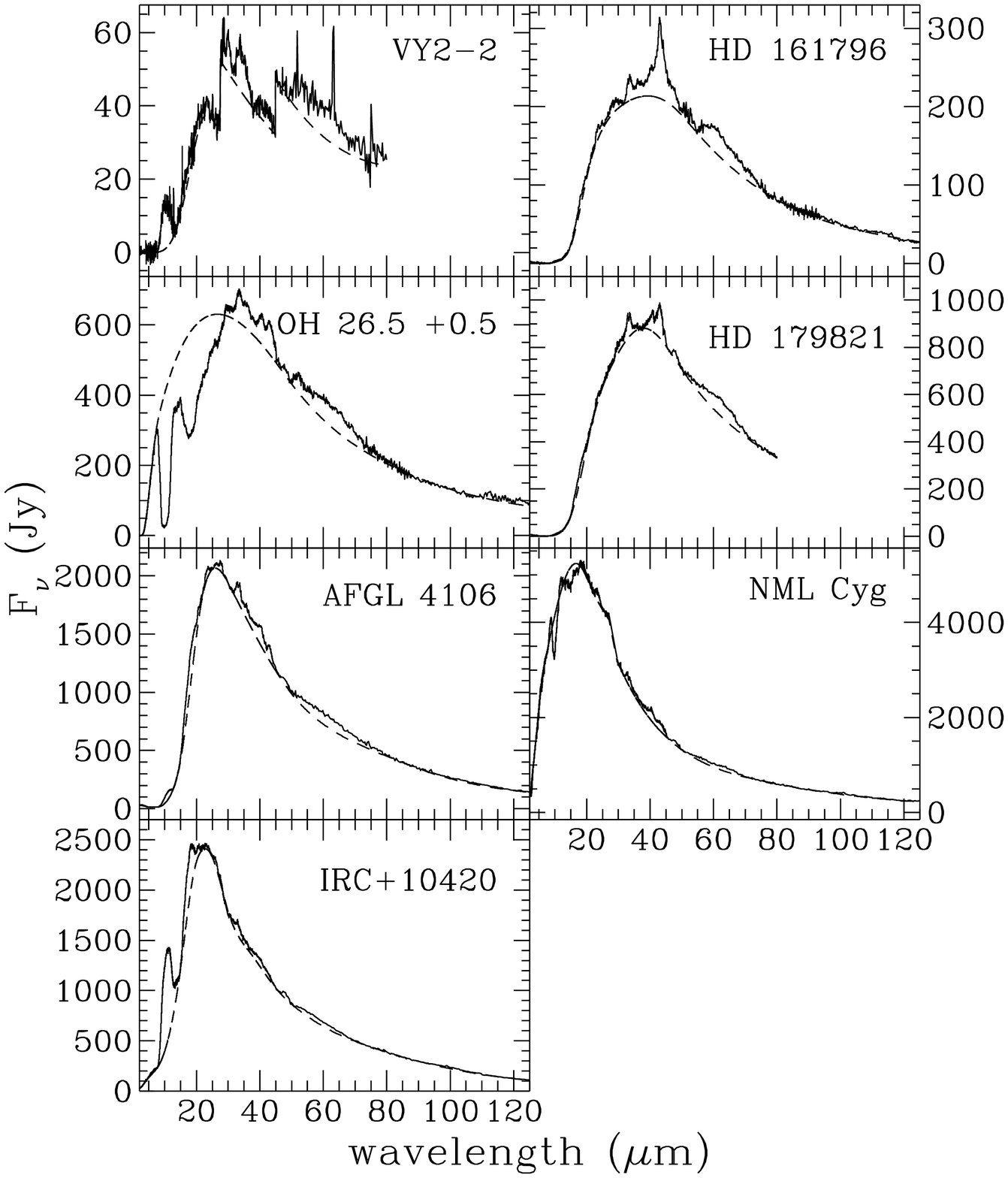,width=143mm}}
\caption[]{\small The spectra of the outflow sources (solid line)
and their eye-ball-spline-fit continuum (dashed line). The jumps
in the spectrum of Vy~2-2 are due to a mispointing of the
satellite.} \label{fig:outflowcont}
\end{figure*}


\subsubsection{SWS and LWS data reduction}

For a description of the SWS instrument and its operating modes we
refer to de Graauw et al. (1996). The spectra used in this study
were reduced using version 7.0 of the SWS offline processing software.
For a description of flux and wavelength calibration
procedures, see Schaeidt et al. (1996) and Valentijn et al.
(1996). We used standard procedures for flatfielding, sigma-clipping
and rebinning of the signals from the 12 detectors
We removed the main fringes in SWS band 3 (12.0 -- 29.5~$\mu$m). The final
resolution ($\lambda/\Delta\lambda$) of the spectrum depends on
the observing mode and the quality of the spectrum, but was
typically around 250 for a speed 1, 300 for a speed 2, 500 for a
speed 3, 750 for a speed 4 AOT01 and 1500 for an AOT06
observation.

The LWS instrument and operating modes are described in Clegg et
al. (1996).  For a description of the flux and wavelength
calibration procedures see Swinyard et al (1996). The LWS spectra
were reduced with the LWS offline processing software version 7.0,
and further processed using ISAP. The reduced and rebinned LWS
spectra of NGC~6302 used in this study were taken from Molster et al.
(2001b); those of OH~26.5+0.6 from Sylvester et al. (1999); those of
AFGL~4106 were taken from Molster et al. (1999b); for HD~44179 from
Hony et al. (in preparation), for HD~161796 from (Hoogzaad et al.,
in preparation), and for MWC~922 from Sylvester (private
communication). We verified that these LWS spectra were reduced in
a homogeneous way. The other LWS spectra were only sigma-clipped
and rebinned and mainly used as a reference for the adjustment of
SWS band 4 and the placement of the continuum (see
section~\ref{sec:cont}).

To match the individual sub-bands with each other, we used
multiplications, when we expected the flux calibration to be the most
important source of error, and a linear shift when we expect the the
dark curents to be the main source for the discrepancy between the
different bands. In all cases we tried to minimize the necessary shifts.

\begin{figure}[t]
\centerline{\psfig{figure=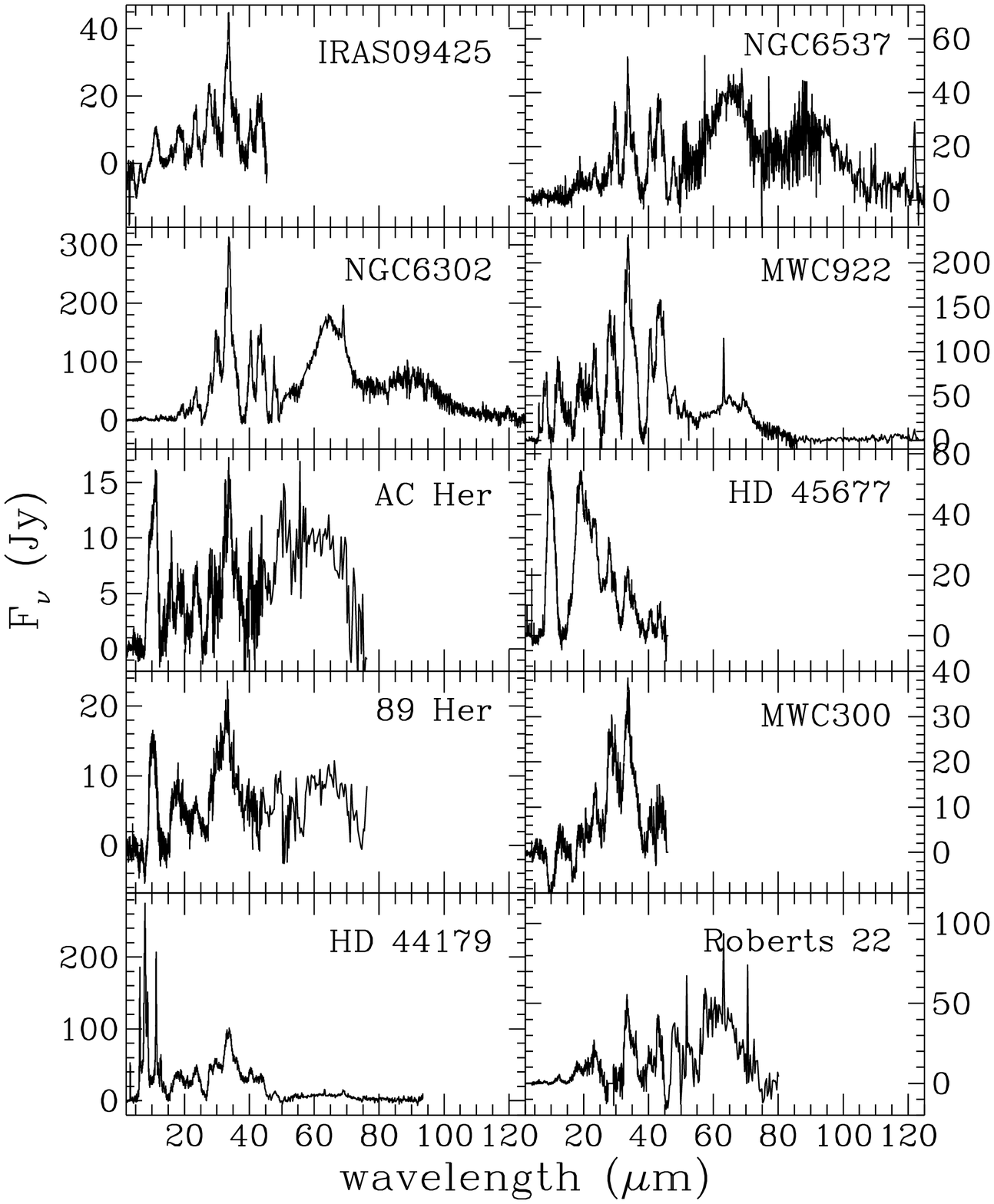,width=88mm}}
\caption[]{\small The continuum subtracted spectra of the disk sources.}
\label{fig:disksub}
\end{figure}

\begin{figure}[t]
\centerline{\psfig{figure=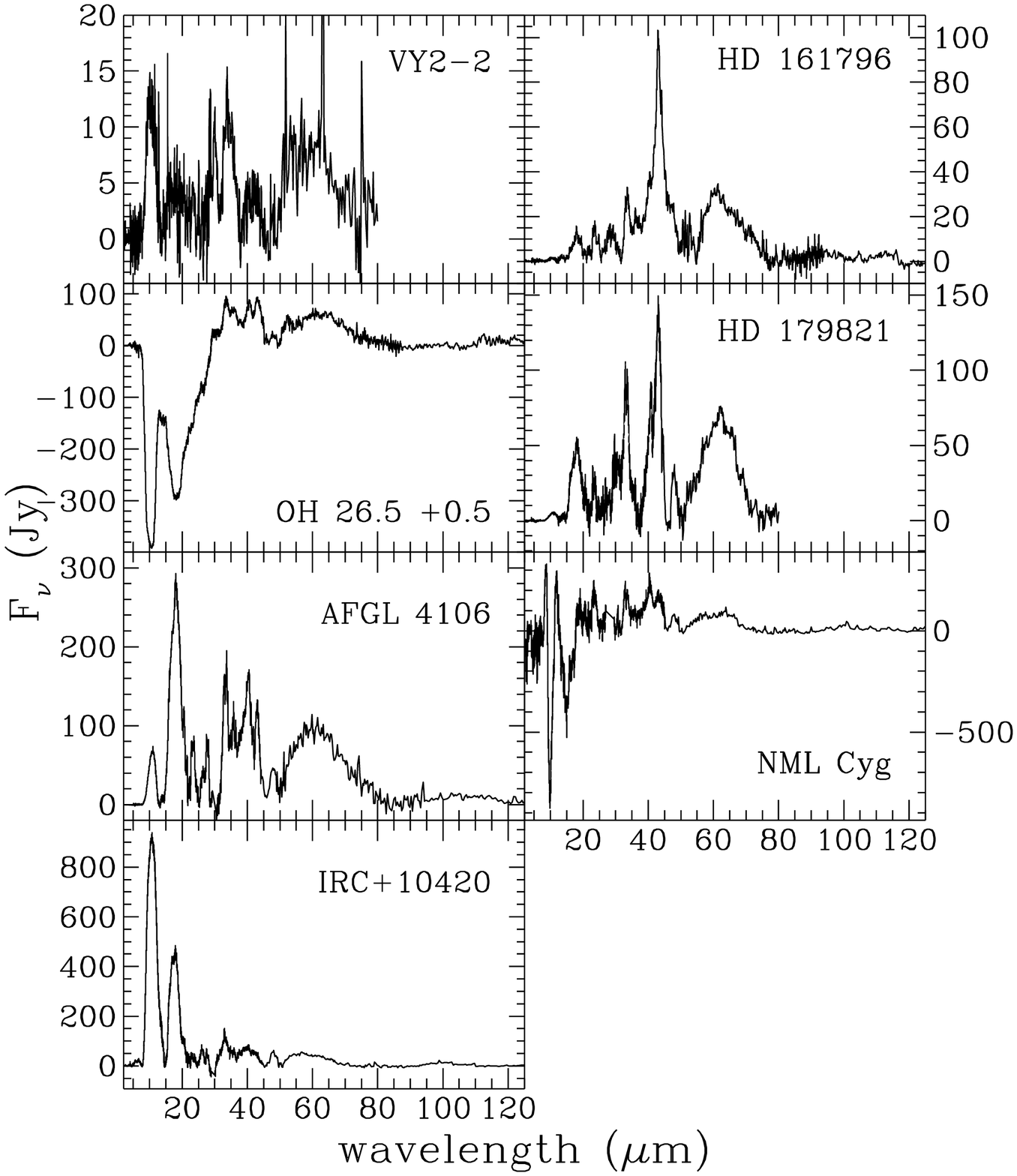,width=88mm}}
\caption[]{\small The continuum subtracted spectra of the outflow sources.}
\label{fig:outflowsub}
\end{figure}


\subsubsection{Determination of an underlying continuum}

\label{sec:cont}

In order to enhance the visibility of the different features we
have defined a continuum for the spectra in
Fig.~\ref{fig:diskcont} and Fig.~\ref{fig:outflowcont}. We have used an
eye-ball spline-fit continuum, maximizing the
continuum and still be smooth (no sudden changes in the slope), both in
$F_{\nu}$ and $F_{\lambda}$. We emphasize that this continuum has
not necessarily a physical meaning but is only used here to enhance the sharp
features on top of the spectrum. In principle, the strength of the features
could be underestimated and there is a possibility that very broad
features are treated as continuum. Whenever possible we tried to
use the whole wavelength range (SWS + LWS) to determine the
placement of the continuum. We have purposely not used (modified) 
blackbody-fits, because most spectra could not satisfactory be fitted by one 
(modified) blackbody (BB). 
This would have forced us to use multiple (modified) 
blackbodies, which would make it as arbitrary as the eye-ball fit and we had 
the feeling that we could be more consistent between the different spectra
bye the eye-ball method than with BB-fitting. Finally, the BB-fitting results
in a quite artificial physical parameter, the temperature, which is not always
directly related to the real temperature of the dust species. 
In order to prevent confusion about this parameter, we prefered the 
meaningless eye-ball continuum.

In two sources, Roberts~22 and Vy~2-2, flux jumps were found due
to a pointing offset of ISO. Pointing offsets result in flux jumps
due to the change in aperture size of SWS with wavelength (see De
Graauw et al. (1996)). In these cases we have fitted separate
continua to each part of the spectrum, where we took into account
the relative slope in the other parts of the spectrum. However,
this may affect the shape of the complexes whose wavelength
coverage extends over sub-bands with different SWS apertures.

The Planetary Nebulae NGC~6537 and NGC~6302 have strong emission
lines that hamper the determination of the dust emission bands. We
decided to remove most of these by subtracting a Gaussian fit to
the (unresolved) lines. For very strong lines (some could be 100
times the continuum level) the subtraction procedure resulted in
excessive noise after removal of the Gaussians. In these cases, we
removed the noisy part of the spectrum.

\subsubsection{The "final spectra": disk and outflow sources}

After some study of the stars in our sample, it became apparent
that there is a wide range in band strengths of the different
sources. We have accordingly ordered the spectra in sequence of
decreasing strength of the 33.6 $\mu$m band 
with respect to its local continuum. As we will show later
(see Sect.~\ref{sec:stars}) this naturally divides the spectra
into objects that have a disk-like distribution of dust, and
objects that have a more spherical dust distribution. We will use
this division into "disk" and "outflow" sources in the remainder
of this paper, and will group our sources accordingly. In
Fig.~\ref{fig:diskcont} and Fig.~\ref{fig:outflowcont} the final
spectra are shown, and Figs.~\ref{fig:disksub} and
\ref{fig:outflowsub} show the continuum subtracted spectra, in
both cases divided into disk and outflow sources.

\subsection{Definition of complexes}

\begin{figure*}[t]
\centerline{\psfig{figure=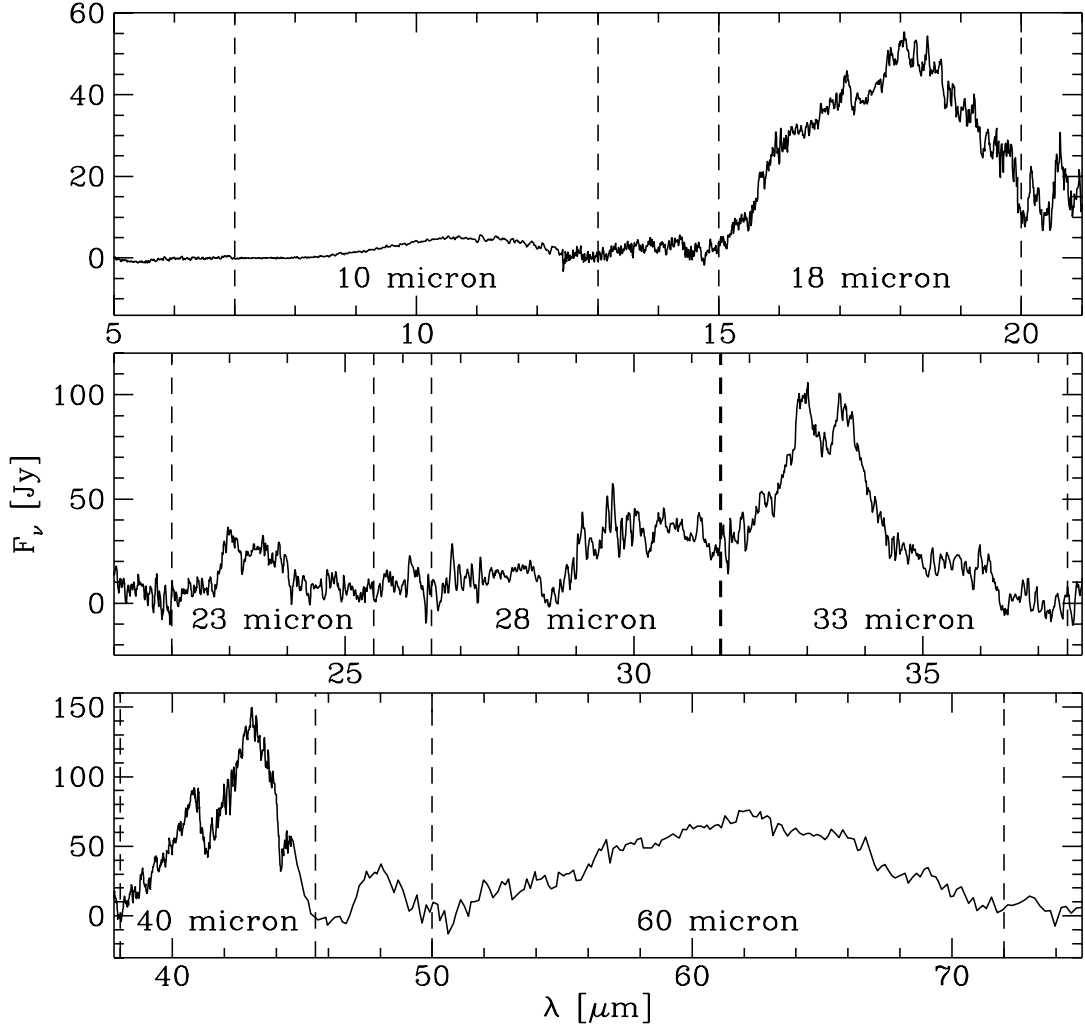,width=143mm}}
\caption[]{\small The continuum subtracted spectrum of HD~179821. We indicated
the range of the different complexes.}
\label{fig:complexdef}
\end{figure*}

A first inspection of the continuum subtracted spectra of our
programme stars indicates that there is a multitude of emission
bands. These bands are not evenly spaced in the spectrum, but tend
to concentrate in a number of wavelength regions. Within these
fairly narrow wavelength ranges, severe blending hampers the
measurement of individual components. As an example we show in
Fig.~\ref{fig:complexdef} the continuum subtracted spectrum of
HD~179821. There is a clear concentration of bands near 10, 18,
23, 28, 33, 40 and 60 $\mu$m.  We therefore decided to define 7
regions, that we will refer to as "complexes", in which features
tend to concentrate, and which we labeled with their central
wavelength: the 10, 18, 23, 28, 33, 40 and 60 micron complex (see
Table~\ref{table:complexes}). We will refer to an individual
emission band within a complex as "feature" or "band". Note that
individual bands can also occur outside a complex, and we will
refer to these as "features" or "bands" as well.

\subsection{Mean spectra}

We have derived `mean' spectra for each complex, by extracting the
wavelength range around and including the complex from the
continuum subtracted spectra for all stars. All these complex
spectra were added, using a weighting factor proportional to the
S/N of the spectrum, to create a `mean' spectrum of each complex.
This was done for both the outflow and the disk-sources.

We could only use 89~Her, AC~Her and HD~45677 to calculate the mean
10 micron disk source complex, because all other disk sources did
not show bands due to oxygen-rich dust. This is due to a
contribution from hot, carbon-rich dust dominating the spectrum,
or because the oxygen-rich dust was too cold to produce
significant emission near 10 $\mu$m. For the outflow sources we
excluded OH~26.5+0.6 and NML~Cyg, because they show the 10 $\mu$m
amorphous silicate band in strong absorption.

The source OH~26.5+0.6 has also been excluded from the average spectra
of the 18 and 23 micron (outflow) complexes because the spectrum is in 
absorption. NML~Cyg shows amorphous silicates in absorption in the 18 micron 
complex,
but the crystalline silicates is already in emission. To avoid problems we
have also excluded this object from the average outlfow spectrum for the
18 micron complex.

In Paper~II an extensive discussion will be presented on
the properties of the mean complexes and bands, for both outflow
and for disk sources. Here we will compare the
individual complex or band positions and shapes with the mean. In
Paper~II we will also present average properties and identify the
bands using laboratory data of several materials. Most of the
bands can be identified with forsterite (Mg$_2$SiO$_4$; the
Mg-rich endmember of the crystalline olivine series) and
enstatite (MgSiO$_3$; the Mg-rich end member of the crystalline
pyroxene series). About 20 percent of the bands still lack an
identification.

\begin{table}
\caption{The definition of the different complexes}
\begin{tabular}{|c|c|c|}
\hline
complex & $\lambda_{\mbox{min}}$ & $\lambda_{\mbox{max}}$ \\
name    &   $\mu$m      &   $\mu$m  \\
\hline
10  & 7     & 13    \\
18  & 15    & 20    \\
23  & 22    & 25.5  \\
28  & 26.5  & 31.5  \\
33  & 31.5  & 37.5  \\
40  & 38    & 45.5  \\
60  & 50    & 72    \\
\hline
\end{tabular}
\label{table:complexes}
\end{table}

\subsection{Measurement of solid state features}

In order to get a consistent view of the strength and position of
the features found in our spectra, we have fitted Gaussians to
features, or multiple Gaussians in the case of complexes, with the
{\em ISAP-line~fit} programme. The fits have been applied to the
original, i.e. non-continuum subtracted, spectrum in
$F_{\lambda}$. In this way we prevent very broad features,
extending over more than one complex, from contributing to the individual
bands.
We determined a local continuum, often a third order polynomial,
for each spectral section and estimated the error on the measured
wavelength, FWHM, peak to continuum ratio and integrated band
flux for these features by multiple independent fits to the (local
continuum subtracted) spectrum. For this purpose we varied the
defined continuum, since this is likely to have the largest
influence on the derived strength of the features. The drawback of this
method is that the uncertainties in the spectrum are not taken
into account. Therefore in a few low S/N cases the derived
1~$\sigma$~errors for the wavelength position (see Appendix~A),
might be too low and should be a factor 2 to 3 higher. Still, for
most cases the wavelength determination is indeed quite accurate.
Sometimes the error in the strength of the features may have been
overestimated, resulting in band strengths that are less than 3
sigma over the noise. However, careful inspection of the
individual cases convinced us of their reality. Finally, we note
that non-Gaussian shapes occur for certain bands, causing a
systematic but reproducible error.

\begin{table*}
\caption[]{\small Overview of the different
features seen in our sample (1=IRAS09425-6040; 2=NGC~6537;
3=NGC~6302; 4=MWC~922; 5=AC~Her; 6=HD~45677; 7=89~Her; 8=MWC~300;
9=Vy~2-2; 10=HD~44179; 11=HD~161796; 12= OH~26.5+0.6; 13=Roberts~22;
14=HD~179821; 15=AFGL~4106; 16=NML~Cyg; 17=IRC+10420). The
wavelengths are in $\mu$m. A `?' indicates an uncertain
detection. Blends are indicated by `\}'. $^1$ indicates that this
features is only found in 1 star. 17.5 and 17.5b indicate
respectively a narrow (enstatite) band and a broad (amorphous
silicate) band. The features at 11.05, 13.5 and 32.97 $\mu$m are
instrumental artifacts. The status of the 13.8 and 14.2 micron
features is unclear, they also seem to suffer from instrumental
artifacts, but a contribution from enstatite cannot be excluded.}
\psfig{figure=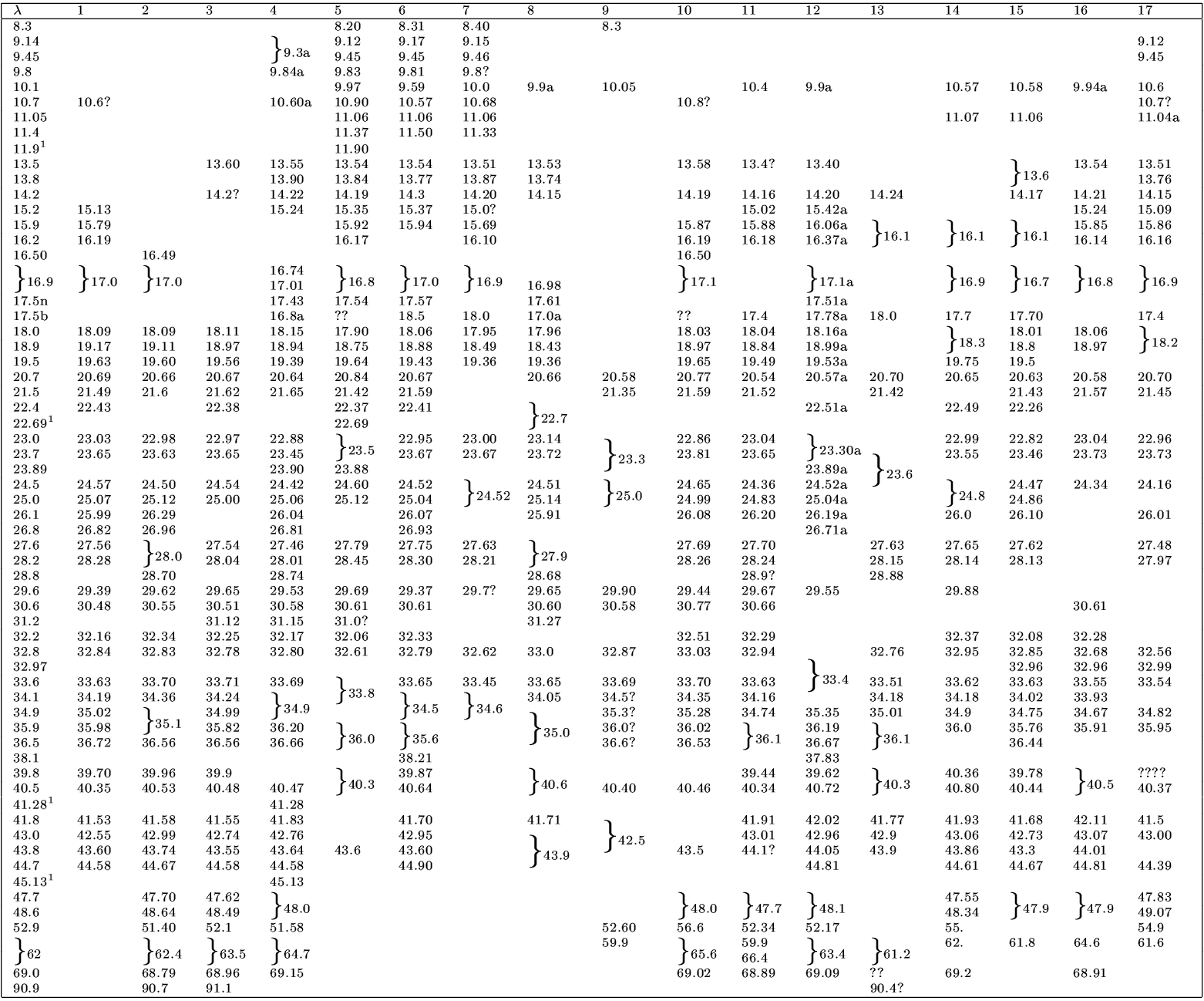,width=180mm,angle=90}
\label{tab:overview1}
\end{table*}


\section{Description of the programme stars}
\label{sec:stars}

Before discussing the bands and complexes, it is useful to
summarize the properties of our programme stars. This is relevant
when one tries to correlate the observed dust spectrum with other
properties of the object. We have searched for correlations
between the relative (to the local continuum) strength of the 
33.6~$\mu$m band (identified with
forsterite, Waters et al. 1996), which is detected in all programme
stars, and other properties of the star or its circumstellar
shell. No obvious correlation could e.g. be found with the
temperature of the star, or with the colour temperature of the
underlying continuum (we will return to these points in
Paper~III). However, a correlation between the geometry of the
circumstellar shell and the (relative) strength of the 
33.6 $\mu$m feature
exists (Molster et al. 1999a). Objects with strong crystalline
silicate emission bands have a highly flattened circumstellar
geometry. This geometry is either derived from direct imaging at
optical or near-IR wavelengths (e.g. in the case of the Red
Rectangle: Osterbart et al. 1997; Monnier et al. 1997), or it
is inferred from the shape of the spectral energy distribution
(SED; e.g. in the case of IRAS09425-6040, Molster et al. 2001a).

Because of these differences we have divided the stars into 
disk and outflow sources and ordered them in decreasing 
33.6 micron band strength.
When discussing individual stars below, present evidence for
either a disk or a more spherical geometry for the dust envelope.

\subsection{Disk sources}
\subsubsection{IRAS09425-6040}
\label{sec:iras09}

This carbon-rich AGB star is one of the most intriguing stars in
our sample. The source is discussed extensively by Molster et al.
(2001a). It was classified as a post-AGB object based on its IRAS
colours. The ISO-SWS spectrum however revealed a J-type
carbon-star at wavelengths below 15~$\mu$m (Molster et al. 2001a),
while at longer wavelengths highly crystalline O-rich dust
dominates the spectrum. It has the highest contrast crystalline circumstellar
dust observed so far in the complete ISO sample (Molster et al.
1999a). The oxygen-rich dust probably originates from a previous
mass loss phase before the star turned C-rich. Somehow this O-rich
material was stored around the star. The IRAS 60 $\mu$m
measurement suggests that the spectral energy distribution is flat
up to that wavelength. Both the shape of the energy distribution
and the complex chemistry point to the presence of a disk instead
of an ordinary outflow. Molster et al. (2001a) propose that
IRAS09425-6040 is a binary and an evolutionary progenitor of the
Red Rectangle.

The SWS spectrum of this source has already
been shown by Molster et al (1999a;2001a).

\subsubsection{NGC~6537}

NGC~6537 is an extreme Type I planetary nebula (PN), which bears a
lot of similarities to NGC~6302.  It has a bipolar outflow
structure, which is probably caused by a disk (Cuesta et al.
1995). He and N are overabundant while C is severely underabundant,
the typical characteristic of Type I PN, but for both NGC~6537 and
NGC~6302 these abundances are more extreme than in other Type I
PN. Both NGC~6537 and NGC~6302 have very high excitation lines, e.g.
[Si VI] (Ashley \& Hyland, 1988) which are not found in other PNe.
NGC~6537 is probably the evolutionary product of the most massive
stars which end as a white dwarf and may have had an initial main sequence 
mass in the range 5-7 M$_{\odot}$.
Shocks are expected to be present due to the interaction of the
fast wind and the disk. The infrared spectrum shows evidence for
both carbon-rich dust (PAHs) and oxygen-rich dust (silicates).

\subsubsection{NGC~6302}

NGC~6302 or the `Butterfly Nebula' is one of the infrared brightest
PN. It has one of the hottest central stars known with T$_{\rm
eff} = 250000$~K (Casassus et al. 2000). Together with the high
abundance of nitrogen and helium in the nebula (Aller et al. 1981)
this points to a massive progenitor. It is a Type I PN, with a
highly bipolar outflow and a thick dusty `disk'.
The spectrum shows highly excited forbidden
emission lines (up to [Si VII] for NGC~6302).
Roche \& Aitken (1986) already detected the PAH-features at 8.6
and 11.3 micron, and Cohen et al. (1989) presented evidence for
the 6.2 and 7.7 micron PAH feature. Both observations indicate the
presence of carbon-rich material. On the other hand Payne et al.
(1988) detected an OH-maser, normally associated with oxygen-rich
environments. This dichotomy in the dust is also present in our
ISO spectrum, where we find the PAH-features and the crystalline
silicates. The presence of an equatorial disk-like structure
has long been known (Meaburn \& Walsh 1980; Lester \& Dinerstein
1984; Rodriguez et al 1985) and can also be seen as a dark (dust)
lane on images of this PN.
The 30 to 45 $\mu$m part of the spectrum has first been published
by Waters et al. (1996), Beintema (1998) showed the complete SWS
spectrum while the LWS spectrum was shown by Barlow (1998).

The full ISO spectrum of NGC~6302 has been discussed by Molster
et al. (2001b).

\subsubsection{MWC~922}

The evolutionary status of MWC~922 is unclear, it was placed among
the unclassified B[e] stars by Lamers et al. (1998). Both a
pre-main sequence (Th\'{e} et al 1994) as well as a post main
sequence status (Voors 1999) have been suggested.
The distance to this object is unknown. The star is projected on
the Ser OB1 association, which is at 1.7 kpc. However, there is no
independent evidence that MWC~922 should be associated with Ser
OB1. Simon \& Dyck (1977) found an infrared excess at 20 and 25~$\mu$m. 
Meixner et al (1999) observed this object at 8.2 and 12.2~$\mu$m 
and marginally resolved it. Its broad spectral energy
distribution and rather high mm continuum flux are not compatible
with a spherically symmetric (continuous) dusty outflow, 
and we classify the star as disk-like. Spectra taken with the Kuiper 
Airborne Observatory (KAO) data show strong PAH emission (Cohen et al.
1989), which usually is found in carbon-rich environments. The ISO
spectrum shows that oxygen-rich material is also present.

The continuum divided LWS part of the spectrum was
already shown by Barlow (1998), while the SWS part was shown by
Voors (1999).

\subsubsection{AC Her}

AC Herculis is an RV Tauri star, with a stable pulsation period of
75.47 days (Zsoldos 1993). The star was
found to be a binary with a period of 1194 days (van Winckel et
al. 1998). The narrow velocity width of the CO rotational line emission
(Jura et al. 1995) suggests Keplerian rotation in a disk rather than
outflow. Also, very strong mm continuum flux (van der Veen et al. 1994)
indicates the presence of large grains, which suggests a long
storage time and therefore a (planet forming?) dust disk. The
infrared spectrum of the binary star AC Her is strikingly similar
to the spectrum of the comet Hale Bopp (Molster et al. 1999a). The
SWS spectrum longwards of 7$\mu$m has already been shown by
Molster et al (1999a) and is briefly discussed by van Winckel et al. (1998).
Recently, the evidence for a disk has been provided by Jura et al. (2000).
They found a dust ring with a radius of approximately 300 AU
in an image taken at a wavelength of 18.7~$\mu$m.

\subsubsection{HD~45677}

HD~45677 is a well studied B2 star whose evolutionary status
is still unclear. A pre-main sequence nature has
often been suggested for this star, however its isolated position
(HD~45677 is not associated with nebulosity) and the absence of
Algol-like variations and `blueing' effect make this
questionable. Lamers et al (1998) also discuss this star and place it 
amongst the `unclassified' B[e] stars. They propose that it is an extreme
example of a classical Be star.
Polarization measurements show that the circumstellar material
is located in a disk (Schulte-Ladbeck et al. 1992). Also, the
strong absorption cores of the Na I {\em D} and Ca II {\em K} lines
indicate the presence of an optically thick disk at zero velocity
(de Winter \& van den Ancker 1997). This disk was already present
before 1950 when a significant disruption took place. After this
event, the star dimmed up to $\approx 2$ mag in 1981 without
significant change in the colours. This is most easily explained
by the production of large (= grey) particles 
(de Winter \& van den Ancker 1997).

The spectrum of this star has already been published
by Voors (1999) and Malfait (1999), who also modeled it.

\subsubsection{89~Her}

89~Her is a high galactic latitude F2~Ibe supergiant with a temperature of
about 6500~K (Waters et al. 1993).
Waters et al. (1993) confirmed the binary hypothesis of
Arrelano Ferro (1984), finding an orbital period of 288.4 days.
The CO(1-0) and CO(2-1) line observations show a narrow ($< 1$ km/sec)
central spike on top of a broader ($\approx 8$ km/sec) weak
component (Likkel et al. 1991). This profile is very different
from the profiles seen from detached shells, and may represent gas
orbiting in a flattened disk-like structure. Several
other arguments for the presence of a disk, such as the lack of
energy balance between the UV and IR, are given by Waters et al.
(1993). Alcolea \& Bujarrabal (1995) imaged 89~Her
in CO(1-0) and found an outer shell, which likely originates from
a heavy mass loss period experienced by the star in the past. This
eruption of mass was probably triggered by the companion by a Roche lobe
overflow or even by a common envelope phase, and ended the
AGB evolution of 89~Her. It is likely that during this period
also the disk was formed.

\subsubsection{MWC~300}

MWC~300 is classified as a B1 Ia$^+$ hypergiant by Wolf \& Stahl (1985).
With a luminosity of $L_* \approx 5 \times 10^5 $L$_{\odot}$ the star is
at 15.5 kpc and about 500 pc above the Galactic plane. Henning et al. (1994)
detected MWC~300 at sub-mm wavelengths and pointed out that for the
derived distance of 15.5 kpc (Wolf \& Stahl, 1985) the total dust
mass in the circumstellar envelope would be in the order of 300
M$_{\odot}$, which seems unrealistically high. Different attempts
were made to spatially resolve this object (Skinner et al. 1993;
Ageorges et al. 1997; Leinert et al. 1997; Pirzkal et al 1997),
but they were all unsuccessful. This indicates that the dust is
indeed circumstellar and not associated with nearby nebulosity. A
more realistic circumstellar dust and gas mass of a few
M$_{\odot}$ would result in a much smaller distance (an order of
magnitude) and therefore a luminosity of about 10$^4$ L$_{\odot}$.
This new luminosity together with its extended atmosphere
characteristics would classify this star as an evolved low mass
star (Voors 1999). Winckler \& Wolf (1989) and Hamann and Persson
(1989) both argue that MWC~300 is surrounded by a (slowly
expanding) disk-like structure and a low density wind near the poles.

\subsubsection{HD~44179}

HD~44179 is the central star in the X-shaped Red Rectangle nebula
(Cohen et al. 1975).
It is an A type supergiant with an effective temperature of about
7500~K in a binary system (P=318 days), surrounded by a
circumbinary disk (Waelkens et al. 1996), which we see (almost)
edge on. The central star is heavily obscured by this disk and
only seen in reflection by scattering lobes below and above the
plane of the disk. The optically thick disk has been imaged with
high resolution in the optical and near-IR by e.g. Roddier et al.
(1995) and Osterbart et al. (1997). The CO(1-0) and
CO(2-1) show very narrow ($\approx 5$ km/sec) line emission (Jura
et al. 1995), atypical for detached AGB remnants.
The mm and cm continuum flux of this source is
rather high and suggests the presence of large (mm-sized) grains
(Jura et al 1997). These grains are likely to be formed in the
long-lived circum-binary dust disk. The similarities with the
disks around young stars lead to speculations about possible
planet formation around this evolved star (Waters et al. 1998).
The detection of a
mysterious dust clump around HD~44179 by Jura \& Turner (1998)
feeds this interesting speculation.
The star has both carbon-rich, as evidenced by the PAH-features, and
oxygen-rich dust, as evidenced by the crystalline silicate features.
The PAHs are predominantly present in the scattering lobes, while
the crystalline silicates are expected to be present in the disk
(Waters et al. 1998).

The SWS part of the spectrum was already shown by Waters et al. (1998).

\subsubsection{Roberts~22}

Roberts~22 is a bipolar reflection nebula, whose evolutionary status has
recently been determined as being post-AGB (see e.g. Sahai et al. 1999).
Allen et al. (1980) found that the central star is
completely obscured by a central dust lane and they determined the
spectral type of this object from the two reflection lobes, which
gave identical spectra (A2 I). At a distance of 2 kpc (Allen et al.
1980; Sahai et al. 1999) its total luminosity
is $\approx 3 \times 10^4$ L$_{\odot}$. From IR and the CO
emission line data the progenitor (AGB) mass-loss rate was about
$10^{-4}$ M$_{\odot}$/yr.
Roberts~22 has a time-variable OH maser (Allen et al. 1980),
which is mainly located in the central waist, but is also seen in
the northern and southern scattering lobes (Sahai et al. 1999).
The velocity distribution of the OH masers might be interpreted as
a rotating disk, which is seen almost edge-on. The fact that the
OH is also seen in the the lobes suggests that the disk is being
disrupted, probably due to the fast wind (450 km/sec) seen in
H${\alpha}$ (also time variable) arising from the central star.
The spectrum of Roberts~22 also shows both PAH-features as well
as the infrared features of crystalline silicates.

Part of the spectrum was shown by Molster et al. (1997).

\subsection{Outflow sources}

\subsubsection{Vy~2-2}

Vy~2-2 is classified as a compact planetary nebula surrounded by a
fossil molecular envelope from the progenitor AGB star (Jewell et
al. 1985 and references therein). Lamers et al. (1998) classified
this star as a compact planetary nebula B[e] star. This very young
PN has both an ionized zone and a neutral, molecular cloud. The
ionized nebula has been resolved as a thin shell extending to
$\approx 0^{''}\!\!\!.5$ both at 15 Ghz (Seaquist \& Davis 1983) and
in H${\alpha}$ (Sahai \& Trauger 1998). The inner radius was
estimated to be $0^{''}\!\!\!.2$. Only the blue shifted OH maser component
is detected in Vy~2-2. This is not in contrast with a homogeneous spherically 
symmetric outflow, because that scenario predicts that the red shifted 
component is absorbed (Seaquist \& Davis 1983).
Although the quality of the spectrum is low, due to mispointing,
we decided to keep it in our sample since it was the only young O-rich PN.

\subsubsection{HD~161796}

HD~161796 is a high galactic latitude F3~Ib supergiant. Skinner et
al (1994) determined a distance of 1.2~kpc and therefore a
luminosity of 3600~L$_{\odot}$. Its photosphere has low
metallicity and an enhanced nitrogen abundance, implying that it is
an evolved Population II object (Luck et al. 1990). The shape of
the very strong CO emission found by Likkel et al. (1991)
resembles the profile of mass-losing AGB stars and confirms 
the post-AGB nature of this object. The expansion
velocity determined from this line is about 12~km/sec, a typical
value for AGB stars. Skinner et al. (1994) resolved
the envelope around this source in the mid-IR (10.5 and 12.5
$\mu$m). They found an expanding dusty equatorial toroid in a
final phase of strongly enhanced, equatorially concentrated mass
loss, which stopped about 240 years ago. During this mass-loss
burst the mass-loss rate was about $3\times 10^{-4}$
M$_{\odot}$/yr. Meixner et al. (1999) confirmed these results.
The continuum divided LWS spectrum was already presented by Barlow (1998).

\subsubsection{OH~26.5+0.6}

OH~26.5+0.6 is an extreme OH/IR star, which shows evidence of two mass-loss
regimes: a superwind phase in which the mass-loss rate is
$\approx 10^{-4}$~M$_{\odot}$/yr which started about 200 years
ago, and an earlier AGB phase with a mass-loss rate of about $\approx
10^{-6}$~M$_{\odot}$/yr (Justtanont et al. 1994, 1996a). The
transition between these two phases was probably very short
($\Delta t < 150$ yr). The total mass lost during the superwind
phase has been estimated to be $\approx 0.1$ M$_{\odot}$
(Justtanont et al. 1996a). Infrared speckle
interferometry at 9.7 $\mu$m (in the silicate feature) gives an
angular size for the circumstellar dust shell of $0^{''}\!\!\!.50
\pm 0^{''}\!\!\!.02$, while the angular size of the dust shell
outside this feature (at 8.7 $\mu$m) is less than $0^{''}\!\!\!.2$
(Fix \& Cobb 1988). This difference is caused by the enhanced
opacity in the 10 $\mu$m silicate feature, therefore the dust seen
at these wavelengths is located at larger radial distances and is
cooler than the dust seen on either side of the feature.

The spectrum of OH~26.5+0.6 has already been shown by
Sylvester et al. (1999).

\subsubsection{HD~179821}

HD~179821 is a G5~Ia supergiant at a distance of about 6 kpc (Zuckerman \& Dyck
1986; Hawkins et al. 1995; Jura \& Werner 1999). At this distance the star
would have
a luminosity of about $3.1 \times 10^5$ L$_{\odot}$, far above the AGB
luminosity limit. If the star is indeed massive, the origin of the dust
envelope was probably the Red Supergiant phase.  In that case, the star may
evolve to the Wolf-Rayet phase, before exploding as a supernova.
The dusty envelope has been resolved at MIR wavelengths (Hawkins
et al, 1995; Jura \& Werner 1999). A ring-like structure was found
with an inner radius of $1^{''}\!\!\!.75$ ($1.6 \times 10^{17}$~cm
at a distance of 6 kpc). This implies that the mass loss burst has
stopped about 1500 years ago. In CO (Bujarrabal et al. 1992) and
NIR scattered light (Kastner \& Weintraub 1995) the dust envelope
was found to extend to at least $18^{''}$, indicating that the
mass loss burst lasted for at least 6000 yr. With a derived gas
mass loss rate of the order of $10^{-3}$ M$_{\odot}$/yr (Kastner
\& Weintraub 1995), about 6 M$_{\odot}$ was lost by the star
during this mass-loss burst. This value is similar to that found
by Jura \& Werner (1999) based on submillimetre measurements, if a
gas-to-dust ratio of 200 is assumed. From infrared imaging (Jura
\& Werner 1999) and maps in CO (Bujarrabal et al. 1992), it is seen
that the gas and dust distribution is not spherical.
Part of the spectrum of this source was already shown by
Waters et al. (1996).

\subsubsection{AFGL~4106}

AFGL~4106 has been discussed by Molster et al. (1999b), and is
a high mass (15 to 20 M$_{\odot}$) binary with two
almost equally luminous stars with temperatures of about
3750~K and  7250~K. The binary is located at a distance
of about 3.3 kpc. During about $4.3 \times 10^{3}$ years,
the more massive star in the system (now the warmer of the two)
had a gas mass-loss rate of about $9
\times 10^{-4}$ M$_{\odot}$/yr, which gives a total expelled mass
of about 4~M$_{\odot}$. This huge mass-loss burst ended about
450 years ago. Mid-IR imaging by Molster et al. (1999b) shows an
asymmetric detached dust shell.

\subsubsection{NML~Cyg}

NML~Cyg is an M6 supergiant at a distance of
1.8 to 2 kpc (Morris \& Jura 1983; Bowers et al. 1983) with a
luminosity of $\approx 5 \times 10^5$~L$_{\odot}$, implying a
main sequence mass of 50 M$_{\odot}$. The present-day mass loss rate
is  between $1.1 \times 10^{-4}$ (from OH and IR; Netzer \& Knapp 1987) 
and $1.8 \times 10^{-4}$M$_{\odot}$yr$^{-1}$ (from CO; Knapp et al. 1982). 
The dust shell has been partially resolved by Fix \& Cobb (1988) in the N-band
(10~$\mu$m). They found an outer radius of $0^{''}\!\!\!.37$, which
corresponds to about $ 10^{16}$~cm. Monnier et al. (1997) found
the same value for their outer dust shell, and detected a second
dust shell, located inside the first one, at about
$0^{''}\!\!\!.125$. The outflow velocity determined from the 1612 MHz
OH maser line is 27.7 km/sec (Bowers et al. 1983).
In the H$_2$O maser map an asymmetry
is found at subarcsecond-scales (Richards et al. 1996). This is
also seen in the OH maser maps on scales of a few arcsec,
indicating that this asymmetry comes from the inside and is not due to
external factors. Parts of the ISO spectrum were
already shown by Justtanont et al (1996b) and Waters et al. (1996).

\subsubsection{IRC+10420}

IRC+10420 is an A5 Ia$^{+}$ hypergiant with a large IR-excess. It is
found to be continuously evolving from a cooler (F8 Ia$^{+}$;
Humphreys et al. 1973) to a hotter spectral type (A5 Ia$^{+}$;
Oudmaijer 1998 and references therein). IRC+10420 is at a distance of $5 \pm
1$~kpc which gives it a luminosity of about $5 \times
10^{5}$~L$_{\odot}$, just below the Humphreys-Davidson limit. From
CO and OH measurements, an outflow velocity of 40 km/sec has been
derived (Nedoluha \& Bowers 1992; Bachiller et al. 1988; Lewis et
al. 1986). Kastner \& Weintraub (1995)
found from their near-IR polarimetric maps that the dust envelope
extends to $9^{''}$, which implies a dynamical age of the dust in
the outer layers of about 5000 years.
The mass loss rate that created the dust shell was of
the order of $10^{-3}$~M$_{\odot}$/yr, and a total of
5~M$_{\odot}$ of gas and dust is present in
the shell: a significant fraction of the 40~M$_{\odot}$ which it
probably had on the main sequence. Bowers (1984) observed this
star in the 1612 and 1667 MHz OH maser lines and found indications
for multiple shells in OH extending out to $4^{''}$. This suggests
that the mass loss went in bursts and was not constant in time.
The circumstellar dust shell
has been resolved in the mid-IR at 8.7, 9.8 and 20.6~$\mu$m (Fix \&
Cobb 1988; Jones et al. 1993; Meixner et al. 1999). In these
images an elliptical structure is found, which Oudmaijer et al.
(1994) attribute to a bipolar outflow, which is beamed into our
direction.

\section{Description of the complexes}
\label{sec:complexes}

In this section we discuss the individual spectra with respect to
the mean spectra and with respect to each other. In Fig.~\ref{fig:10mud} to
Fig.~\ref{fig:60muo} we show the emission (and in a few cases the absorption)
complexes of the individual stars
and compare them with the mean spectra for every complex. In these figures we
indicate the noise level and the wavelength (spread) of the individual
features. The noise level is the mean 3$\sigma$ noise level, but might change
with wavelength in one complex; e.g. due to a significant change in continuum
level or to changes in SWS-band and therefore detector characteristics. These
effects are particularly present in the 10 and 28 micron complexes. Here,
we limit the discussion to the featuress that we are confident are real. In
Appendix~A we discuss some more dubious detections and artifacts.

\subsection{The 10 micron complex}

\begin{figure*}[th]
\centerline{\psfig{figure=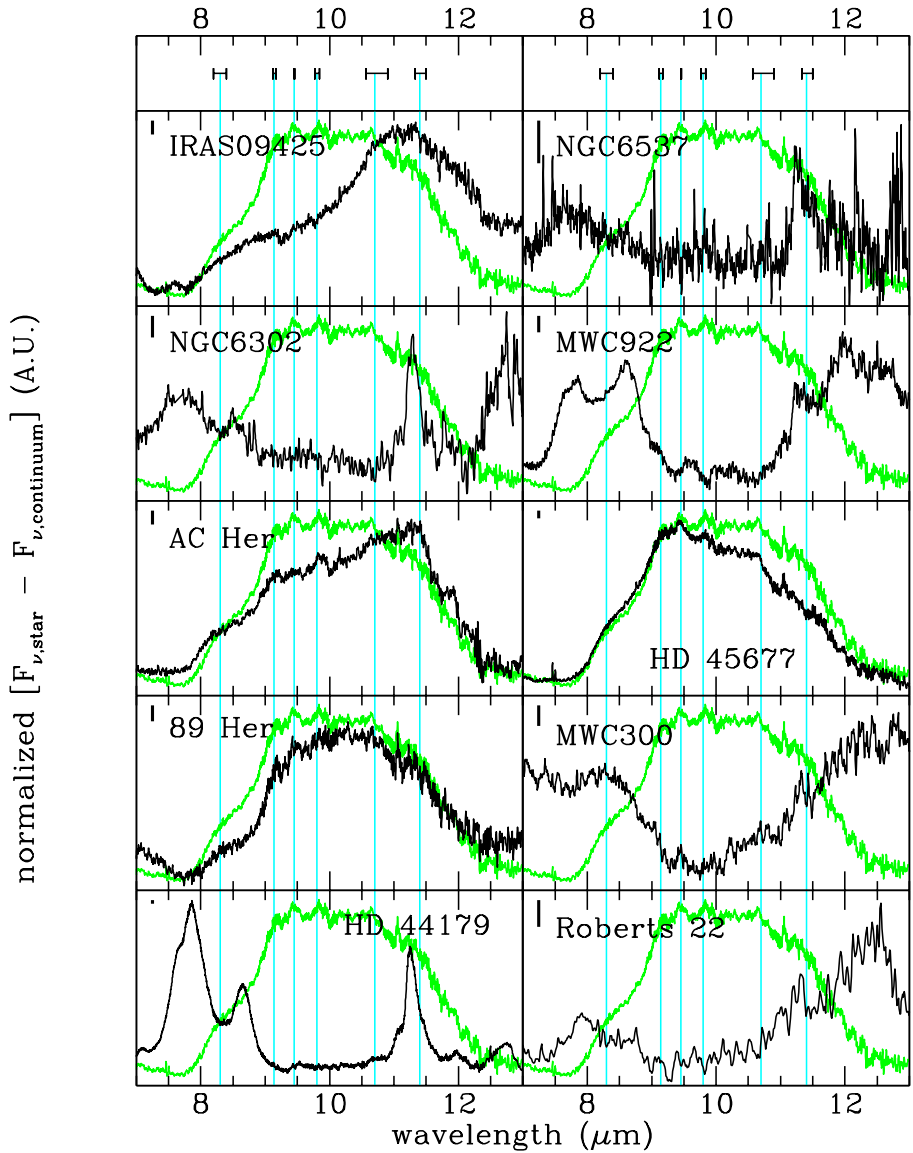,width=120mm}}
\caption[]{\small The 10 micron complex for the disk sources, together with
the mean disk spectrum (gray line). The thin
gray vertical lines indicate the mean peak positions of the features found,
while their range is indicated by the errorbar on top of the
plot. The thick line in the upper left corner in each panel
indicates the mean noise level.} \label{fig:10mud}
\end{figure*}

\begin{figure*}[th]
\centerline{\psfig{figure=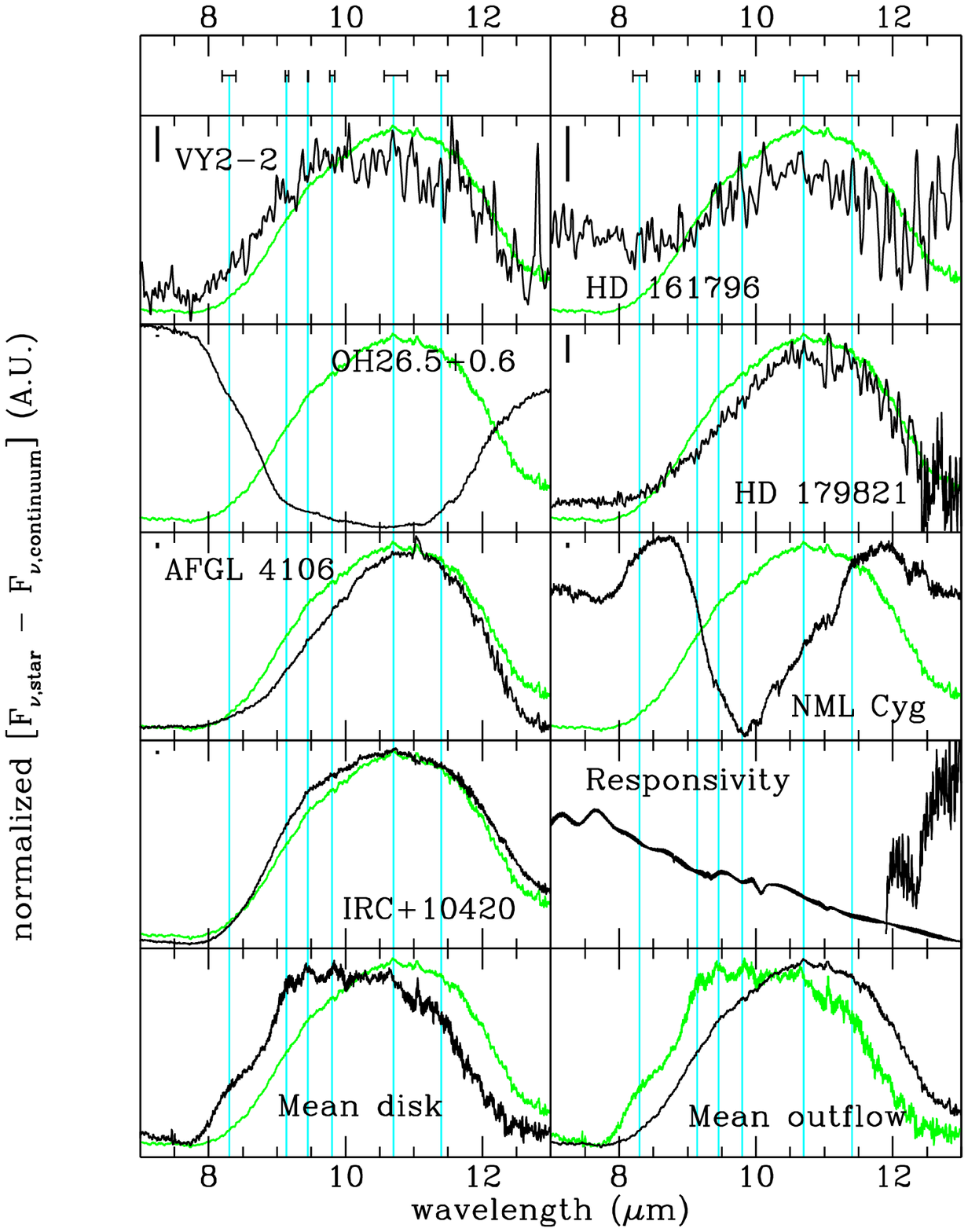,width=120mm}}
\caption[]{\small The 10 micron complex for the outflow sources,
plus the mean spectra and the responsivity profile of the
relevant bands in this wavelength region (SWS-band 2C and 3A).
The gray line is the mean outflow spectrum, except in the lower
right corner where it is the mean disk spectrum.
The vertical gray lines indicate the mean peak positions of the features
found, while their range is indicated by the errorbar on top of
the plot. The thick line in the upper left corner in each panel
indicates the mean noise level.} \label{fig:10muo}
\end{figure*}

The 10 micron complexes of the disk and outflow sources are shown in
Figs.~\ref{fig:10mud} and~\ref{fig:10muo}.

\subsubsection{Disk sources}
\label{sec:disk10}

{\em IRAS09425-6040}: The 10 micron complex of IRAS09425-6040 is
still dominated by the carbon star and its present outflow,
therefore C-rich molecules are found in absorption and prominent
SiC emission is seen in the 10 micron complex. No clear evidence
for crystalline material has been found.

{\em NGC~6537}: The 10 micron complex of NGC~6537 is dominated by PAH
features, and no evidence for (crystalline) silicate features has
been found.

{\em NGC~6302}: The 10 micron complex of NGC~6302 is dominated by
PAH features, and no evidence for (crystalline) silicate features
has been found.

{\em MWC~922}: The 10 micron complex of MWC~922 is dominated by PAH
features. Around 10 micron, absorption features seem to be present, which
line up very well with the position of crystalline silicate
emission features in other stars. Since crystalline silicates are
not abundant in the ISM, this must be circumstellar material, 
It is one of the few sources
where crystalline silicates are found in absorption in the 10
micron complex.

{\em AC Her}: The 10 micron complex of AC Her is one of the best
examples of the presence of crystalline silicates. All crystalline
silicate features are found in this source (in emission). This
complex is well fitted by only crystalline silicates (paper~III),
but an amorphous component cannot be excluded.

{\em HD~45677}: Crystalline silicate features are found in the 10
micron complex of HD~45677. The 10 micron amorphous silicate
feature peaks at shorter wavelengths than for e.g. AC Her.

{\em 89~Her}: 89~Her is the third source showing a clear evidence of
crystalline silicates in its 10 micron complex. Still, the
weakness of the 8.3 micron feature suggests that the complex is
dominated by the 10 micron amorphous silicate feature. The peak
position is red shifted with respect to the ISM absorption
feature (which is at 9.7~$\mu$m),
an indication of the presence of large grains.

\begin{figure}
\centerline{\psfig{figure=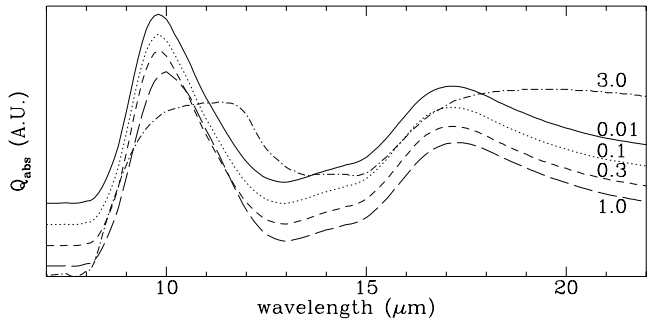,width=85mm,angle=270}}
\caption[]{The Q$_{\rm abs}$ for spherical amorphous olivine grains
(MgFeSiO$_4$; Dorschner et al. 1995) for different grain sizes:
0.01 $\mu$m (solid line), 0.1$\mu$m (dotted line),
0.3$\mu$m (short dashed line), 1.0$\mu$m (long dashed line), 3.0$\mu$m
(dashed dotted line).
The curves were scaled to get an equal strength around 10 $\mu$m and
then offset from each other.}
\label{fig:shift}
\end{figure}

{\em MWC~300}: The 10 micron
amorphous silicate absorption feature is remarkable. There
seems to be some substructure in the 10 micron complex, which
might be explained by crystalline silicate features in emission.
However, apart from the prominent 8.3 micron feature, these are
always seen only in one scan direction and are therefore not
trusted. The 8.3 micron feature is very similar to the ones in
AC~Her, 89~Her and HD~45677, all sources which have crystalline silicate
emission in the 10~micron complex. In MWC~300 it seems that there
is amorphous silicate absorption. The E(B-V) to this star is 1.19
(Voors 1999) which corresponds to an A$_V$ of 3.7, assuming a
normal reddening law. This would correspond to a $\tau_{9.7} =
0.25$ using the extinction law of Sandford et al. (1995). Based on the ISO
spectra we derived $\tau_{9.7} = 0.13 \pm 0.03$. The interstellar
extinction curve presented by Roche \& Aitken (1984) would give
$\tau_{9.7} = 0.2$, which still above the observed $\tau_{9.7}$.
The lower value from the ISO data suggests that the shape and
strength of this feature is not only due to interstellar
absorption. A natural explanation would be that the interstellar
absorption profile is filled in by circumstellar emission. The
central wavelength of this broad circumstellar emission feature is
likely shifted from the 9.7 $\mu$m interstellar absorption peak,
because the observed profile does not look like the typical
interstellar absorption profile. This shift to longer wavelengths
might be explained by large grains (see Fig~\ref{fig:shift}).

{\em HD~44179}: The 10 micron complex of HD~44179 is dominated by
PAH-features. No crystalline silicates seem to be present. A
possible crystalline silicate feature might be seen around 10.8 $\mu$m,
however at about the same position a feature is seen in genuine
PAH-sources, and attributed to a PAH cation.

{\em Roberts~22}: The 10 micron complex is dominated by PAH
features. The noise in the spectrum prevents us from reaching a
conclusion about the presence or absence of crystalline silicate
features.

\subsubsection{Outflow sources}

{\em Vy~2-2}: The 10 micron complex of Vy~2-2 is dominated by the
amorphous silicate feature, which is relatively broad. This
feature starts at relatively blue wavelengths which suggests that
the 8.3 micron feature is present. Because of the noise level
(increasing with wavelength) no statements could be made about the
presence of other crystalline silicate features. At 12.8 $\mu$m
the [Ne II] line is found.

{\em HD~161796}: As in all outflow sources, the 10 micron feature in
HD~161796 is dominated by the 10 micron amorphous silicate feature.
The signal to noise level is too low to detect the crystalline
silicate features.

{\em OH~26.5+0.6}: The 10 micron complex is dominated
by amorphous silicate absorption. Some structure seems
present, however the absorption pattern cannot be matched with the
features of crystalline silicates seen in other stars. Emission is
not expected here, since at longer wavelengths we do see
crystalline silicate features in absorption. The origin of the
substructure therefore remains unclear.

{\em HD~179821}: The 10 micron complex
is dominated by the red-shifted amorphous silicate
feature. The shape and position resembles the feature of AFGL~4106
where it has been attributed to the presence of large grains
(Molster et al. 1999b).

{\em AFGL~4106}: This complex is characterized by the 10 micron
amorphous silicate feature. The sharp feature at 11.06 $\mu$m is
instrumental; no correction, as in the other sources, has been
applied to this feature. There is no evidence for crystalline
silicate features.

{\em NML~Cyg}: The 10 micron feature is dominated by 
amorphous silicate absorption. The emission wings, at both sides
together with the location of the center of the absorption profile
suggest that it is mainly self absorption. A (small) contribution
from interstellar extinction cannot be excluded or confirmed. The
substructure at 9.39, 9.54, 10.07, 10.33, 10.76 and 11.0 $\mu$m is
due to gas-phase NH$_3$ (Yamamura private comm.).

{\em IRC+10420}: The 10 micron feature is dominated by
amorphous silicate emission. It peaks at a wavelength
significantly offset from the usual 9.7 micron feature. This might be due
to large grains, as in AFGL~4106. Fix and Cobb (1988) also
suggested the presence of large grains to explain the 10 over 20
micron amorphous silicate ratio (see Fig.~\ref{fig:shift}).
There are indications of
substructure in the amorphous silicate feature, which is also seen
in stars with prominent 10 micron crystalline silicate features.
Interestingly enough no 11.3 micron feature (attributed to
forsterite) is detected. This suggest that the forsterite
abundance and/or temperature is probably very low, in contrast
to the enstatite abundance. It should be noted that in the
rest of the spectrum the forsterite features are weak or not
detected.

\subsection{The 18 micron complex}

\begin{figure*}[th]
\centerline{\psfig{figure=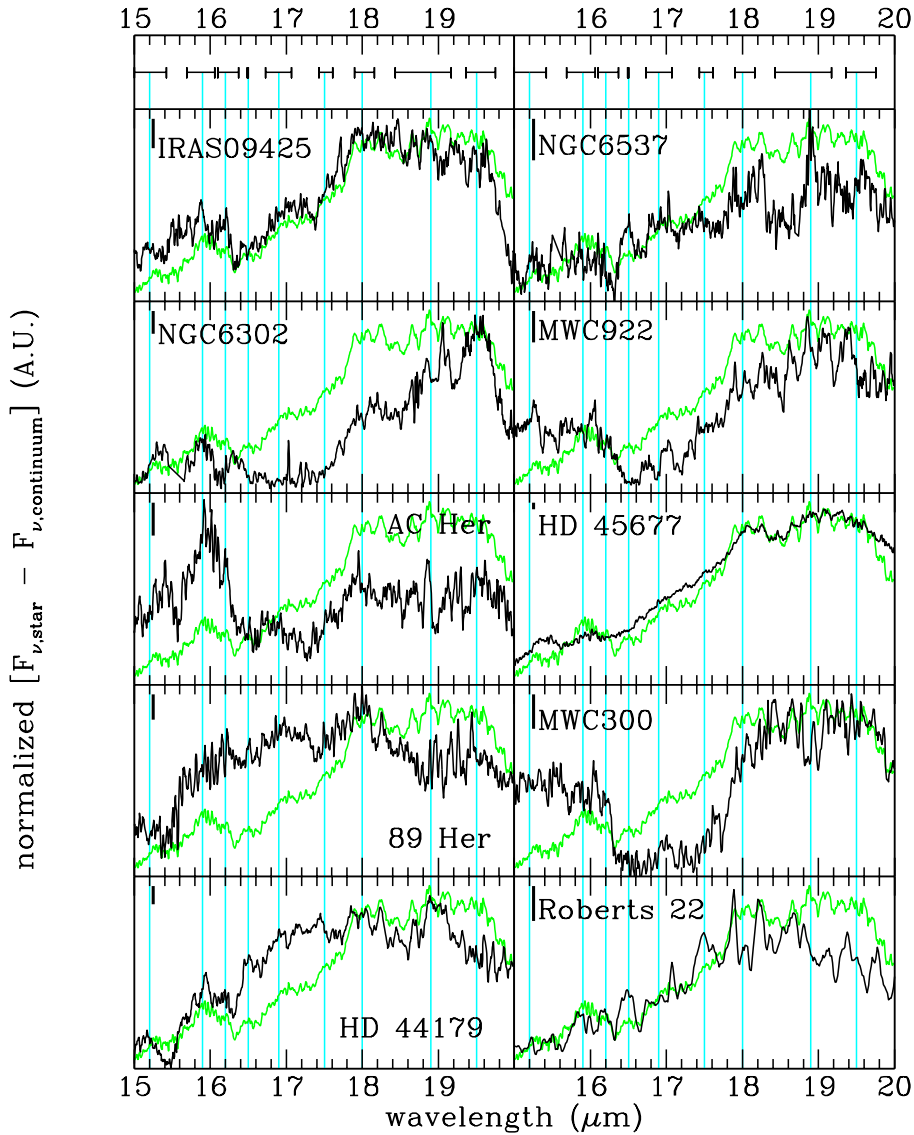,width=120mm}}
\caption[]{\small The 18 micron complex for the disk sources, together with
the mean disk spectrum (gray line). The vertical
gray lines indicate the mean peak positions of the features found,
while their range is indicated by the errorbar on top of the
plot. The thick line in the upper left corner in each panel
indicates the mean noise level.}
\label{fig:18mud}
\end{figure*}

\begin{figure*}[th]
\centerline{\psfig{figure=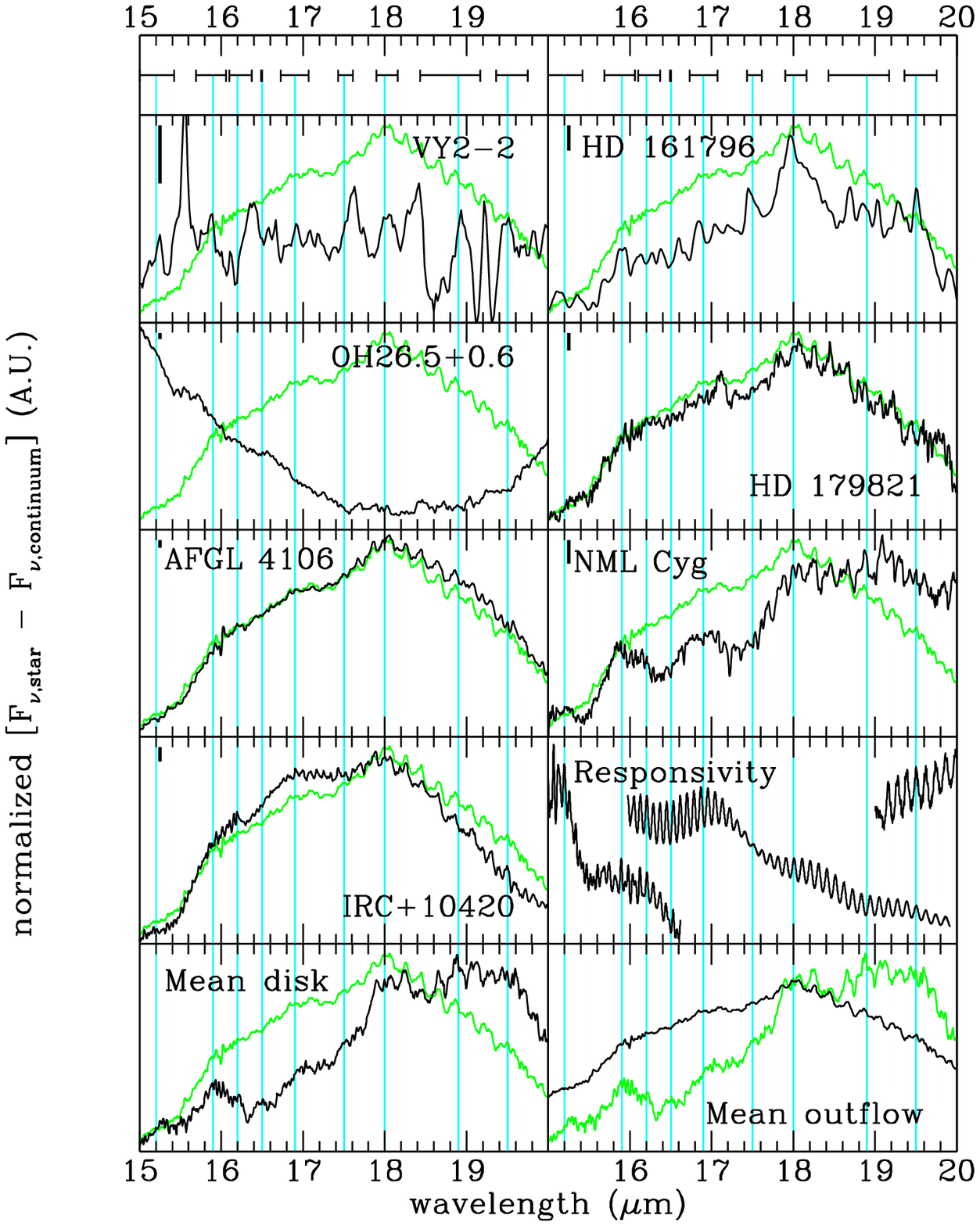,width=120mm}}
\caption[]{\small The 18 micron complex for the outflow sources,
plus the mean spectra, and the responsivity profile of the
relevant bands in this wavelength region (SWS-band 3A, 3C and 3D).
The gray line is the mean outflow spectrum, except in the lower
right corner where it is the mean disk spectrum.
The vertical gray lines indicate the mean peak positions of the
features found, while their range is indicated by the errorbar on
top of the plot. The thick line in the upper left corner in each
panel indicates the mean noise level.}
\label{fig:18muo}
\end{figure*}

The 18 micron complexes of the disk and outflow sources are shown in
figures \ref{fig:18mud} and \ref{fig:18muo}.

\subsubsection{Disk sources}

{\em IRAS09425-6040}: The 18 micron complex of IRAS09425-6040 is
quite similar to the mean 18 micron complex of the disk sources,
both with regard to the positions of the features as well as in the strength of
the features.

{\em NGC~6537}: Because of the low dust temperature, the 18 micron
complex is less prominent than in other sources. Still, most
prominent features are easy to recognize. The 18.1 micron feature
is relatively strong compared to the other features. The 18.9 and
19.5 micron features are nicely separated. The sharp feature at
18.88 $\mu$m is unidentified; see the Appendix for more details on
this feature.

{\em NGC~6302}: The 18 micron complex of NGC~6302 is dominated by
crystalline silicates.
The 19.5 micron feature is relatively strong compared to the 18.1~micron 
feature.

{\em MWC~922}: Compared to the mean disk spectrum, MWC~922 shows a
drop around 16.5$\mu$m. It is quite similar to what has been found
for MWC~300. In Sec~\ref{sec:disk10} it has been argued, that the 10 micron 
complex of MWC~300 is likely to be filledin interstellar absorption.
MWC~922 has an E(B-V)$\approx 2$ (Voors 1999), which is higher than
for MWC~300. It is not known which fraction is due to interstellar
and which is due to circumstellar reddening, but this high value
leaves the possibility of filled in interstellar absorption open.
One would expect to see evidence for this hypothesis in the 10 micron
complex. Unfortunately that complex is dominated by PAH features,
which will fill in any absorption profile and make it more
difficult to recognize, since these features are located on the
slopes of the silicate absorption profile, masking its presence. The
absorption by crystalline silicates in the 10 micron complex
suggest that circumstellar amorphous silicates will also be in
absorption at 10 and 18~$\mu$m. Another indirect argument for
interstellar absorption in the 18 micron complex is the peak
wavelength of the absorption feature. It is much bluer than the
peak wavelength of the amorphous silicates found in emission in
the disk sources.

This is the only star where the 17 micron feature is clearly
divided into a 16.7 and 17.0 micron feature. In other sources where
these features are found, they are severely blended. A closer look
at the mean disk spectrum indicates that this feature is a blend
of 2 features, even if one excludes MWC~922 from this mean
spectrum. It is also found in the two independent observations of MWC~922, so
all these arguments make us confident about the reality of these
two features. The 19.5 micron feature is the bluest found in our
sample. There seems to be weak structure on top of the 18.1 micron
feature, which is also found in HD~44179.

{\em AC Her}: The 18 micron complex is dominated by the 15.3, 15.9 and 16.2
micron features. These features are relatively strong because of the
high temperature the crystalline silicates around
this source attain.
As in MWC~300 and 89~Her, the 18.9 micron feature is rather
blue-shifted. This object has a low E(B-V) due to interstellar
extinction ($\approx 0.1$). This low value of the interstellar
E(B-V), the strength of the 15.9 and 16.2 micron features, and the fact
that the spectrum returns to the same level around 15 $\mu$m,
suggests that the spectral structure around 17 $\mu$m in this object is caused
by the 16.2 and 18.1 micron features. Note that in this source the
spectrum drops at 18 $\mu$m, while in MWC~300 and MWC~922 the
spectrum still rises longwards of 18 $\mu$m, indicating that in
the latter sources the 18.1 micron feature is on a slope.

{\em HD~45677}: The 18 micron complex of HD~45677 is very similar to
the mean disk spectrum. It shows evidence for the presence
of amorphous silicates. On top of the amorphous silicate feature,
weak crystalline silicate features are found. Among the disk sources, it is
the object with the most distinct presence of amorphous silicates. The
amorphous silicate feature peaks at a rather red wavelength, which
might be an indication of relatively large grains. However, there
is only a small shift in the 10 micron complex, which suggest that
the grains are relatively small. Therefore a compositional difference
of the dust particles is more likely to explain the red peak
position. We can also not exclude that there are contributions
from other materials, such as simple oxides and/or other silicates.

{\em 89~Her}: The 18 micron complex of 89~Her shows a broad
emission plateau, unlike most disk sources. This is evidence
for the presence of amorphous silicates. Its profile is
rather broad, with a prominent blue wing. On top of this amorphous
feature, the crystalline features are present. The 18.0 micron
feature is relatively strong. Similar to MWC~300 and AC Her, the 18.9
micron feature is severely blue-shifted (towards 18.49 $\mu$m in
this source), which makes the 19.5 micron feature rather
prominent. The 16.9 micron feature is more pronounced than in most
spectra.

{\em MWC~300}: The 18 micron complex of MWC~300 is characterized by
(interstellar) silicate absorption together with circumstellar
silicate emission. The circumstellar amorphous silicate emission
profile is red-shifted with respect to the normal (=interstellar)
amorphous silicate profile. This might indicate that large grains
are present, a fact which can also be inferred from the high mm
continuum flux (Henning et al. 1994). The different features of
crystalline silicates are visible in the spectrum, although
influenced by interstellar absorption. The clearest example
of a feature influenced by interstellar absorption, is the 18.0
micron feature. In most sources it is one of the strongest
features in the 18 micron complex, while in MWC~300 it is rather
weak, since it has to fill up part of the interstellar absorption
feature. The removal of the interstellar absorption profile is
likely to reveal a prominent 18.0 micron feature in MWC~300 too.
The shape of the 18.0 micron feature is also influenced by the
blue-shifted 18.9 micron feature, a shift which was also found in
AC~Her and 89~Her. The 19.5 micron feature is prominently present
and rather broad, as is also seen for the AC~Her and 89~Her.
In the spectrum of MWC~922 the often blended 16.9
micron feature is split into two sharp features at 16.7 and 17.0
$\mu$m. In the spectrum of MWC~300 only the sharp 17.0 micron
feature is present, while the 16.7 micron feature is missing.

Crystalline silicates also influence the shape of this
complex. The 18.1 and 16.2 micron features are located at the
positions of the rise in the `absorption' feature. This also leads
to a change in the feature appearance and is another proof of
silicate emission together with interstellar absorption. We
exclude the possibility of no absorption at all and only emission
features, because this would lead to a broad emission plateau
between the 10 and 18~micron region, which has not been observed
in other stars. Narrow absorption like profiles are also found
in the spectra of MWC~922, AC Her, IRAS09425-6040 and NGC~6302 (see
Fig.\ref{fig:18mud}). However, in most of these cases it is caused
by strong 16.2 and 18.1 micron features. This is evidenced by
the decrease of the flux (in the continuum subtracted spectra) on
the blue side of the blend of the 15.9 and 16.2 micron features.
This drop is not seen in MWC~300 nor probably MWC~922.

{\em HD~44179}: It cannot be excluded that part of the features in
this complex originate from a carbon based chemistry. The 18
micron complex differs in several respects from the mean disk
spectrum. The 18.9 micron feature is the strongest seen in this
sample. The spectrum around 15 $\mu$m is influenced by CO$_2$
absorption at 13.9, 14.9, 15.3 and 16.2 $\mu$m. Without these
absorptions the feature at 15 $\mu$m would be very similar to the
one found in 89~Her. The 15.9 and 16.2 micron features are rather
prominent, although not as strong as in AC Her. The 16.5 micron
feature, probably a PAH-feature (Van Kerckhoven et al. 2000),
is present in both the AOT01 and the AOT06 spectrum and
seems therefore rather secure. The only other star where this
feature has been detected is NGC~6537.
Also, longwards of 16.5~$\mu$m these two stars look quite similar, with similar
features at 16.9, 18.0 and 18.9 $\mu$m. The 3 features found
around 18~$\mu$m (at 17.8, 18.0 and 18.2~$\mu$m) are seen in both
the AOT01 and the AOT06 spectra as well as in the up and the down
scans and there is not much doubt about their reality. The 19.5
micron feature is rather weak in HD~44179.

{\em Roberts~22}: The 18 micron complex is rather noisy,
preventing the measurement of the individual components. The
gentle slope at the blue side of the amorphous silicate feature
suggests the presence of some crystalline silicates around 16
$\mu$m. Other crystalline silicate features might be present
but the noise prevents their detection.

\subsubsection{Outflow sources}

{\em Vy~2-2}: Unfortunately no interesting features could be
confirmed in this wavelength range, because of the low signal to
noise ratio. The emission line at 15.56 $\mu$m is [Ne~III].

{\em HD~161796}: The 18 micron complex of HD~161796 is dominated by
the 18.1 micron feature. However, one should be careful with the
interpretation of this feature (see Appendix for more details).
The 19.5 micron feature is relatively strong.

{\em OH~26.5+0.6}: The 18 micron complex of OH~26.5+0.6 is an
absorption spectrum. Both the amorphous and the crystalline
silicates are seen in absorption. We can recognize most of the features
which are normally seen in emission. All absorption features are
slightly red-shifted, we speculate that this red-shift is due to a
temperature effect (Bowey et al. 2000). It is clear from the absorption
profile that crystalline silicates only play a minor role in the
spectrum (at these wavelengths).

{\em HD~179821}: This complex lies on a very steep slope. It shows
the characteristic outflow source spectrum. It much
resembles AFGL~4106 and IRC+10420, two other massive stars. It
is dominated by the 18 micron amorphous silicate feature. On top
of this feature, weak crystalline silicates can be detected.

{\em AFGL~4106}: The 18 micron complex of AFGL~4106 is very similar
to the mean outflow spectrum, a broad amorphous silicate feature
with weak crystalline silicate features on top. No
differences between the mean spectrum and AFGL~4106 were found.

{\em NML~Cyg}: The 18 micron amorphous silicate absorption is less
prominent than for similar high mass star
IRC+10420. This might indicate that the 18 micron amorphous
silicate band is just in the regime between self-absorption and
emission. Whatever the reason, it makes the crystalline silicates
more apparent. Most crystalline silicate features are clearly
seen. Still, the 19.5 micron feature seems absent.

{\em IRC+10420}: The 18 micron spectrum is very similar to the
mean outflow spectrum. A strong amorphous component is seen with some weak
crystalline silicate features on top of it.

\subsection{The 23 micron complex}

\begin{figure*}[th]
\centerline{\psfig{figure=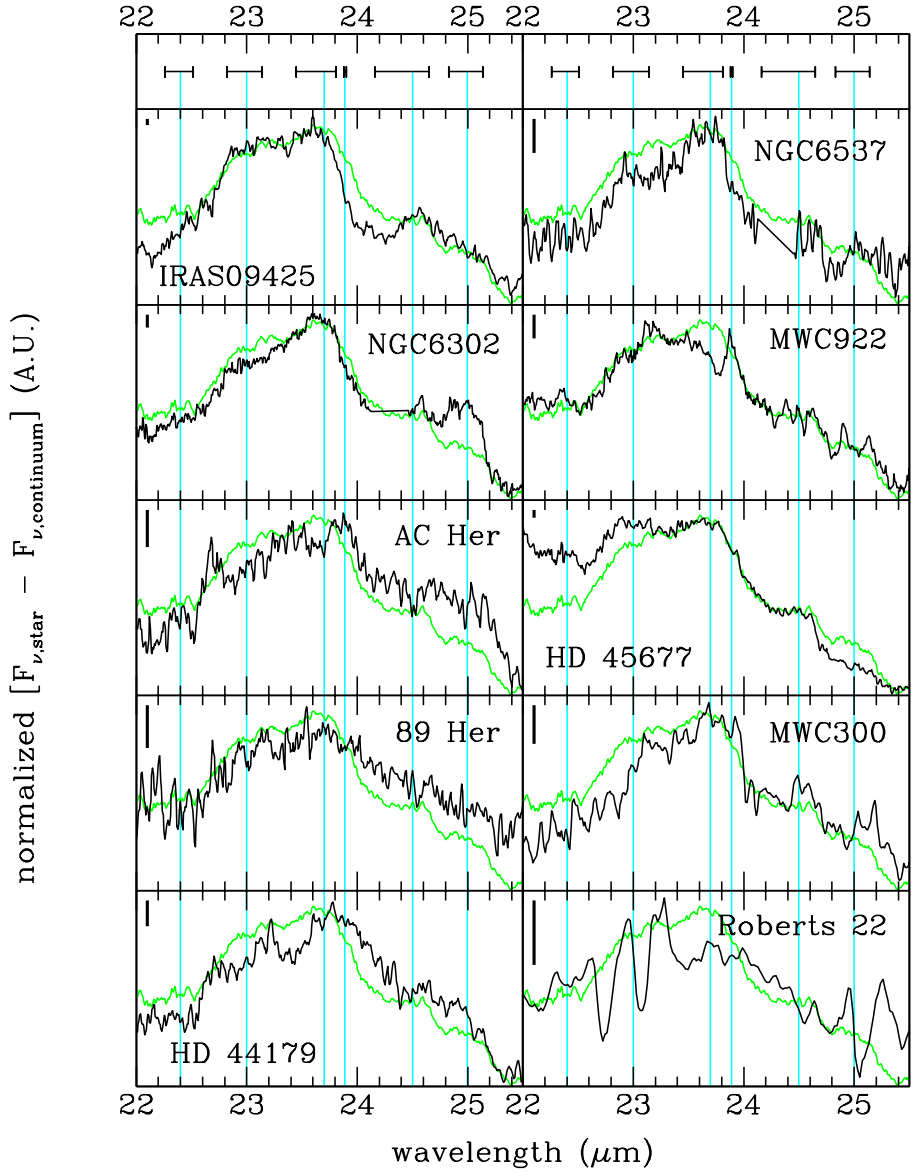,width=120mm}}
\caption[]{\small The 23 micron complex for the disk sources, together with
the mean disk spectrum (gray line). The vertical
gray lines indicate the mean peak positions of the features found,
while their range is indicated by the errorbar on top of the
plot. The thick line in the upper left corner in each panel
indicates the mean noise level.} \label{fig:23mud}
\end{figure*}

\begin{figure*}[th]
\centerline{\psfig{figure=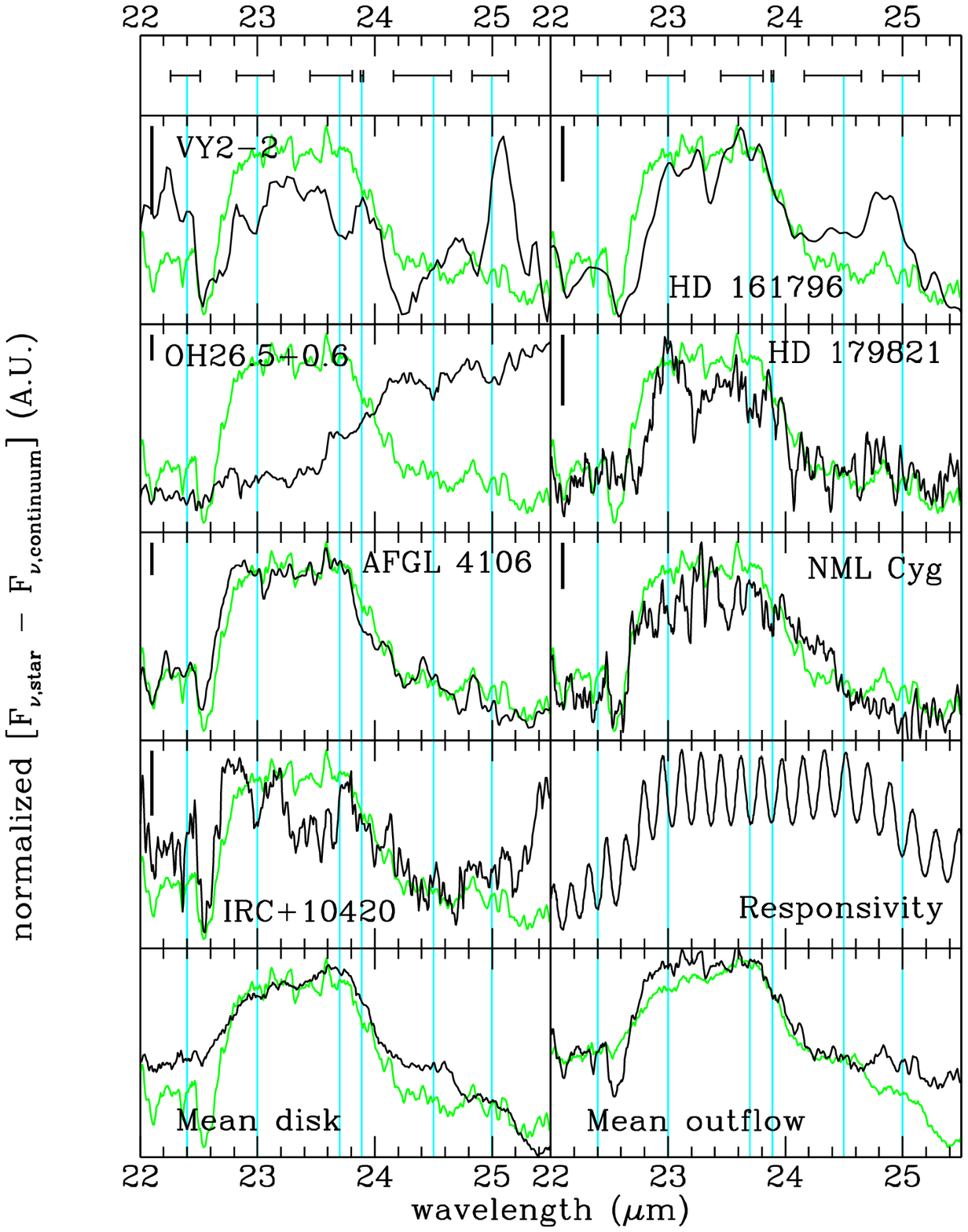,width=120mm}}
\caption[]{\small The 23 micron complex for the outflow sources,
plus the mean spectra, and the responsivity profile of SWS-band 3D.
The gray line is the mean outflow spectrum, except in the lower
right corner where it is the mean disk spectrum.
The vertical gray lines indicate the mean peak positions of the features
found, while their range is indicated by the errorbar on top of
the plot. The thick line in the upper left corner in each panel
indicates the mean noise level.} \label{fig:23muo}
\end{figure*}

The 23 micron complex spectra of the disk and outflow sources are presented in
figures \ref{fig:23mud} and \ref{fig:23muo}.

\subsubsection{Disk sources}

{\em IRAS09425-6040}: As for the 18 micron complex, the 23 micron
complex in this source is also
quite similar to the mean 23 micron complex for the disk sources.
There are some differences: The slope to the short wavelength side
continues to shorter wavelengths than in the mean disk spectrum
and the plateau does not show the step-like structure seen in
the mean (disk) spectrum. The 24.5 micron feature is relatively
strong.

{\em NGC~6537}: The 23 micron complex of NGC~6537 is characterized
by a very strong 23.7 micron feature and relatively weak 23.0
micron feature. Although the 24.5 micron feature is severely
influenced by the [Ne V] emission line, the step-like structure is
visible in the plateau.

{\em NGC~6302}: The main differences between the mean 23 micron
complex and the 23 micron complex of NGC~6302, is found in the
strength of the 23.0 and 25.0 micron features. The 23.0 micron
feature is slightly weaker than average while the 25.0 micron
feature is significantly stronger.

{\em MWC~922}: The most striking aspects of the 23 micron complex
of MWC~922 is the strong 23.9 micron feature and thus the relative
weakness of the 23.7 micron feature. Also the 23.1 micron feature
is relatively strong. However, there is no correlation between
these two features; e.g., compare MWC~922 with AC~Her.

{\em AC Her}: The 23 micron complex differs in several respects
from the mean spectrum. The feature at 22.69 $\mu$m has a
remarkable shape: it is sharp and rather prominent. It appears very
similar to a feature in HD~44179, but there it is weaker. However,
there are some questions about the reality in AC Her (see the Appendix).
If this feature can be attributed to the 23.0 micron feature it is
the most blue-shifted, which implies that also the 23.7 micron
feature is significantly blue-shifted. Alike to MWC~922, there is a
23.89 micron feature present. The 24.5 micron feature in the
spectrum of AC Her is not as prominent as in the mean spectrum.

{\em HD~45677}: The 23 micron complex of HD~45677 is rather similar
to the mean 23 micron disk complex. Still there are two subtle
differences. The 25.0 micron feature is weaker in HD~45677 than
in the mean spectrum, while at the blue side of this complex the
23.0 micron feature is stronger than in the mean spectra. This is
exactly the opposite of what is found in NGC~6302.

{\em 89~Her}: The 23 micron complex of 89~Her is rather smooth and
few individual structures can be recognized. The broadness of this
complex is similar to the complexes, including the plateau, seen
in other sources. The width of the features seems larger than in
the average disk spectrum and this may cause severe blends. Still,
better signal-to-noise will probably unveil similar structures to those
found in the other sources,

{\em MWC~300}: The 23 micron complex attracts attention because of
the absence of the red shifted 23.0 micron feature. Two of the
components of the 25 micron plateau are quite clear in this
object.

{\em HD~44179}: Where the 18 micron complex looks quite similar to
NGC~6537, the 23 micron complex is rather different. It has a
rather prominent 23.0 micron feature and the slope at the red side
of the 23.7 micron feature is very gentle. There are indications
for the step-function of the plateau but the noise level prevents
strong conclusions. The 23.0 micron feature is rather prominent
and blue-shifted with respect to the mean spectrum.

{\em Roberts~22}: The 23 micron complex is dominated by noise and
by fringes. The strong structures around 23~$\mu$m and 25~$\mu$m
are due to constructive interference of the fringes in the up and
down scan, and the relatively flat structure around 24~$\mu$m is
due to destructive interference of the fringes. The width of the
whole structure is quite similar to the mean spectrum width. This
suggests that the complex is present but not much can be said
about the structure of this complex.

\subsubsection{Outflow sources}

{\em Vy~2-2}: The spectrum in this wavelength range also suffers
from a lot of noise. The comparison with the mean outflow spectrum
suggests the presence of the 23.0 and 23.7 micron blend.

{\em HD~161796}: Apart from the 23.1 and the 23.7 micron
features, the most striking feature is the 25.0 micron feature. It
is rather blue-shifted with respect to the 25.0 micron
features in other sources. Its presence is also confirmed in the rev071 
dataset, although the feature there seems less blue-shifted and more in
line with the other 25.0 micron features.

{\em OH~26.5+0.6}: As in the case of the 18 micron complex, the
crystalline features appear as weak absorption features on top of
the wing of the amorphous silicate absorption.
The blend of the 23.0 and
23.7 micron features is rather narrow compared to the mean
spectrum. The 23.89 micron feature seems present in absorption.

{\em HD~179821}: As for all outflow sources, the 25 micron plateau
is also very weak in this source. The feature at 23.0 micron
feature is surprisingly strong and the 23.7 micron feature rather
broad. The sharp rise of the 23.0 micron feature, which is typical
for the outflow sources, is evident here.

{\em AFGL~4106}: The 23 micron complex is somewhat remarkable. The
25 micron plateau is, as for most outflow sources, rather weak
and ends already before 25 $\mu$m. The features around 23 $\mu$m
produce a very flat plateau. Some structure can be detected, but
together with IRAS09425-6040 it shows the flattest structure in
our whole sample.

{\em NML~Cyg}: The 23 micron complex is interesting in two ways.
The 25 micron plateau is almost absent. However, the blend of the
other features is the broadest seen in our sample. In this respect
the spectrum is somewhat similar to the 23 micron complex of
IRC+10420.

{\em IRC+10420}: The 23 micron complex is weak with respect to
the continuum and therefore affected by noise. Still we could
identify the different features. The complex is also quite broad
and alike to what is seen for AFGL~4106 and NML Cyg.

\subsection{The 28 micron complex}

\begin{figure*}[th]
\centerline{\psfig{figure=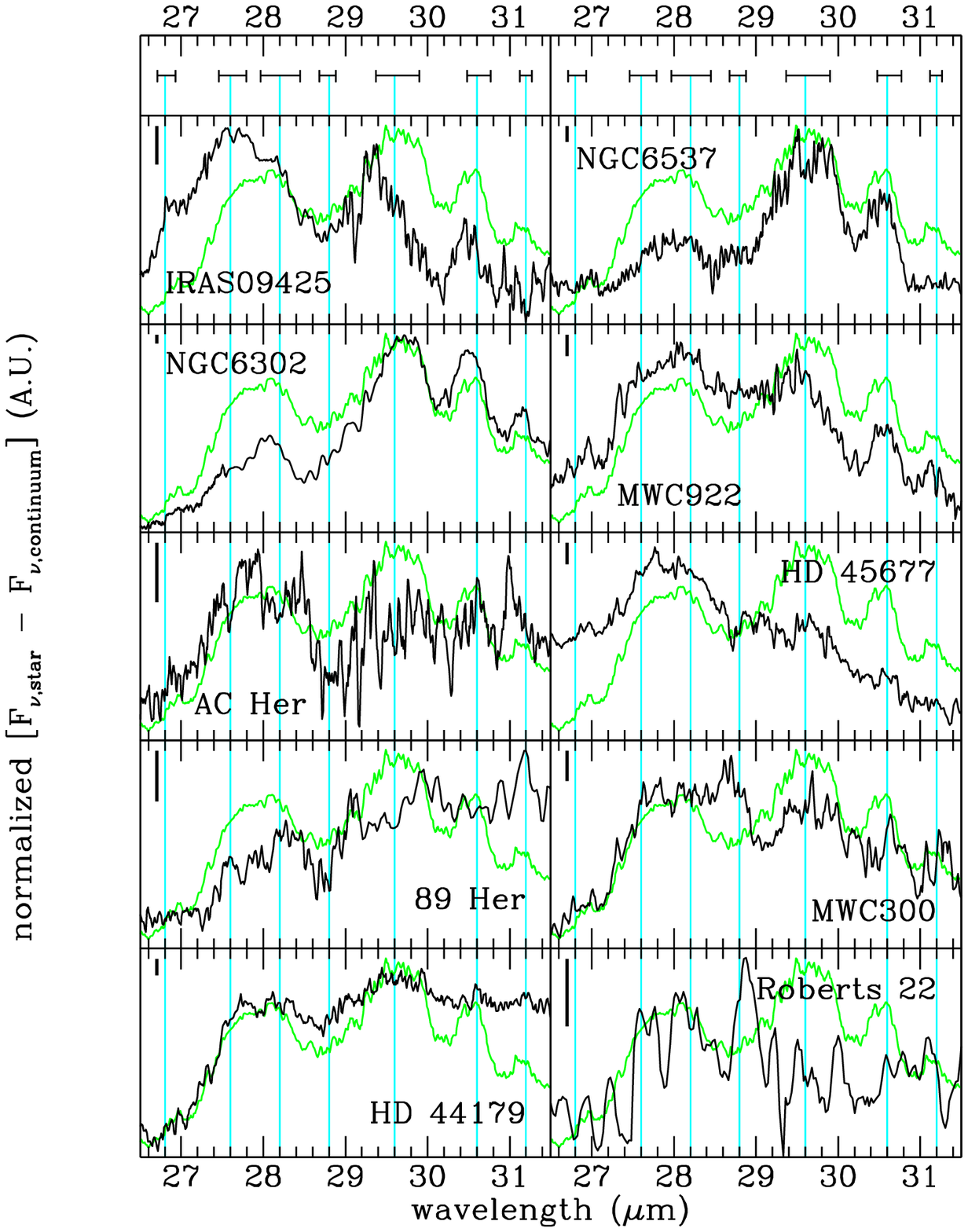,width=120mm}}
\caption[]{\small The 28 micron complex for the disk sources, together with
the mean disk spectrum (gray line). The vertical
gray lines indicate the mean peak positions of the features found,
while their range is indicated by the errorbar on top of the
plot. The thick line in the upper left corner in each panel
indicates the mean noise level.} \label{fig:28mud}
\end{figure*}

\begin{figure*}[th]
\centerline{\psfig{figure=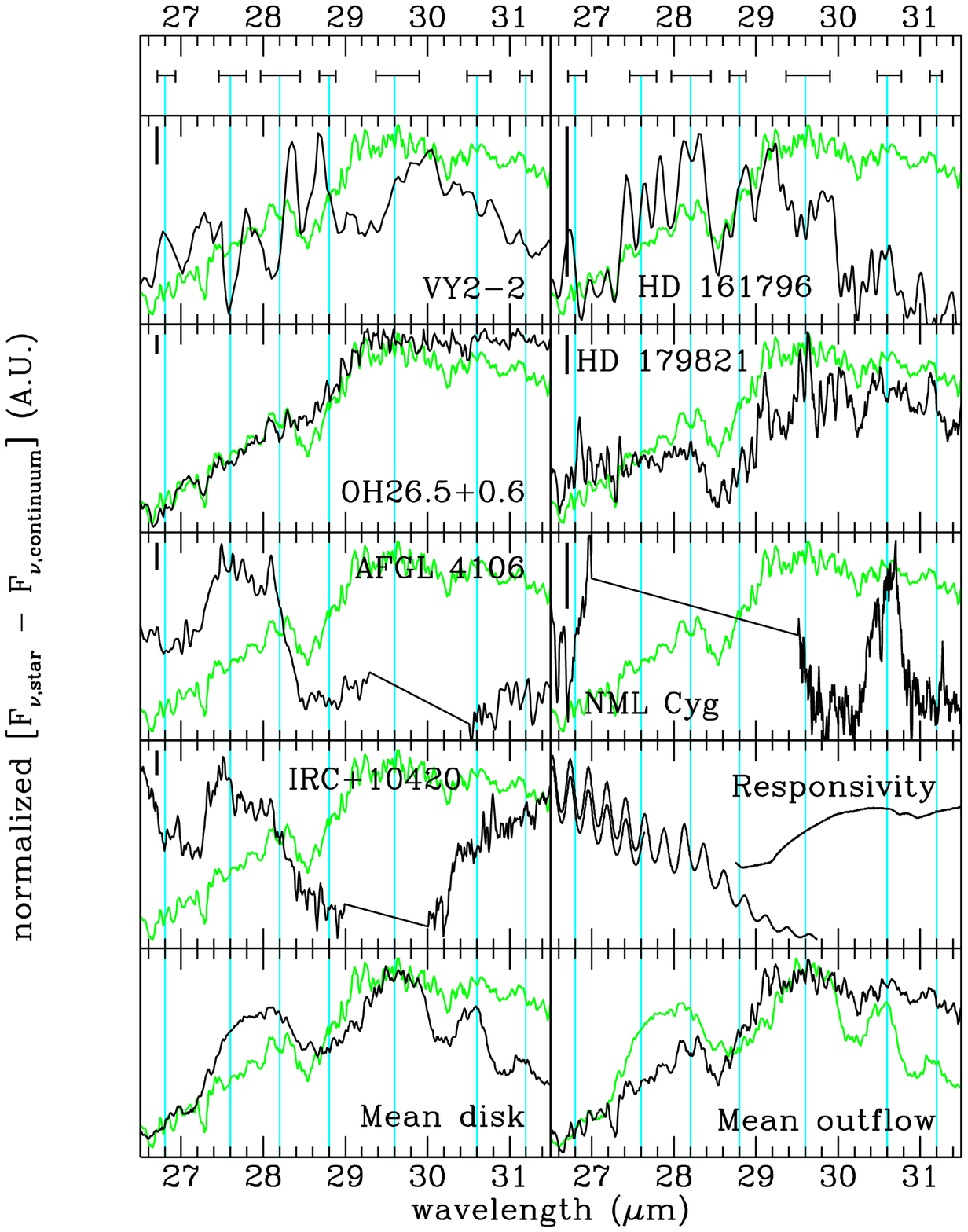,width=120mm}}
\caption[]{\small The 28 micron complex for the outflow sources,,
plus the mean spectra, and the responsivity profile of the
relevant bands in this wavelength region (SWS-band 3D, 3E and 4).
The gray line is the mean outflow spectrum, except in the lower
right corner where it is the mean disk spectrum.
The vertical gray lines indicate the mean peak positions of the features
found, while their range is indicated by the errorbar on top of
the plot. The thick line in the upper left corner in each panel
indicates the mean noise level.} \label{fig:28muo}
\end{figure*}

The 28 micron complex spectra of the disk and the outflow sources are
plotted in figures \ref{fig:28mud} and \ref{fig:28muo}.

\subsubsection{Disk Sources}

{\em IRAS09425-6040}: The strength of its 27.6 micron feature is
an intriguing aspect of the 28 micron complex. It is also very broad and
blends with the 26.9 and 28.2 micron features. The 30.6 micron
feature is relatively strong. There is a hint of the 31.2 micron
feature in the rev084 data. However, this cannot be confirmed by
the rev254 data because of the noise level in this area. The
29.6 micron feature peaks at the very blue side of the feature in
the mean disk spectrum.

{\em NGC~6537}: One of the most interesting aspects of the 28
micron complex in NGC~6537 is the absence of the 31.2 micron
feature. The 29.6 and 30.6 micron features are both very strong
relative to the 27.6 and 28.2 micron features.

{\em NGC~6302}: The 29.6 and 30.6 micron features are very strong
compared to other sources, but very similar to NGC~6537. In
contrast to NGC~6537, the 31 micron feature is prominent in this
source. The strength ratio of the 27.6~micron feature to the
28.2~micron feature is rather low compared to the average.

{\em MWC~922}: The 27.6 and 28.1 micron features are relatively
strong in this star. Compared to other spectra, the 29.6~micron
feature `misses' intensity at the long wavelength side. If the
29.6 micron feature is indeed a blend of two features, then the reddest
feature is severely depressed in MWC~922. The 30.6 and
31.2~micron features are clearly present and relatively strong.

{\em AC Her}: The 28 micron complex is rather noisy, especially
longwards of 29 micron, where band 4 starts, but all features seem
present. The 27.6 and 28.2 micron feature are still well visible
and rather strong. The 29.6 and 30.6~micron features are hardly
detectable above the noise. The sharp and strong peak at 31.0
$\mu$m is questionable. Still, there are indications of an
underlying 31.2 micron feature.

{\em HD~45677}: The 28 micron complex of HD~45677 is interesting due
to the strong 27.6 and 28.2 micron features. The ratio between the
flux of these two features and the 29.6~micron feature is the
highest found in our sample. The 30.6 micron feature is also weak,
therefore the dust species that causes the prominent 29.6 and 30.6
micron features in NGC~6302 and NGC~6537 must be only a minor
component in HD~45677.

{\em 89~Her}: The 28 micron complex is dominated by the rise of
the broad 33 micron band. On top of this feature are the 27.6 and
28.2 micron features. We have possibly found a broad structure around
29.7 $\mu$m and no indications for a 30.6 micron feature. The 33
micron band together with the low S/N for Band~4 make this a
problematic part of the spectrum.

{\em MWC~300}: MWC~300 is again a source where the 27.6 and 28.2
micron features are stronger than the 29.6 and 30.6 micron
features. The 28.8 micron feature is present in this source, but
the sharp peak (at 28.7 $\mu$m) on top of the 28.8 micron feature
has significant uncertainties (see Appendix). Neglecting this
peak, the remaining feature lines up very well with the weak
feature found in NGC~6537.

{\em HD~44179}: The 28 micron complex is, like that of 89~Her, dominated by
the onset of the broad 33 micron band. On top of the blue rise of
this broad band, the 27.6 and 28.2 micron features are rather
prominent (in contrast, for example, to NGC~6302). The 29.6
micron feature seems less pronounced and also the 30.6 micron
feature is just above the noise level. The 31.2 micron feature
is not detected.

{\em Roberts~22}: Because of the flux jumps around band 3E, it
was difficult to subtract a reliable continuum for band 3E.
We have tried to maintain the relative slope after a jump in the
flux. In this way we were able to subtract the continuum, but the
relative strength of the features in the different bands
(3D, 3E and 4) in this wavelength range is severly influenced by
these flux jumps. We measured the bands in this wavelength range
using a local continuum. We found the 27.6, 28.2
and 28.8 micron features, all three of which were seen in both the
rev254 and the rev084 spectra. The 28.8 micron band is interesting
since it is the most red-shifted one from the four stars where
this band is found.

\subsubsection{Outflow Sources}

{\em Vy~2-2}: The beginning of band 4 is less effected by noise.
Band 4 has a larger aperture and is therefore less influenced by
the mispointing. The 29.6 and 30.6 micron features can be
identified.

{\em HD~161796}: The noise is rather severe in this part of the
spectrum. Still, some features can be identified. The blend of 27.6
and 28.2 micron features is present. The 29.6 micron feature is
detected but has an extension at the short wavelength side, in
both the up and down scan, which is not seen in the rev071 data.
This leaves some uncertainty on the reality of this extension.

{\em OH~26.5+0.6}: This is the complex where the absorption
spectrum of this source goes into an emission spectrum.
The 29.6 micron feature,
the first feature in this star which appears in emission, is
rather blue-shifted. No 30.6 or 31.2~micron feature is found.

{\em HD~179821}: The 27.6 and 28.2 micron features are relatively
weak with respect to the 29.6 micron feature. Whether this is a
typical outflow source characteristic is difficult to say, because the
mean outflow spectrum is rather influenced by this spectrum (see
paper~II).

{\em AFGL~4106}: because of badly corresponding up and down scans
we have removed the first part of band 4 (29 to 30.5 $\mu$m).
Still, the 27.6 and 28.2 micron features stand out very
prominently, with sharp drops on either side of the blend.

{\em NML~Cyg}: The end of band 3D and the beginning of Band 3E
suffer from spectral leakage from shorter wavelengths, therefore
no features are measured in this wavelength range. The 30.6 micron
feature is prominent in this spectrum, while, on the other hand,
there is no indication of the 31.2~micron feature.

{\em IRC+10420}: Part of the spectrum has been removed due to
deviating up and down scans. The 27.6 and 28.2 micron features are
clearly present, but rather blue-shifted.
As in many outflow sources, there is no evidence for a 30.6 or
31.2~micron feature.

\subsection{The 33 micron complex}

\begin{figure*}[th]
\centerline{\psfig{figure=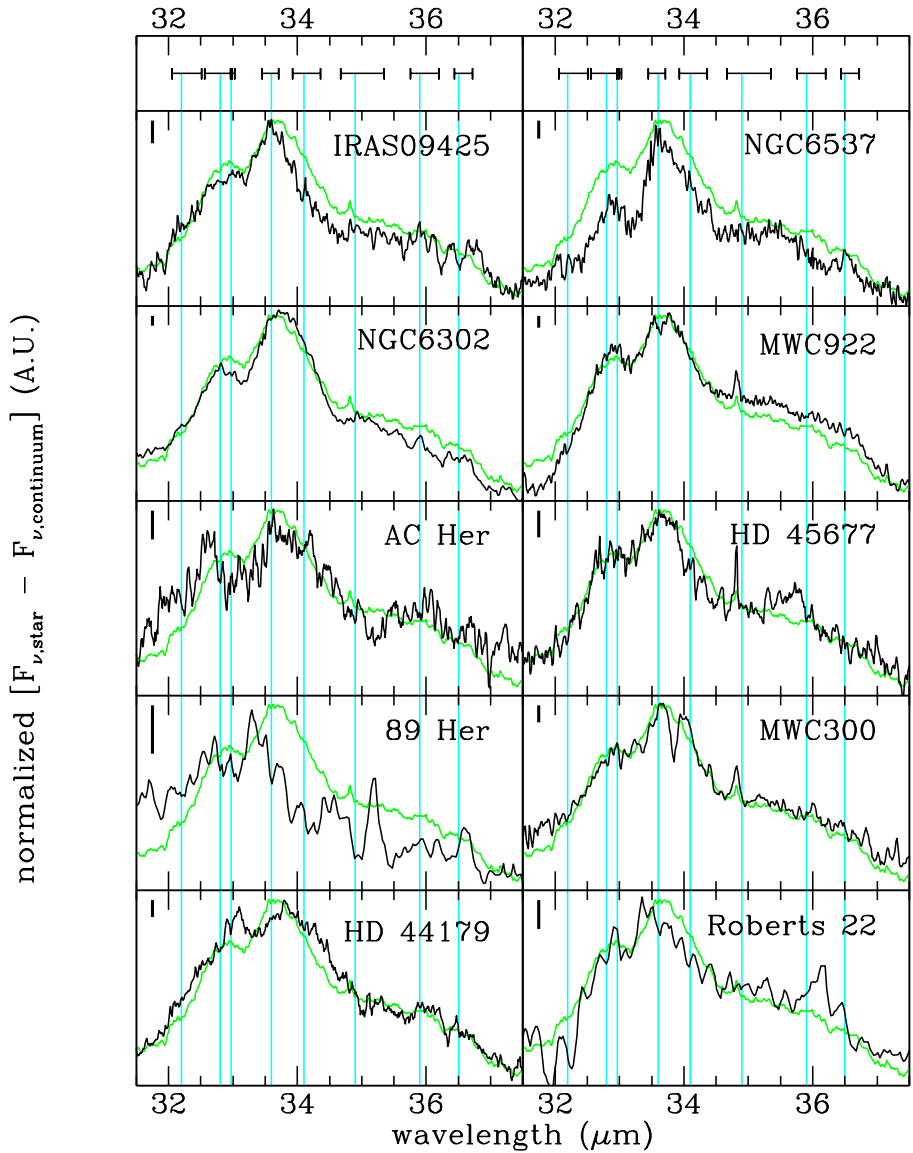,width=120mm}}
\caption[]{\small The 33 micron complex for the disk sources, together with
the mean disk spectrum (gray line). The vertical
gray lines indicate the mean peak positions of the features found,
while their range is indicated by the errorbar on top of the
plot. The thick line in the upper left corner in each panel
indicates the mean noise level.} \label{fig:33mud}
\end{figure*}

\begin{figure*}[th]
\centerline{\psfig{figure=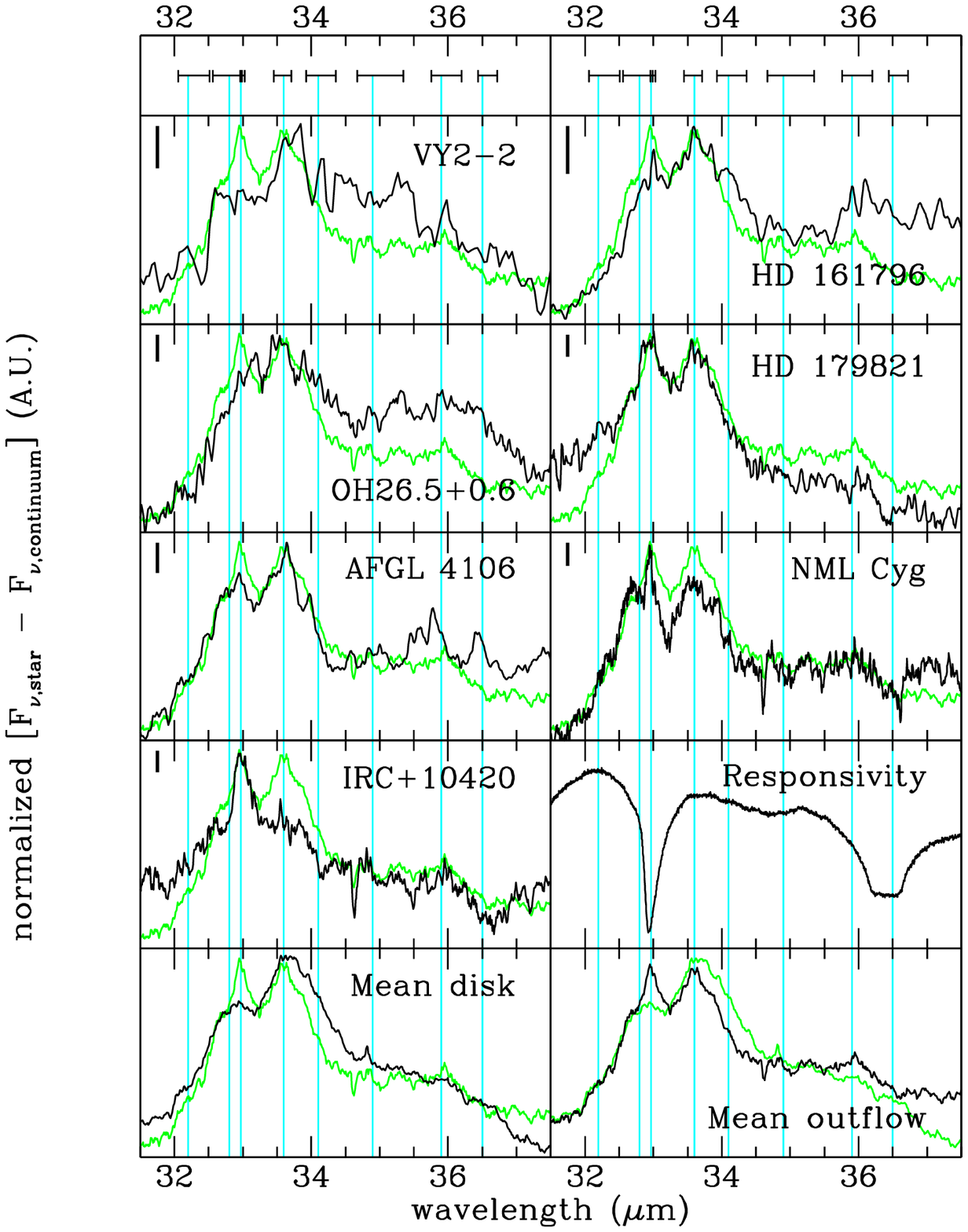,width=120mm}}
\caption[]{\small The 33 micron complex for the outflow sources,,
plus the mean spectra and the responsivity profile of SWS-band 4.
The gray line is the mean outflow spectrum, except in the lower
right corner where it is the mean disk spectrum.
The vertical gray lines indicate the mean peak positions of the features
found, while their range is indicated by the errorbar on top of
the plot. The thick line in the upper left corner in each panel
indicates the mean noise level.} \label{fig:33muo}
\end{figure*}

The 33 micron complexes of the disk and the outflow sources are
plotted in figures \ref{fig:33mud} and \ref{fig:33muo}.

\subsubsection{Disk Sources}

{\em IRAS09425-6040}: The 33 micron complex follows the mean
spectrum very well. There are two major discrepancies, the 32.2
micron feature is stronger than in the average spectrum and the
plateau drops more steeply at the end, which might be an indication of
the 36.5 micron forsterite feature. It should be noted that the
peak to continuum ratio of the 33.6 feature is the highest
found so far in the ISO database.

{\em NGC~6537}: The 33 micron complex of NGC~6537 is characterized
by sharp features. The 33.6 micron feature is very prominent The
peak caused by the 32.8 micron feature is rather sharp, and much
weaker than the 33.6 micron peak. The plateau ends at almost
37~$\mu$m due to the presence of a 36.5~micron feature. The
34.1~micron feature is seen as a small plateau on the 33.6~micron
feature.

{\em NGC~6302}: In general, the complex in this source is quite
similar to the mean complex structure. However, the 32.8~micron
feature is weaker than average in this source.

{\em MWC~922}: The 33 micron complex of MWC~922 is characterized by
a prominent plateau. The weak structure (at about 33.5 $\mu$m) on
top of the 33.6 micron structure is likely the [S III]
emission line. The 32.8~micron feature is slightly stronger than
average. The feature at 34.8~$\mu$m is the [Si II] line.

{\em AC Her}: It should be noted that the band 4 data is of low
quality, although a lot of spurious points (glitches and
glitch-tails) have already been removed by hand. As in the 23
micron complex there is some doubt about one of the most prominent
features, the one at 32.6 $\mu$m (see for more details the
Appendix). We note that the 33 and 23 micron
complexes have a similar appearance. In the 23 micron spectrum
of this source, a sharp feature has also been found at the short
wavelength side.

{\em HD~45677}: The 33 micron complex of HD~45677 is that of a
typical disk source. It has a rather strong 33.6 micron feature
and a relatively weak 32.8 micron feature. The main difference is
found in the strength of the 35.9 micron feature, which is much
stronger in this source than in the mean spectrum. As for
NGC~6537, the red side of the 33.6 micron feature shows evidence for the
34.1 micron feature. The plateau extends beyond 37 $\mu$m.

{\em 89~Her}: Interpretation of the 33 micron complex suffers
from the low flux
levels. The three main features (the 32.8 and 33.6 micron features
and the plateau) are present. Any substructure is difficult to
quantify due to the low S/N. The 32.8 and 33.6 micron features are
rather blue-shifted.

{\em MWC~300}: The overall shape is very similar to the mean
spectrum. The 33.6 and 34.1 micron bands are clearly split in this
source, and the structure nicely correlates with the structure seen in
AFGL~4106. The plateau extends to almost 37~$\mu$m and the sharp
feature at 34.81 $\mu$m is the [Si II] line. The 32.8 micron
feature is weak but clearly present.

{\em HD~44179}: The 33 micron complex can be described as a very
broad feature peaking at 33.6 $\mu$m, a sharp 32.8 micron feature
and a plateau extending to wavelengths larger than 37~$\mu$m. This
very broad feature is a blend of the 33.6 and 34.0 micron
features. In the 35 micron plateau the different components can be
identified.

{\em Roberts~22}: The general shape is roughly equal to the mean
disk spectrum. We masked the unreliable narrow (0.1 $\mu$m) peak
at 33.3 $\mu$m (see Appendix for more details), but still the 33.6
micron feature peaks at rather blue wavelengths like 89~Her,
which shows a similar complex. The 36.0 micron feature, which is
mostly found in outflow sources, is rather strong. It was
already stated before that this source shows a lot of outflow
characteristics. On the other hand, the strength ratio of the 32.8
micron feature to the 33.6 micron feature is more like that of the disk
sources.

\subsubsection{Outflow Sources}

{\em Vy~2-2}: The 33 micron complex suffers also from a low
signal-to-noise ratio, but the general structure can be
distinguished. The ratio of the 32.8~micron feature to the
33.6~micron feature is rather low in comparison with the rest of
the outflow sources. The plateau is relatively strong.

{\em HD~161796}: The 33 micron complex of HD~161796 is more like
the disk sources than the outflow sources. The 33.6 micron feature
is stronger than the 32.8 micron feature. Also the gentle slope at
the red side of this last blend resembles the disk sources better
than the outflow sources. On the other hand, the 36.0 micron
feature is more typical of the outflow sources.

{\em OH~26.5+0.6}: The 33 micron complex of OH~26.5+0.6 is rather
similar to that of HD~161796; especially the broadness of the whole
structure. The main difference between these two stars is found in
the 35 micron plateau. The 36.0 micron feature seems present in
both sources, but there is some extra absorption around 35.0
$\mu$m in OH~26.5+0.6.

{\em HD~179821}: The 33 micron complex is characterized by the two
relatively sharp features at 32.8 and 33.6~micron
and, as in the 23 micron complex, a weak plateau. The
33.6 micron feature is one of the sharpest in our sample. We note
that there seems to be no evidence for the 36.5 micron feature in both
datasets. The strength ratio of the 32.8 to 33.6 micron features
suggests that this star is massive and likely a post-red
Supergiant.

{\em AFGL~4106}: At first sight, the 33.6 micron feature
looks much stronger than the 32.8 micron feature, in reality they
do not differ so much, because of the presence of a broad
35~micron band in the spectrum of AFGL~4106. This is in accordance with
most other outflow sources, where equal strengths for these two
features are found. The broad underlying 35~micron band might
be similar to the one seen in 89 Her and HD~44179. While it seems
to start at a longer wavelength than in the other two sources,
this might be an artifact because the data around 30~$\mu$m
have been ignored in this analysis.

{\em NML~Cyg}: The spurious 32.97 micron feature gives the 33
micron complex its unique appearance. Apart from this detector respons
structure the 32.97 micron feature looks normal.
The 34.1 micron feature is very clear. Because the 33.6 and 34.1~micron
features are found in both the up and down scans of the AOT06 and
the AOT01 data, we are confident about their reality. This makes
this source one of the stars where they are most clearly separated.
At 34.6~$\mu$m, the OH maser pump line can be found in absorption as was
already reported by Justtanont et al. (1996b).

{\em IRC+10420}: The 33 micron complex is dominated by the strong
32.97 micron feature, which is predominantly seen in high flux
sources, both in the carbon-rich as well as in the oxygen-rich
dust environments. Therefore we assume that it is an artificial
feature caused by the detector responsivity for high fluxes.
Underneath this feature, the usual 33~micron complex is visible,
although at a rather low level. The 34.1~micron feature seems rather
strong with respect to the 33.6~micron feature, although this last
feature is also influenced by the presence of the 32.97~micron feature.
Alike to the features found in the 28 micron complex, these features
are also blue-shifted with respect to the mean spectrum. In the 35
micron plateau, in which the different components can be detected,
the 34.6 $\mu$m OH maser pumping line is nicely detected in
absorption (see Sylvester et al. 1997).

\subsection{The 40 micron complex}

\begin{figure*}[th]
\centerline{\psfig{figure=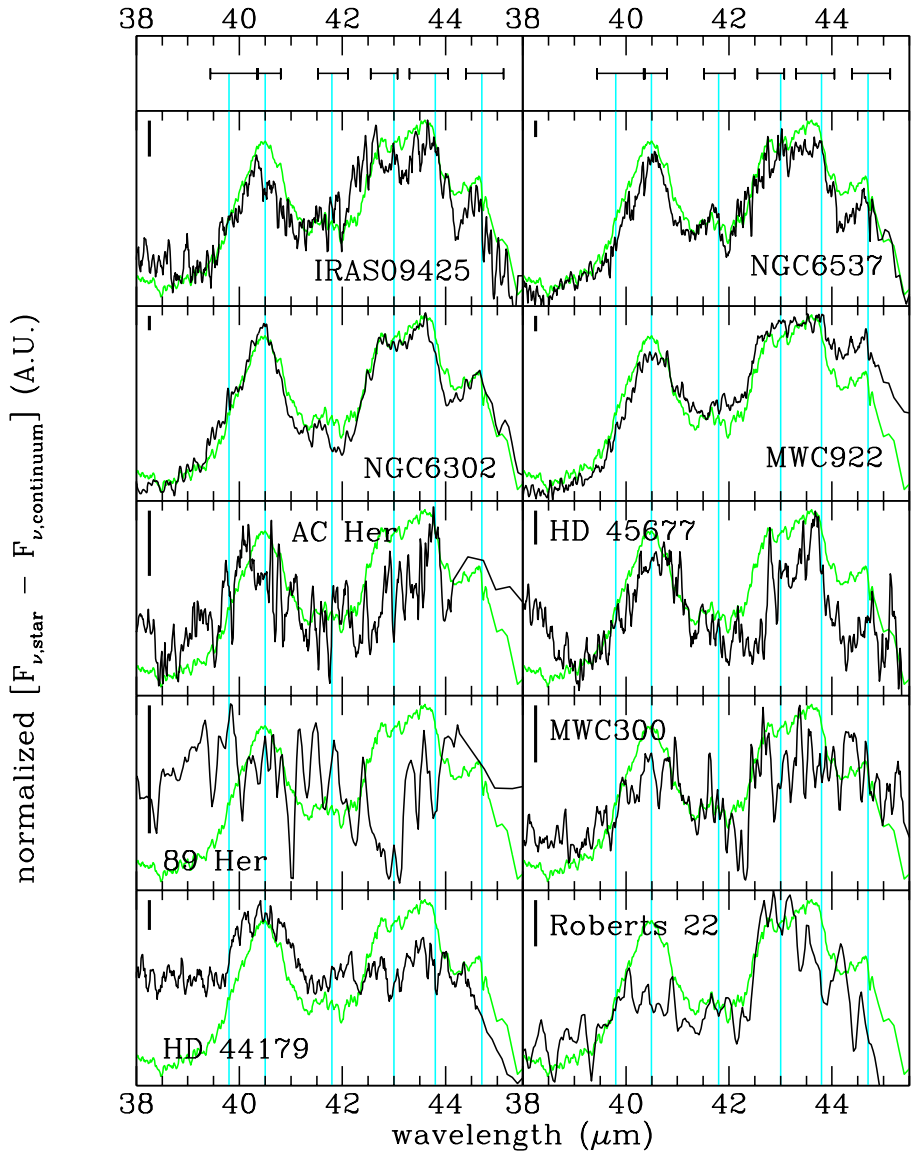,width=120mm}}
\caption[]{\small The 40 micron complex for the disk sources, together with
the mean disk spectrum (gray line). The vertical
gray lines indicate the mean peak positions of the features found,
while their range is indicated by the errorbar on top of the
plot. The thick line in the upper left corner in each panel
indicates the mean noise level.} \label{fig:40mud}
\end{figure*}

\begin{figure*}[th]
\centerline{\psfig{figure=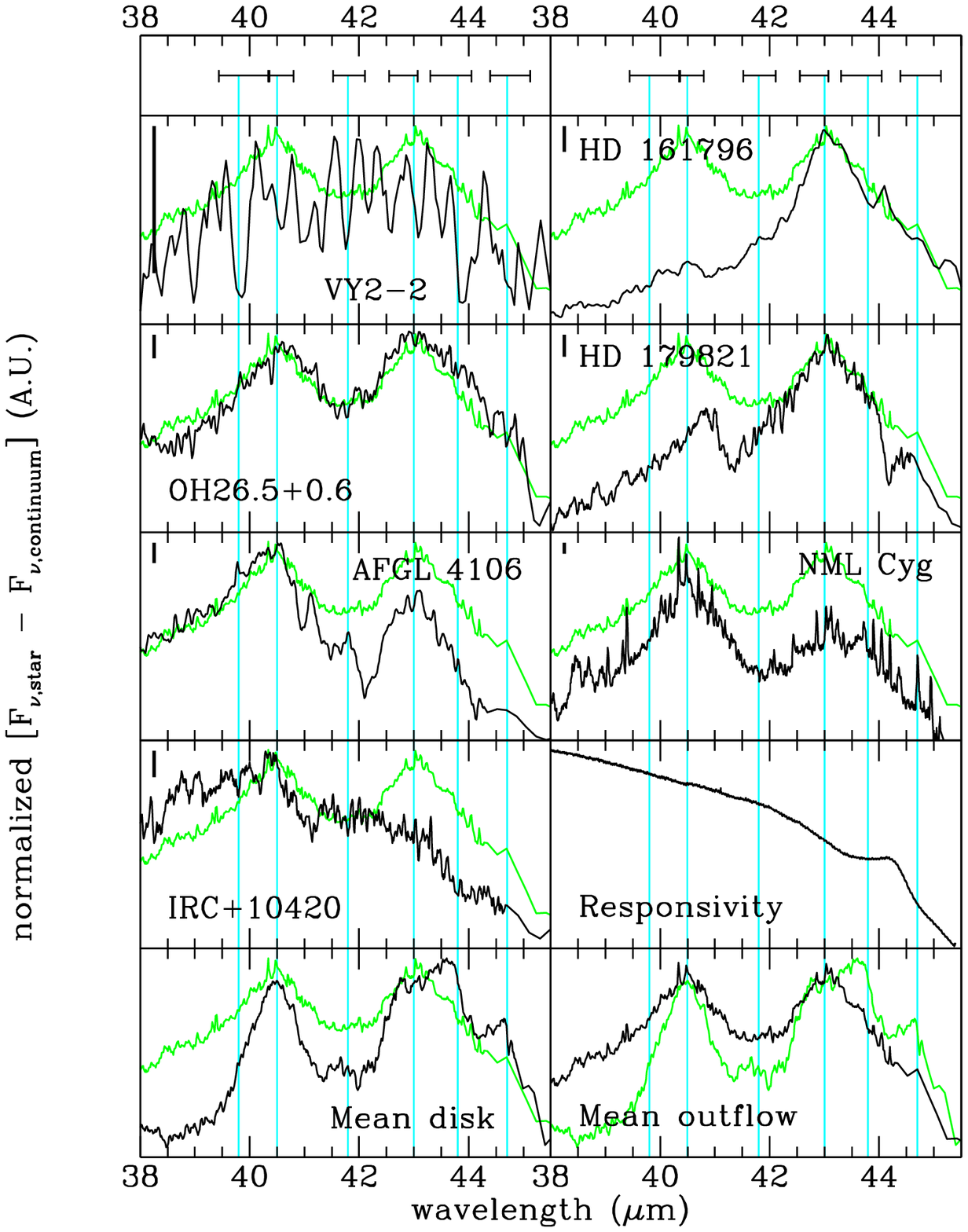,width=120mm}}
\caption[]{\small The 40 micron complex for the outflow sources,
plus the mean spectra and the responsivity profile of SWS-band 4.
The gray line is the mean outflow spectrum, except in the lower
right corner where it is the mean disk spectrum.
The vertical gray lines indicate the mean peak positions of the features
found, while their range is indicated by the errorbar on top of
the plot. The thick line in the upper left corner in each panel
indicates the mean noise level.} \label{fig:40muo}
\end{figure*}

The 40 micron complexes of the disk and outflow sources are
plotted in figures \ref{fig:40mud} and \ref{fig:40muo}.

\subsubsection{Disk Sources}

{\em IRAS09425-6040}: The 40 micron complex is similar to the mean.
The most prominent
difference is in the strength of the 43.0 micron feature, which is
also very blue shifted. This suggests a significant (cold) crystalline
water-ice contribution. This would be strange for a disk source
and it is very unfortunate that this claim cannot be confirmed by
a spectrum of the 60 micron complex.

Another difference is the long wavelength side of the 40.5 micron
feature. The short wavelength side of this feature fits the
mean spectrum very well, and peaks at a rather blue
wavelength. This feature might be a blend (see Paper~II).
While noisy, this might imply that in IRAS09425-6040 the short
wavelength component is dominant over the long wavelength
component.

{\em NGC~6537}: The 40 micron complex is rather similar to the mean
disk complex. The main difference is that the 43.8 micron feature
is somewhat weaker than on average. The 41.7 micron feature is
rather pronounced in this star.

{\em NGC~6302}: The 40 micron complex is also rather similar to the
mean disk spectrum. The features in NGC~6302 seem a little sharper
than in the mean spectrum. Because of the sharpness of the
features, the enstatite and crystalline water ice peaks around
43~$\mu$m are nicely separated.

{\em MWC~922}: The 40 micron spectrum of MWC~922 is rather similar
to the mean disk spectrum. All features in the mean spectrum are
also seen in MWC~922, and the feature strength ratios are
comparable. Only the 44.7 micron feature seems somewhat stronger,
but this might be due to the continuum subtraction.

{\em AC Her}: The noise in this part of the spectrum makes it very
difficult to say more about this feature than that the 40.5 micron
feature and the 43.8 micron feature are present. There seems some
evidence for the 43.0 micron crystalline water ice feature, which
is supported by the LWS data, showing evidence for
the 60 micron feature. The 44.7 micron feature is, also visible in
the LWS spectrum.

{\em HD~45677}: The 40 micron complex does show the usual features,
i.e. the 40.5, 41.8, 43.0, 43.8 and 44.7 micron features. The 43.0
and 43.8 micron feature are rather narrow. The 43.0 micron feature
is relatively weak. Unfortunately, there is no LWS data to check
for the presence and shape of the 60~micron feature.

{\em 89~Her}: The 40 micron complex is completely dominated by
noise and therefore no conclusions can be drawn.

{\em MWC~300}: The 40 micron complex of MWC~300 suffers from the low
flux level in this part of the spectrum. Still we can disentangle
the 40.5 micron feature, the  blend of the 43.0 and 43.8 micron
features and the 41.8 micron feature.

{\em HD~44179}: The 40 micron complex of HD~44179 is interesting
because it shows a prominent 40.5 micron feature, but the 43.0 and
43.8 micron features are much less pronounced. The steep slope at
the red side suggests that they are present, but part of the
spectrum between the 40.5 and 43.0 micron features must have been
filled in. This might be due to an exceptionally strong 41.8
micron feature, but the width and wavelength of this
feature as measured in the other stars indicate that it is not
sufficient to solely account for the closing of the gap between
the 40.5 and 43.0 micron features.

{\em Roberts~22}: The 40 micron complex is dominated by the 43.0
micron feature, indicating that there is a lot of crystalline
water ice. In this respect the spectrum looks somewhat similar to
that of HD~161796, although the relative strength of the 40.5 micron
feature in Roberts~22 is much stronger than in HD~161796.

\subsubsection{Outflow Sources}

{\em Vy~2-2}: The low signal to noise ratio in this part of the
spectrum prevents us from reaching firm conclusions about the structure in
this complex. It seems that the 40.5 micron feature and the 43.0 +
43.8~micron blend can be distinguished.

{\em HD~161796}: The 40 micron complex is completely dominated by
a very strong 43.0 micron (crystalline water ice) feature. There
might be a red-shifted 43.8 micron feature in the strong 43.0
feature, but this is difficult to quantify. Most of the features found in
other 40 micron complexes are present in this complex. There is an
indication that the 40.5 micron feature is in fact a blend of 2
separate bands. Closer inspection of the spectrum suggests that
there is a broad band underlying the strong crystalline H$_2$O
band. A suitable candidate for this feature might be amorphous
H$_2$O ice. This helps the fitting of the 40.5 micron band
(Paper III).

{\em OH~26.5+0.6}: The spectrum of OH~26.5+0.6 looks very similar to the mean
outflow spectrum. Where the spectra of OH~26.5+0.6 and HD~161796
were very similar for the 33 micron complex, they look quite
different in the 40 micron complex. The strength ratio of the 40.5
and 43.0 micron features is dramatically different. Still, as for
all outflow sources, the 43.0 micron feature is stronger than the
43.8 micron feature. The 44.7 micron feature is also weakly
present in this spectrum.

{\em HD~179821}: Like HD~161796, the spectrum of HD~179821 is dominated by the
43.0 micron crystalline water ice feature. However, the 40.5 micron
feature is much stronger, rather red-shifted, and shows a very
gentle slope on the blue side. The 43.8 enstatite feature produces
some recognizable structure.

{\em AFGL~4106}: The 40.5 micron feature looks strong, but again
this is partly due to the broad feature peaking around
35~$\mu$m. As for HD~179821, the 40.5~micron feature has a gentle
slope; a characteristic feature of outflow sources. The 43.8
micron feature is weak and almost completely overwhelmed by the
43.0 micron water ice feature.

{\em NML~Cyg}: The general shape of the complex is quite similar
to what has been found for other sources. The 40.5 micron feature
is relatively strong with respect to the 43.0 + 43.8 micron blend.
However this is seen in more sources, with HD~44179 as the most
extreme case. The gas-phase rotational water lines, seen in
emission, are most striking in this wavelength region (Justtanont
et al. 1996b). The modeling of these lines is outside the scope of
this paper. Besides gas phase water, crystalline water ice is also
present, as evidenced by the presence of the 43.0 micron feature.

{\em IRC+10420}: The 40 micron complex suffers more from the
noise than the other complexes. Still the main features can be
distinguished. The 42.8 micron feature is difficult to
disentangle, but there seems hardly any 43.6 micron feature.

\subsection{The 60 micron complex}

\begin{figure*}[th]
\centerline{\psfig{figure=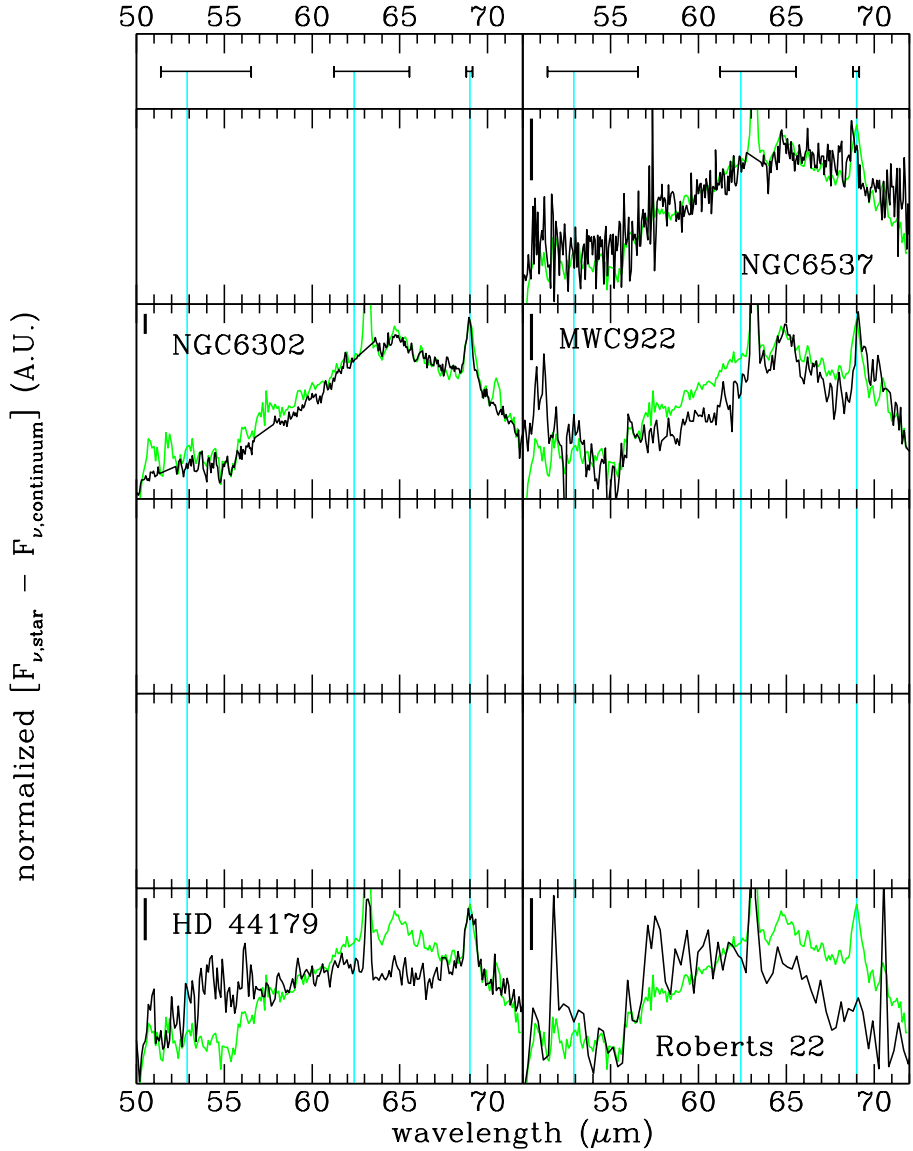,width=120mm}}
\caption[]{\small The 60 micron complex for the disk sources, together with
the mean disk spectrum (gray line).
The vertical gray lines indicate the mean peak positions of
the features found, while their range is indicated by the
errorbar on top of the plot. The thick line in the upper left
corner in each panel indicates the mean noise level.}
\label{fig:60mud}
\end{figure*}

\begin{figure*}[th]
\centerline{\psfig{figure=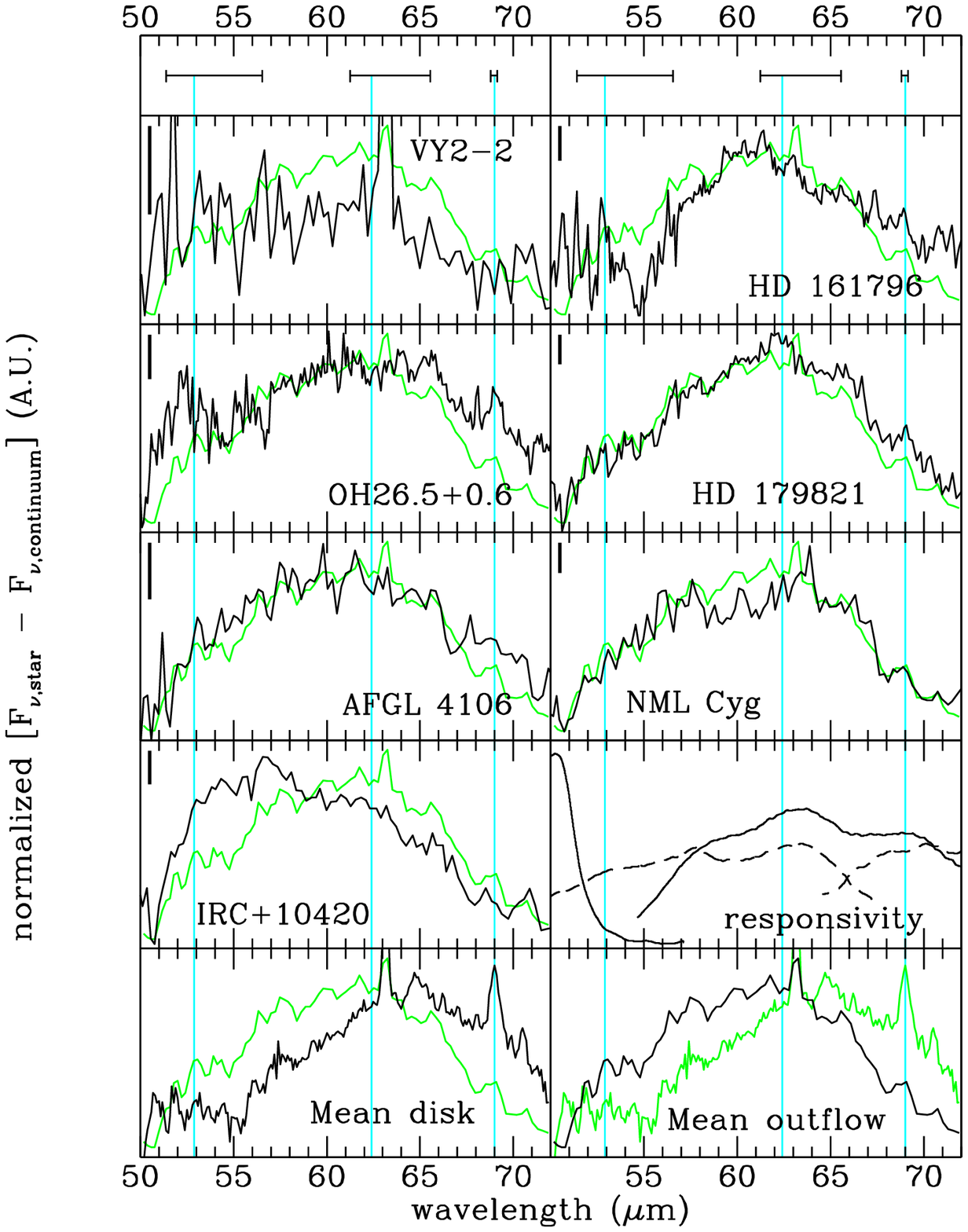,width=120mm}}
\caption[]{\small The 60 micron complex for the outflow sources, plus
the mean spectra and the responsivity profiles of LWS band 1
(solid line), band 2 (dashed line), band 3 (solid line) and band 4
(dashed line). The gray line is the mean outflow spectrum, except in the lower
right corner where it is the mean disk spectrum.
The vertical gray lines indicate the mean peak positions of the features
found, while their range is indicated by the errorbar on top of
the plot. The thick line in the upper left corner in each panel
indicates the mean noise level.} \label{fig:60muo}
\end{figure*}

The 60 micron complexes of the disk sources and the outflow sources
are plotted in figures \ref{fig:60mud} and \ref{fig:60muo}.

\subsubsection{Disk Sources}

{\em NGC~6537}: The 60 micron complex is rather smooth and peaks
at about 65 $\mu$m. The 69 micron forsterite feature is clearly
present. The gentle rise at the blue side of this feature
indicates that crystalline water ice is present, as was already
indicated by the 43.0 micron feature in the 40 micron complex.
Another dust component is necessary to explain the peak position
and the red side of this feature, and diopside might be a good
candidate for this (Koike et al. 2000). The forbidden emission lines
were removed by hand, but in the original spectrum the following lines were
found: [O~III] at 51.79~$\mu$m, [N~III] at 57.31~$\mu$m and 
[O~I] at 63.16~$\mu$m. It is interesting to note that
all lines are blue shifted (in both scan directions). We probably
see here the fast outflow from the star. Based on the shifts, the
outflow velocity in our direction is $135 \pm 30$~km/sec. This is comparable to
the velocity found by Cuesta et al. (1995). They
came up with a model, based on [N~II] profiles at 6584 \AA, which predict
a velocity of $180 \pm 40$~km/sec in the direction of the observer 
($400 \pm 100$~km/sec total outflow velocity).

{\em NGC~6302}: The 60~micron complex is very similar to the mean
disk spectrum and lines up pretty well with those of NGC~6537 and MWC~922.
Compared to MWC~922, NGC~6302 shows a little more intensity on the
blue side, which probably indicates that there is a larger contribution
from crystalline water ice as is expected from the 40
micron complex. The 69.0 micron forsterite feature is very prominent and
quite narrow.

{\em MWC~922}: The 60 micron complex of MWC~922 peaks at very long
wavelengths, implying that crystalline water ice does not play an
important role at these wavelengths. This is in contrast with the
40 micron complex, where a band at 43.0 micron has been found.
In Paper~III we will show that the 40 micron complex of this source
can nicely be fitted with only enstatite, confirming the absence of
crystalline water ice.

The 69 micron feature is very prominent and seems broadened at its
red side, which might be an indication for another dust component.
The rather sharp 65 micron feature seems flanked by another feature
at 66.5~$\mu$m.

{\em HD~44179}: The 60 micron complex is rather flat and does not
show much structure, apart from the prominent 69 micron feature
and the [O~I] line at 63.18 $\mu$m. Because of this broad flat complex it
is likely that there is only a small amount of crystalline water
ice together with an extra dust component radiating at these
wavelengths.

{\em Roberts~22}: While most disk sources do peak at a wavelength
around 65~$\mu$m, Roberts~22 seems to peak around 60~$\mu$m. This
is very much alike to the outflow sources and HD~161796 in particular. The
63.18~$\mu$m [O I] line is also visible. In correspondence with
the outflow sources, there is hardly any 69 micron feature.

\subsubsection{Outflow Sources}

{\em Vy~2-2}: The 60 micron complex is typical of an outflow
spectrum. It peaks around 60-62~$\mu$m and the 69 micron feature
is absent.

{\em HD~161796}: The 60 micron spectrum is characterized by the
crystalline water ice structure. It peaks at $\approx 60 \mu$m,
which is much bluer than the disk sources, but in better agreement
with the expectation that there is crystalline water ice around
this star. There is a hint of a 69 micron feature. We have
measured it, but better data seem necessary for this feature.

{\em OH~26.5+0.6}: The 60 micron complex has 2 peaks, one at about
60~$\mu$m, probably due to crystalline water ice, and one at
65~$\mu$m, more like MWC~922. The 69 micron feature is clearly
present, which is somewhat unexpected for an outflow source.

{\em HD~179821}: The 60 micron complex peaks at rather blue
wavelengths, indicating the presence of a lot of crystalline water
ice; as was already concluded from the 40~micron complex.

{\em AFGL~4106}: The 60 micron complex peaks around 60~micron. This
is typical for sources which are dominated by crystalline water ice,
and in agreement with the 40~micron complex spectrum.

{\em NML~Cyg}: The 60 micron feature is very similar to the mean
outflow spectrum, indicating that most of this feature is formed
by emission from crystalline water ice.

{\em IRC+10420}: The 60 micron complex peaks at very short
wavelength, indicating that there is crystalline water ice and a
relatively low abundance of crystalline silicates, as was also
found in the other complexes.

\subsection{Remaining Bands}

We chose  not to show the individual remaining bands of our
programme stars. However, in Paper~II we show the average band
profiles of the disk and outflow sources of the most prominent
remaining bands, near 14, 21, 26, 48 and 90~$\mu$m.

\subsubsection{Disk Sources}

{\em IRAS09425-6040}: The (instrumental) features around 14~$\mu$m
are weak or absent in IRAS09425-6040. The 20.7 and 21.5~micron
features are clearly distinguishable in the spectrum. The
26.1~micron feature is also rather prominent. We unfortunately have
no data for features at wavelengths longer than 45~$\mu$m.

{\em NGC~6537}: As in IRAS09425-6040, the (instrumental) features
around 14 $\mu$m are weak or absent in NGC~6537. The 20.7 and 21.5
micron feature are present. The 26.1 micron feature is one of the
reddest in our sample. The 47.7 and 48.6 micron features are
separately identified. Finally there is clear evidence for the
presence of a 91 micron feature, with the [O~III] line at 88.35 $\mu$m.

{\em NGC~6302}: NGC~6302 has very nice 20.7 and 21.5 micron
features. The 26.1 micron feature is obscured by the presence of a
strong emission line ([O IV]) and its presence cannot be confirmed
at the moment. The 48 micron features is split up into its separate
components, i.e. the 47.7 and 48.6 micron features. The 91 micron
feature is present and even rather prominent.

{\em MWC~922}: The 3 (instrumental) features around 14~micron are
all detected in this source. The 20.7 and 21.5~micron are clearly
present and are only slightly less prominent than in IRAS09425-6040.
The 26.1~micron feature is also rather prominently present.

{\em AC Her}: The 20.5 and 21.5 features are more blended than in
other stars and the final result looks somewhat similar to the
responsivity profile. This might point to a problem in the
dark-current subtraction. However, there are some remarkable
differences between the data and the responsivity at the edges of
this structure and also the same kind of profile is found in the
rev106 data. Therefore, we believe it is real.

{\em HD~45677}: The 3 features around 14 $\mu$m, the 13.5, 13.8
and 14.2 micron features, are clearly present in the spectrum. The
20.7 and 21.5 micron feature are rather prominent and almost as
strong as the 23 micron complex.

{\em 89~Her}: The (instrumental) features around 14 $\mu$m are all
present, but the 20.7, 21.5 and 26.1 micron features are not
detected.

{\em MWC~300}: The (responsivity) features around 14~$\mu$m
are all present in this source. This is one of the few sources
where the 21.5 micron feature is not present, while the 20.7
micron feature is rather prominent. This is evidence that these
two features are (at least partially) caused by different dust
species. The 26.1 micron feature is present and rather prominent.

{\em HD~44179}: The 13.5 and 14.2 micron features are found, but
might be due to carbon rich material that also peaks at these
wavelengths (Hony et al. 2000). There is no indication of a
13.8 micron feature. There is a weak 26.1 micron feature present.
In the LWS regime a 48 micron feature is found, but it could not
be resolved into the two separate components. No 91 micron feature
could be detected because the spectrum was unreliable at these
wavelengths.

{\em Roberts~22}: Interestingly the 14.2 feature seems present but
without the 13.6 feature.
The 26.1 micron feature seems absent and there is also no
indication for features around 48~$\mu$m. The presence of the
other feature are uncertain and will be discussed in the Appendix.
There is some uncertainty over the presence of a 91 micron
feature.

\subsubsection{Outflow Sources}

{\em Vy~2-2}: The features around 14 $\mu$m and the 26.1 micron
feature are not detected. Whether this is due to the noise level
or that they really are absent is difficult to say. A comparison
with other stars suggests the first possibility. While the 20.7 and
21.5~micron features are somewhat uncertain, they fit the average
profile quite well. Nevertheless, the noise around these wavelengths
makes them suspect.

{\em HD~161796}: The 21.5 micron feature is affected by the
differences seen in the up and down scan. Comparison with the
rev071 and other data suggests that the 21.5 micron feature is
probably real, while other features in this wavelength region are
probably not. The two features around 48 micron are found as a
blend.

{\em OH~26.5+0.6}: Although all features are in absorption, the 3
features around 14~$\mu$m appear in emission. This strengthens our
suspicion that these features are instrumental. The 20.7 micron
feature is thought to be in absorption, although it is rather
blue-shifted with respect to the emission spectra from other
sources. The 26.1 micron feature might have some self-absorption.

{\em HD~179821}: There are indications of a broad feature around
13 micron, which is also seen in HD~161796 and AFGL~4106. However, it
is more likely that it is just continuum and that the 18 micron
amorphous silicate feature already starts around 14.5 $\mu$m. The
14 micron (responsivity) features were not found in this source. A
20.7 micron feature has been found, without a 21.5 micron
feature. Also the 26.1 micron feature has been found. The 2
features around 48 $\mu$m were nicely split and therefore measured
separately.

{\em AFGL~4106}: There is a broad feature around 13 $\mu$m which
is also seen in HD~161796 and HD~179821. It is thought to be just
continuum and probably induced by an early onset of the 18~micron
amorphous silicate feature. On top of this broad structure, we see
the (instrumental) features around 14 $\mu$m. The 20.7 and 21.5
micron features are found as well as the 26.1 micron feature,
which is rather broad. We also detect the 48 micron blend.

{\em NML~Cyg}: The 13.6 and 14.2 micron features are blended with
absorption features from HCN at 14.02 $\mu$m and an unknown broad
absorption at 13.25 $\mu$m. These features are seen in all
datasets. The 20.7 and 21.5 micron features were detected. The 48
micron blend has also been detected (and measured).

{\em IRC+1042}: The features around 14 $\mu$m are all found.
The 20.7 and 21.5 micron feature are also present, although the
first one is the broadest one found in our sample. The 26.1
micron feature is weak but present. The 47.7 and 48.6 micron
feature are also found, with the last one, in contrast to what has
been found for other silicate features in this star, the most
red-shifted found in our sample.

\section{Discussion and conclusions}

We have made an extensive study of the infrared spectra of O-rich
environments of evolved stars.  The spectra show a remarkably rich
variety of features, most of which never seen before.  We have
defined seven complexes at 10, 18, 23, 28, 33, 40 and 60 $\mu$m.  Each
of these complexes is a convolution of several contributions, which
vary independently, giving each spectrum a unique appearance.  We have
derived average shapes for the complexes. The large
sample enabled us also to identify weak structures.  We have quantified
the properties of the complexes and of the components that make up these
complexes in terms of peak position, band strength and band width.  We
find that the band strength correlates with the geometry of the
circumstellar environment, and have accordingly conducted a systematic
separate inventory of the bands and complexes for disk sources and outflow
sources.  We have presented a detailed description of the shapes of the
bands and complexes of the individual stars, 
and how the spectra of individual objects differ from the 
average complex shapes. In Paper II, we will discuss, compare and identify
the average complexes.

The richness of the spectra undoubtedly contains important information
related to the conditions in which the grains were formed and their
thermal history.
The crystalline
nature of the carriers of most of the narrow bands (see also papers II
and III) allows a detailed chemical analysis of these components, in a
way which is unprecedented.  In this way, it may become possible to
trace the evolution and processing of grains from their place of birth
(in the outflows of evolved stars), through the interstellar medium, into
star forming regions and proto-planetary disks.  For instance, Molster
et al. (1999a) suggested that in the disk sources low temperature
crystallization takes place.  The crystallization of the silicates in
outflow sources is likely to have taken place close to the star, i.e.
high temperature crystallization.  Therefore the difference found in the
spectra and thus dust properties might have to do with the main
crystallization process in these sources.

The data set presented in this study will serve as a starting point for
more detailed investigations relating to the identification of the
carriers of the bands, and for correlation studies.  These will be
presented in papers II and III. In Paper~II we will discuss the average
complexes and will identify the carriers of the bands using laboratory
spectra of a variety of plausible components. In Paper~III we will study
the correlations between the wavelengths and strengths of the bands, and
derive fits to the bands using simple radiative transfer and
laboratory spectra.

\vspace{1.0cm}
\noindent{\bf Acknowledgements.}\\
FJM wants to acknowledge the support from NWO under grant 781-71-052 and under
the Talent fellowship programm. LBFMW acknowledges financial support from an 
NWO `Pionier' grant.

\clearpage

\section{Appendix; Notes on individual stars}

\subsubsection{IRAS09425-6040}

We have 2 SWS AOT01 observations for this star, and unfortunately no LWS data.
We have slightly tilted band 2A so that the different subbands were nicely connected
with each other.

There is C$_2$H$_2$ absorption at 7.8~$\mu$m, also HCN absorbs around these wavelengths.
The change in slope around 10.64 $\mu$m suggests the presence of a 10.7~micron
feature. However, no other features are found, which makes it somewhat doubtful.
The 11.3 SiC feature prevents the detection of
the 11.4 micron feature associated with forsterite. But, because
of the absence of structure at shorter wavelengths, it is likely not present.

Because of the noise level, the 15.3, 15.9 and 16.2 micron
features are measured as one, but they seem all present in this star.
There are indications in the spectrum that there is a 17.5 micron feature,
but this feature has not been measured separately.
There might be an underlying amorphous silicate feature, but it
has not been measured.
The 24.0, 28.8 and 34.1 micron features might be present but have
not been measured separately. There is a hint of the 31.2~micron feature in 
the rev084 data. However, this cannot
be confirmed on the basis of the rev254 data because of the noise level in
this area.

\subsubsection{NGC~6537= HD~312582= Hen~2-340= IRAS18021-1950}

For NGC~6537 we have 2 SWS and 2 LWS spectra. Unfortunately, the second SWS spectrum
was mispointed and could hardly be used, even to check uncertain features
in an independent dataset.

The origin of the sharp feature at 18.88 $\mu$m is unknown. A rise in the spectrum
is found in both the up and the down scan. A similar feature is seen in AC~Her,
but only in one of the two datasets. No other stars show this feature,
including NGC~6302, which is almost
a twin to this source. Therefore, we still have some doubts about this feature.
Independent of the status of this feature, it influenced our measurement
of the broad 18.9 micron feature underneath.
The 24.5 micron feature is severely influenced by the [Ne V] emission
line.
The 27.6 and 28.2 micron features are measured as one.

\subsubsection{NGC~6302= Butterfly Nebula= HD~155520= Hen~2-204=
IRAS17103-3702= CD-36$^{\circ}$11341}

Since NGC~6302 was one of the calibration sources, a lot of data was available.
We had an AOT01 and a complete wavelength coverage in AOT06,
together with 7 LWS datasets.
The SWS pointings were slightly offset from the centre.
Therefore, we multiplied the whole SWS
spectrum to fit the LWS spectrum. The final spectrum is
in agreement with the IRAS broad band fluxes.

The 13.6 and 14.2~micron features seem to be present in the AOT01
dataset, while in the AOT06 dataset only the 14.2 seem to be present.
However, strong atomic lines at these wavelengths make a clear
detection of these features impossible.
The spectrum of NGC~6302 shows features at 15.3 and 15.8~$\mu$m.
However, the strong [Ne III] line in between these features influences
them. The difference in the up and down scans
also made this a problematic area. The feature at 16.3~$\mu$m feature is only
seen in band 3C of the AOT01 speed 4 dataset, and not in band 3A of the same 
dataset. These last 3 features are also not visible in the AOT06 spectrum, 
therefore we decided not to include them.
The sharp feature at 19.1~$\mu$m is found in both the up and down scans.
Also the AOT06 spectrum shows evidence for a narrow feature at that wavelength.
The source of this feature is unknown. It seems too narrow for a dust feature.
We cannot exclude a forbidden emission line, although it is quite broad
with respect to the other lines in this wavelength region and we don't have
a clear identification yet. 
The 24.5 micron feature is difficult to measure since there is another [Ne V]
line close to it.
Also for the 26.1 micron feature we face the problem of a rather strong line
on top of it ([O IV]), which prevents us from getting a
reasonable estimate for this feature.
If the 40.5 micron feature is a blend of 2 features, as is suggested in
paper~II, these two features nicely blend into one structure
peaking at 40.5~$\mu$m.

\subsubsection{MWC~922= Hen~3-1680= AFGL2132= IRAS18184-1302}

The SWS spectrum of MWC~922 is derived from a combination of 2 AOT01 speed 2
observations.
Features were treated as real if they were seen in both observations.

The 9.1 and 9.5
micron absorption features seem to form a blend, this might
be caused by a known instrumental feature at about 9.35 $\mu$m.
The 9.8 micron feature is very clear and seen in both datasets (up
and down scan), the same holds for the 10.6 micron feature.
The 11.3 micron feature, if present, is completely overwhelmed
by the 11.3 micron PAH feature.

As discussed in the text, the 18 micron complex might suffer from interstellar
absorption. It was impossible to disentangle the circumstellar amorphous silicate
feature from the interstellar absorption and the circumstellar crystalline
silicate emission and therefore it has not been measured.
The (interstellar) absorption by amorphous silicates prevents
us from measuring the features around 16~$\mu$m. Based on a
comparison with other stars,
these features seem to be present, but there remains uncertainty about their
strength and FWHM.
The 18.0, 18.9 and 19.5 micron features are clearly present.
The small structure (3 features) on top of the 18.1 micron feature
is located at the same position as in HD~44179.
The presence of a weak feature at 17.43 $\mu$m is uncertain, since
the two datasets do not fully agree around this wavelength.
The feature at 33.5~$\mu$m is the [S III] line.
No separate 36.0 micron feature can be distinguished in the plateau.
The 47.7 and 48.6~micron features were difficult to disentangle and
therefore measured as one. The [O I] line is visible at 63.18~$\mu$m.
There is a small extension at the long wavelength side of the 69.0~micron
feature. It might be caused by an unknown dust component, or by a
population of relatively warm forsterite grains, which will
emit at a slightly longer wavelength (Bowey et al. 2000).

\subsubsection{AC~Her= HD~170756= SAO~86134= IRAS18281+2149}

For AC~Her, we have two SWS and one LWS datasets. No special
treatments were necessary during data reduction.

The feature at 18.85 micron is only seen in the rev520 dataset
and not in the rev106 data. Although a similar feature has been found
in NGC~6537, it makes this feature questionable.
There is a hint in the spectrum of the 22.4 micron feature and we
have measured it, but the reader should be aware of the noise level
in this part of the spectrum.
The strong feature at 22.69 $\mu$m looks rather prominent and seems also to be
present in HD~44179, where this feature is seen in both the AOT01 as well
as the AOT06 data. However, there is no indication of this feature in the
rev106 spectrum of AC~Her.
It should be noted that the responsivity drops around that
wavelength. It is clear that this feature should be treated with caution.
The reality of the sharp peak around 31.0 $\mu$m is questionable.
This noise peak sits probably on top of a genuine 31.2 micron feature, which
seems also present in the rev106 spectrum.
The shape of the underlying 31.2~micron feature corresponds rather well with
the mean shape of the 31.2 micron feature.
The prominent feature at 32.61 $\mu$m is present in both the up and down scans
of rev520 (although much more prominent in the up scan). However, nothing is
seen in the rev106 dataset and we have some doubt about its
reality. Nevertheless, it should be noted that its wavelength position is
not atypical for a 32.8 micron feature.
The 33.7 and 34.1 micron features are measured together.
In the rev520 dataset, there is an indication of the presence of a 35.9 micron
feature and the absence of the 34.9 micron feature. However, it seems
vice versa for the rev106 data, where a 34.9 micron feature seems present
but hardly anything at 35.9 $\mu$m. It is clear that the noise level in
both datasets play tricks on us here.
We interpret this as evidence for the presence of the 35 micron
plateau although it is not clear where this plateau ends.

\subsubsection{HD~45677= FS~CMa= MWC~142= SAO~151534= AFGL~5195=
IRAS06259-1301}

For HD~45677 we only had one SWS dataset and unfortunately no LWS data.
Band 2 B has been rotated to let all the bands nicely match.

In general, the features of HD~45677 are rather sharp compared to the other
stars. The 34.1 and 34.9 micron features are both present but
blend in this object and are therefore measured as one. The same applies
for the 35.9 and 36.5 micron features.
The sharp line at 34.81~$\mu$m is due to [Si II].

\subsubsection{89~Her= V441~Her= BD+26$^{\circ}$3120= HD~163506=
 IRAS17534+2603= AFGL~2028= SAO~85545}

For 89 Her, we also have redundancy in the SWS domain with two datasets.
For the LWS wavelength range, we have only one dataset.
To match the different bands with each other,
we are forced to slightly tilt band~2B.

There is uncertainty about the 15.2 micron feature. The structure seen at
these wavelengths seems real, because it is found in both the rev082 and the rev518
data. However, it is rather blue-shifted and also very broad. Therefore, the
question arises whether the feature found at 15.0~$\mu$m in this object is
the same as the 15.2 micron feature.

We have fitted the smooth 23 micron complex with 3 features but these were
not well constrained. Therefore caution should be taken in interpreting the
results of this complex, even though the errors might look small.

The 31.2 micron feature is not detected. The structure that is seen around
this wavelength is noise.

\subsubsection{MWC~300= V431~Sct= AFGL~2170= IRAS18267-0606}

For MWC~300 we unfortunately have only one SWS spectrum and
no long wavelength data.
The substructure found in the 10 micron complex, which might be attributed to
crystalline silicates, is always seen in only one scan and we therefore
do not trust its appearance. We have therefore not measured them.
There might be structure around 16 $\mu$m, but this is severely
blended by interstellar silicate absorption and has therefore not been
measured. The amorphous silicate feature detected at 17.03~$\mu$m is
probably due to
interstellar absorption and not a circumstellar amorphous
silicate emission feature, which has therefore not been measured.
The up and down scans differ for the sharp peak at 28.68~$\mu$m; in one it is 
present, but in the other it is not. But the drop around 28.9~$\mu$m is seen in
both the up and down scan, confirming the presence of the 28.8~micron feature.
There is an indication of the 32.2~micron feature but at a very low level.
Therefore it has not been measured.
The up and down scans do not correspond at 42.5 micron. Therefore, we have
removed this feature before measuring the 40 micron complex.
The shallow slope of the blue side of the 40.5 micron feature suggests the
presence of a 39.8 micron feature.
The presence of the 44.7 micron feature can neither be proven nor rejected.

\subsubsection{HD~44179= Red Rectangle= BD-10$^{\circ}$1476= IRAS06176-1036=
AFGL~915= SAO~151362}

For this source we have an AOT01 spectrum and some parts done
with a higher resolution (AOT06). Also an LWS spectrum is available.

The feature at 17.07~$\mu$m is likely to be a blend of the 16.7 and
17.0~micron features. The reality of the 20.7 micron feature is uncertain since
it is very broad in the AOT01 spectrum (although seen in both the up and
down scans), but nothing is seen in the AOT06 spectrum. Therefore, caution
should be taken in interpreting this feature in this star. The structure
around 23.2~$\mu$m is not found in all datasets and differs between the
up and the down scans. Therefore, we masked this part before measuring the
complex. Comparing the 28 micron complex with 89 Her suggests that
the 26.8~micron feature is also present on the blue side of
the 27.6~micron feature.

\subsubsection{Roberts~22= Hen~3-404= Wray~15-549= AFGL~4104=
OH284.18-0.79= IRAS10197-5750}

When ISO observed Roberts~22 with the SWS it was offset, and a lot of flux was
missed. Increasing apertures gave also flux jumps between the different
subbands. In total, we have 2 SWS and 1 LWS spectrum for this source.

We have measured the 18 micron amorphous silicate band. However, it is
likely that the 18.0, 18.9 and 19.5 micron features are present and
are therefore included in this measurement, but the noise level prevents firm
conclusions about their presence. The 15.9 and 16.2 micron features are
measured together. The 23 micron complex has been measured as one feature,
since it was too noisy to separate the different features. The 28~micron
complex suffers from flux jumps between the different bands. Therefore, there
is some uncertainty about the continuum in band 3E. Still, we were able to
measure the features in this band since we used a local continuum for these
measurements. However their relative strength is questionable.
All 3 features found in band 3E were seen in both the rev084
as well as the rev254 datasets. Before measuring the 33~micron complex, we
masked the narrow (0.1~$\mu$m) peak at 33.3~$\mu$m, since it was
only found in one scan direction and not in the rev084 data.
The 35.9 + 36.5~micron features were measured as one, alike to the
39.8 + 40.5~micron features.
The 20.7 and 21.5 micron features are found in rev254, where their
shape resembles the average shape of these features quite well.
However, because nothing is found in the rev084 dataset, and because of the
low S/N in this part of the spectrum, these features are suspect.
There is some concern about the data beyond 80~$\mu$m. Therefore,
the detection of the 90 micron feature in Roberts~22 is uncertain.

\subsubsection{Vy~2-2= IRAS19219+0947}

Vy~2-2 is the other source on which ISO was not centered correctly.
This resulted in flux jumps and intensity losses. We had one SWS and one LWS
spectrum. The whole SWS spectrum suffered from the mispointing of the
satellite. This reduced the flux level and therefore the signal to noise ratio.

The reality of the feature at 11.6 $\mu$m is severely doubted, since the up and
down scans do not agree with each other. It is therefore a good indicator of
the noise-level in that part of the spectrum. The amorphous silicate features,
clearly present in the 10 micron complex, could not be detected
at the 18 micron complex, likely due to the noise level.

The feature at 25.1 $\mu$m is too broad to be a forbidden emission line.
However it also seems narrower than the normal 25.0~micron feature.
The noise in this part of the spectrum is quite severe, therefore we have
some doubts about this feature. If other data confirm this to be the
25.0~micron feature, it would be the strongest 25.0~micron feature in the
sample. The overall intensity from 24.2 to 25.2~$\mu$m suggests that the
plateau is present and this has been measured (excluding the uncertain
strong peak at 25.1~$\mu$m).

There were a lot of spurious features in band 4, which were removed by
hand during the data reduction.
There are hints in both the up and down scans for the different
features in the 35 micron plateau, therefore we have measured them.
However, the signal to
noise ratio in this region makes these features somewhat uncertain
and therefore no firm conclusions should be drawn based on these
features alone.
Also the 40 micron complex suffers from the high noise level.
The features seen in the spectrum provided by the ISO-data archive in Vilspa
are probably due to glitches and glitch tails, which were found in this
part of the spectrum. We have removed these spurious points by hand.
We have tried to split the 40 micron complex
into a 40.5 micron feature and a 43.0 + 43.8 micron blend and measured
the result.

\subsubsection{HD~161796 = V814~Her= BD+50$^{\circ}$2457= IRAS17436+5003=
AFGL~5384= SAO~30548}

We got a nice dataset for HD~161796, including an AOT06 observation of band 4.
In the final spectrum we attached this AOT06 spectrum between 29.03 and
43.17~$\mu$m.

A broad feature around 13 micron may be present, which is also seen in 
AFGL~4106 and HD~179821, however in the latter spectra it resembles continuum 
with an early rise of the 18 micron silicate feature, starting at about 
14.5~$\mu$m. The 18 micron complex is located on the slope of a steep spectrum 
and there is some uncertainty about the strength of the 18 micron feature.
The up and down scans do not completely agree around this feature.
Still the peak position and FWHM agree very well
with the feature seen in the mean outflow spectrum. So we conclude that
it is likely that there is a feature at this wavelength, but its
strength is somewhat controversial.
Because of the presence of the crystalline silicate features,
it is more difficult to measure the amorphous
silicates. This dust feature also hampers from our method of measurement.
We fitted Gaussians while this feature is not a Gaussian.
The small feature which is seen at 17.5~$\mu$m is noise.
The shape of the 20.5 micron feature is similar to the shape in
the mean spectrum, however its strength is similar to
the features at 22.0 and 22.4 $\mu$m, for which the up and down scans
differ significantly. Therefore, the detection of the 20.5 micron
feature is only marginal.
Because of the reasonable uncertainty about the 22.4 micron feature, it has
not been measured.
The 28.8 micron feature might be present but one should take into account
that we only measured it after masking the spurious feature at 29.2 $\mu$m.
It should be noted that the 30.6 micron feature is found in the AOT06 data and
in the rev342 data but nothing is seen in the rev071 dataset.
The spectrum between 29.03 and 43.17 $\mu$m is from the AOT06 observation.
The 35.9 and 36.5 micron features have been measured as a blend.
The 35.1 micron feature is probably somewhat underestimated and blue-shifted.
The status of the feature at 44.09~$\mu$m is not completely sorted out.
It is seen in all data sets, but the up and down scans do often differ
at this point. For now, we assume that it is the result of a blend of the
43.8 and the 44.7 micron features.

\subsubsection{OH~26.5+0.6= V347~Sct= IRAS18348-0526= AFGL~2205}

For OH~26.5+0.6 we got 2 LWS and one SWS dataset.

Up to 28 $\mu$m all features are in absorption.
The amorphous 10 micron feature suffers from a high degree of absorption
and no Gaussian profile is fitted to it, the centre of the absorption is
at $9.9 \pm 0.1$~$\mu$m. The substructure in the 10 micron absorption band
might be real, it is seen in both the up and down scans, but the positions
do not match with other sources in our sample.
The 23.0 and 23.7  micron features are measured as a blend.
It is not completely clear whether the 27.6 and 28.2 micron features
are present or not. They are located at the border
where the spectrum goes from absorption to emission.
It seems that there is some excess emission at these
wavelengths, but no clear peak has been seen.
This suggest that they suffer from self absorption, which
makes it impossible to measure them with Gaussians. So we have not
measured these features, although they seem to be present.
The 33 micron complex without the plateau has been measured as one feature,
since it was not possible to make sure how to separate the
different features, for which are indications for their presence.
Although there is an indication of 2 different features
in the 60 micron complex of OH~26.5+0.6 we have measured it as one.
There seems to be self-absorption in the 26.1 micron feature, which hampers the
measurements of this feature.
The 48 micron blend has again been measured as one due to the noise level
in that part of the spectrum.

\subsubsection{HD~179821 = V1427~Aql= BD-00$^{\circ}$3679= IRAS19114+0002=
AFGL~2343= SAO~124414}

Our data products of HD~179821 consist of two SWS and one LWS spectrum.

The 18 micron complex lies on a very steep slope, which makes it difficult
to identify and measure the different features.
The 18 micron amorphous silicate feature might be about 10\%
too strong since it has been measured including the crystalline silicate
features.
The 15.9 and 16.2 micron features were measured together,
since it appeared very difficult to separate them.
The 17.0 micron feature is very weak, but its width corresponds
well with the mean feature shape.
The detection of the 18.9 micron feature is uncertain, it has therefore
been measured together with the 18.0 micron feature.
The 24.5 and 25.1 micron features have been measured as one.
The 40.5 micron feature shows a relatively shallow slope towards
the shorter wavelengths, this is fitted with an extra
Gaussian, which is attributed to the 39.8 micron features.
This extra Gaussian influences the strength
and probably also the derived width of the 40.5 micron feature.
The sharp drop around 67~$\mu$m might have to do with the border between
two different bands in the LWS spectrum. We remind the reader that we have not
performed any shift of the sub-bands for this LWS spectrum.

\subsubsection{AFGL~4106= HD~302821= CPD-58$^{\circ}$2154= IRAS10215-5916}

There are three SWS datasets for AFGL~4106 and one LWS dataset.
We had to rotate band 3E to match band 3D and 4.

The 18 micron amorphous silicate feature might be about 10\% - 20\%
too strong, since it has been measured including the crystalline silicate
features. Again some features have been measured as a blend, such as the
15.9 + 16.2 micron features, the 16.7 + 17.0 micron features and
the 47.7 + 48.6 micron blend.
There are slight indications of a 22.4 micron feature, however we decided
not to measure it because of the noise level, which makes this feature
rather uncertain.

\subsubsection{NML~Cyg= V1489~Cyg= IRC+40448= OH80.8-1.9= AFGL~2650}

For NML Cyg there was a nice dataset available, with 4 SWS,
of which one was an AOT06 observation of band 4, and 3 LWS observations. 
We have included this high
resolution spectrum in our final spectrum. Band 2A has been tilted to fit the
other bands.

The 8.3 micron feature might be present, but is difficult to separate from the
amorphous silicate feature which still peaks at these wavelengths.
The structure around 10.07~$\mu$m coincides with a feature in the
responsivity and is therefore not trusted. The structures around 11.05~$\mu$m
are due to the apparently not completely correctly removed 11.1~micron
responsivity feature.

The 13.6 and 14.2 micron features are influenced by molecular absorption 
features at 13.25 and 14.02 $\mu$m.

We assumed that there was no 13.8 micron feature and that therefore the flux
at that wavelength was continuum. The features were measured with
respect to that continuum.

The 16.9 and 17.2 micron features are measured as one. If there is a very
weak 19.5 micron feature it has been measured together with the 18.9 micron
feature. At 14.96 $\mu$m gas phase CO$_2$ absorption has been detected.

The 18 micron amorphous silicate could not be extracted from the data, it
is probably between emission and self absorption.
The 24.5 micron feature is rather blue-shifted and strong compared to the
other 24.5 micron features. It gives NML Cyg the broadest appearance
for this complex of all sources. We cannot exclude that this feature
is not the same as the other 24.5 micron features.

The end of band 3D and the beginning of Band 3E are not OK due to leakage,
therefore no features are measured in this wavelength range
There might be a 39.8 micron feature present, however this is difficult to
measure with so many water lines present in that region.
Therefore we measured it as one feature, neglecting the presence of the
water lines. This might lead to an overestimation of the flux level of the
features.

The 47.7 and 48.6 micron features were measured as a blend.

\subsubsection{IRC+10420= V1302~Aql= IRAS19244+1115= AFGL~2390}

The last source in our sample is IRC+10420. We got complete spectral
coverage in the SWS and LWS wavelength range. Part of the SWS-spectrum was
done with an AOT06 which is included in our final spectrum.
The infrared spectrum of IRC+10420 shows some structure in the 10 micron
complex which lines up quite nicely with crystalline silicates in other
sources. The weak 10.6 micron feature is uncertain, since it is not consistent
with the feature seen in the mean disk spectrum feature. The 11.3~micron
feature is consistent with a feature seen in AC Her but very weak.

The 16.2 micron feature is somewhat unsure, since it is clearly seen in
band 3c (up and down scan), but not in band 3A (neither in the up nor in the 
down scan).
The amorphous feature has been measured without the crystalline features and
the strength is therefore somewhat overestimated, however this
effect is negligible compared to the uncertainty in continuum level.
The 18.0 + 18.9 micron features are measured as one.

Band 3D was not corrected for leakage problems in the calibration sources,
because the flux level in IRC+10420 is so high that
it is very likely that also here leakage has its influence.
Band 3E is tilted to fit the AOT06 and LWS on one side and the rest of the data
at the other side.
The 20.7 micron feature is the broadest one found in our sample, still the data
seem valid, since the up and down scans show exactly the same trend.

The up and down scans of the AOT06 spectrum differ severely between 29.3 and 
30~$\mu$m, probably due to memory effects for this high flux source.
Therefore we have removed this part of the spectrum, since also the AOT01 
suffered from the same problems in this part of the spectrum.

There is a possibility of a 39.8 micron feature however the different
datasets contradict each other around these wavelengths. Therefore we have
measured this feature as a 40.5 micron feature.


\begin{thebibliography}{}

\bibitem[]{}
 Ageorges N., Eckart A., Monin J.-L. and M\'{e}nard F., 1997, A\&A 326, 632
\bibitem[]{}
 Alcolea J., Bujarrabal, 1985, A\&A 303, L21
\bibitem[]{}
 Allen D.A., Hyland A.R. and Caswell J.L., 1980, MNRAS 192, 505
\bibitem[]{}
 Aller L.H., Ross J.E., O'Mara B.J. and Keyes C.D., 1981, MNRAS 197, 95
\bibitem[]{}
 Arrelano Ferro A., 1984, PASP 96, 641
\bibitem[]{}
 Ashley M.C.B. and Hyland A.R., 1988, ApJ 331, 532
\bibitem[]{}
 Bachiller R., G\'{o}mea-Gonz\'{a}lez J., Bujarrabal V. and Martin-Pintado J., 1988, A\&A 196, L5
\bibitem[]{}
 Barlow M.J., 1998, A\&SS, 255, 315
\bibitem[]{}
 Beckwith S.V.W.,  Sargent A.I., Chini R.S., Guesten R., 1990, AJ 99, 924
\bibitem[]{}
 Beintema D.A., 1998, A\&SS, 255, 507
\bibitem[]{}
 Bowers P.F., 1984, ApJ 279, 350
\bibitem[]{}
 Bowers P.F., Johnston K.J. and Spencer J.H., 1983, ApJ 274, 733
\bibitem[]{}
 Bowey J.E., Lee C., Tucker C. et al., 2000, in proceedings of
 ISO beyond the peaks: The 2nd ISO workshop on analytical spectroscopy,
 eds. A. Salama, M.F.Kessler, K. Leech \& B. Schulz, ESA-SP 456, 339
\bibitem[]{}
 Bujarrabal V., Alcolea J. and Planesas P., 1992, A\&A 257,701
\bibitem[]{}
 Casassus S.P., Roche P.F. and Barlow M.J., 2000, submitted to MNRAS
\bibitem[]{}
 Clegg P.E., Ade P.A.R., Armand C., et al., 1996, A\&A 315, L38
\bibitem[]{}
 Cohen M., Anderson C.M., Cowley A., et al., 1975, ApJ 196, 179
\bibitem[]{}
 Cohen M., Tielens A.G.G.M., Bregman J. et al., 1989, ApJ 341, 246
\bibitem[]{}
 Cohen M., Barlow M.J., Sylvester R.J. et al., 1999, ApJ 513, L135
\bibitem[]{}
 Cuesta L., Phillips J.P. and Mampaso A., 1995, A\&A 304, 475
\bibitem[]{}
 Dorschner J., Begemann B., Henning Th. et al., 1995, A\&A 300, 503
\bibitem[]{}
 Fix J.D. and Cobb M.L., 1988, ApJ 329, 290
\bibitem[]{}
 de Graauw Th., Haser L.N., Beintema D.A. et al., 1996, A\&A 315, L49
\bibitem[]{}
 Hamann F. and Persson S.E., 1989, ApJS 71, 931
\bibitem[]{}
 Hawkins G.W., Skinner C.J., Meixner M.M. et al., 1995, ApJ 452, 314
\bibitem[]{}
 Henning Th., Launhardt R., Steinacker J. et al., 1994, A\&A 291, 546
\bibitem[]{}
 Hony S., Van Kerckhoven C., Peeters E., et al., 2000 in proceedings of
 ISO beyond the peaks: The 2nd ISO workshop on analytical spectroscopy,
 eds. A. Salama, M.F.Kessler, K. Leech \& B. Schulz, ESA-SP 456, 63
\bibitem[]{}
 Humphreys R.M., Strecker D.W., Murdock T.L. and Low F.J., 1973, ApJ 179, L53
\bibitem[]{}
 Jewell P.R., Schenewerk M.S., Snyder L.E., 1985, ApJ 295, 183
\bibitem[]{}
 Jones T.J., Humphreys R.M., Gehrz R.D. et al., 1993, ApJ 411, 323
\bibitem[]{}
 Jura M., Balm S.P., Kahane C., 1995, ApJ 453, 721
\bibitem[]{}
 Jura M., Turner J. and Balm S.P., 1997, ApJ 474, 741
\bibitem[]{}
 Jura M. and Turner J., 1998, Nature 395, 144 (1998)
\bibitem[]{}
 Jura M. and Werner M.W., 1999, ApJ 525, L113
\bibitem[]{}
 Jura M., Chen C., Werner M.W., 2000, ApJ 541, 264
\bibitem[]{}
 Justtanont K., Skinner C.J. and Tielens A.G.G.M., 1994, ApJ 435, 852
\bibitem[]{}
 Justtanont K., Skinner C.J., Tielens A.G.G.M. et al., 1996a, ApJ 456, 337
\bibitem[]{}
 Justtanont K., de Jong T., Helmich F.P. et al.,1996b, A\&A 315, L217
\bibitem[]{}
 Kastner J.H. and Weintraub D.A., 1995, ApJ 452, 833
\bibitem[]{}
 Kessler M.F., Steinz J.A., Anderegg M.E. et al., 1996, A\&A 315, L27
\bibitem[]{}
 Knapp G.R., Phillips T.G., Leighton R.B. et al., 1982, ApJ 252, 616
\bibitem[]{}
 Koike C., Tsuchiyama A., Shibai H. et al., 2000, A\&A 363, 1115
\bibitem[]{}
 Lamers H.G.J.L.M., Zickgraf F-J., de Winter D. et al, 1998, A\&A 340, 117
\bibitem[]{}
 Leinert C., Richichi A. and Haas M., 1997, A\&A 318, 472
\bibitem[]{}
 Lester D.F. and Dinerstein H.L., 1984, ApJ 281, L67
\bibitem[]{}
 Lewis B.M., Terzian Y. and Eder J., 1986, ApJ 302, L23
\bibitem[]{}
 Likkel L., Forveille T., Omont A., and Morris M., 1991, A\&A 246, 153
\bibitem[]{}
 Luck R.E., Bond H.E. and Lambert D.L., 1990, ApJ 357, 188
\bibitem[]{}
 Malfait K., 1999, {\em PHD~ thesis}, Katholieke Universiteit Leuven
\bibitem[]{}
 Meaburn J. and Walsh J.R., 1980, MNRAS 191, 5P
\bibitem[]{}
 Meixner M., Ueta T., Dayal A. et al, 1999, ApJS 122, 221
\bibitem[]{}
 Molster F.J., Waters L.B.F.M., de Jong T. et al, 1997,
 in ``Planetary nebulae'', Proc. IAU180,
 Habing H.J. and Lamers H.J.G.L.M. Lamers (eds), Kluwer, Dordrecht, p 361
\bibitem[]{}
 Molster F.J., Yamamura I., Waters L.B.F.M. et al., 1999a, Nature 401, 563
\bibitem[]{}
 Molster F.J., Waters L.B.F.M., Trams N. et al., 1999b A\&A 350, 163
\bibitem[]{}
 Molster F.J., Yamamura I., Waters L.B.F.M. et al., 2001a, A\&A 366, 923
\bibitem[]{}
 Molster F.J., Lim T.L., Sylvester R.J. et al., 2001b A\&A 372, 165
\bibitem[]{}
 Molster F.J., Waters L.B.F.M., Tielens A.G.G.M., 2002a accepeted by A\&A
 (Paper II)
\bibitem[]{}
 Molster F.J., Waters L.B.F.M., Tielens A.G.G.M. et al., 2002b, accepted A\&A
 (Paper III)
\bibitem[]{}
 Monnier J.D., Bester M., Danchi W.C. et al., 1997, ApJ 481, 420
\bibitem[]{}
 Morris M. and Jura M., 1983, ApJ 267, 179
\bibitem[]{}
 Nedoluha G.E. and Bowers P.F., 1992, ApJ 392, 249
\bibitem[]{}
 Netzer N. and Knapp G.R., 1987, ApJ 323, 734
\bibitem[]{}
 Osterbart R., Langer N. and Weigelt G., 1997, A\&A 325, 609
\bibitem[]{}
 Oudmaijer R.D., 1998, A\&AS 129, 541
\bibitem[]{}
 Oudmaijer R.D., Geballe T.R., Waters L.B.F.M. and Sahu K.C., 1994, A\&A 281, L33
\bibitem[]{}
 Payne H.E., Phillips J.A. and Terzian Y., 1988, ApJ 326, 368
\bibitem[]{}
 P\'{e}rez M.R., Grady C.A., van den Ancker M.E. et al., 1993,
 in ``Frontiers of Space and Ground-Based Astronomy, The Astrophysics
 of the 21st Century'', Longair M.S., Kondo Y. and Wamstekker W. (eds.),
 Kluwer, Dordrecht, The Netherlands, p563
\bibitem[]{}
 Pirzkal N., Spillar E.J., and Dyck H.M., 1997, ApJ 481, 392
\bibitem[]{}
 Richards A.M.S., Yates J.A. and Cohen R.J., 1996, MNRAS 282, 665
\bibitem[]{}
 Roche P.F. and Aitken D.K., 1984, MNRAS 208, 481
\bibitem[]{}
 Roche P.F. and Aitken D.K., 1986, MNRAS 221, 63
\bibitem[]{}
 Roddier F., Roddier C., Graves J.E. and Northcott M.J., 1995, ApJ 443, 249
\bibitem[]{}
 Rodriguez L.F., Garcia-Baretto J.A., Canto J. et al, 1985, MNRAS 215, 353
\bibitem[]{}
 Sahai R. and Trauger J.T., 1998 AJ 116, 1357
\bibitem[]{}
 Sahai R., Zijlstra A., Bujarrabal V. and te Lintel Hekkert P., 1999, AJ 117, 1408
\bibitem[]{}
 Sandford S.A., Pendleton Y.J., Allamandola L.J., 1995, ApJ 440, 697
\bibitem[]{}
 Schaeidt S.G., Morris P.W., Salama A. et al, 1996, A\&A 315, L55
\bibitem[]{}
 Schulte-Ladbeck R.E., Shepherd D.S., Nordsieck K.H. et al., 1992,
 ApJ 401, L105
\bibitem[]{}
 Seaquist E.R. and Davis L.E., 1983, ApJ 274, 659
\bibitem[]{}
 Simon T. and Dyck H.M., 1977, AJ 82, 725
\bibitem[]{}
 Skinner C.J., Meixner M.M., Hawkins G.W. et al., 1994, ApJ 423, L135
\bibitem[]{}
 Skinner S.L., Brown A., Stewart R.T., 1993, ApJS 87, 217
\bibitem[]{}
 Swinyard B.M., Clegg P.E., Ade P.A.R. et al., 1996, A\&A 315, L43
\bibitem[]{}
 Sylvester R.J., Barlow M.J., Liu X.W., et al., 1997, MNRAS 291, L42 
\bibitem[]{}
 Sylvester R.J., Kemper F., Barlow M.J. et al., 1999, A\&A 352, 587
\bibitem[]{}
 Th\'{e} P.S., de Winter D., P\'{e}rez M.R., 1994, A\&AS 104, 315
\bibitem[]{}
 Valentijn E.A., Feuchtgruber H., Kester D.J.M. et al., 1996, A\&A 315, L60
\bibitem[]{}
 Van Kerckhoven C., Hony S., Peeters E., et al., 2000, A\&A 357, 1013
\bibitem[]{}
 Van Winckel H., Waelkens C., Waters L.B.F.M. et al, 1998, A\&A 336, L17
\bibitem[]{}
 van der Veen W.E.C.J., Waters L.B.F.M., Trams N.R. and Matthews H.E.,
 1994, A\&A 285, 551
\bibitem[]{}
 Voors R.H.M., 1999, {\em PHD~ thesis}, University of Utrecht
\bibitem[]{}
 Waelkens C., Van Winckel H., Waters L.B.F.M. and Bakker E.J., 1996, A\&A 314, L17
\bibitem[]{}
 Waters L.B.F.M., Waelkens C., Mayor M. and Trams N.R., 1993, A\&A 269, 242
\bibitem[]{}
 Waters L.B.F.M., Molster F.J., de Jong T. et al., 1996, A\&A 315, L361
\bibitem[]{}
 Waters L.B.F.M., Waelkens C., Van Winckel H. et al., 1998, Nature 391, 868
\bibitem[]{}
 Winckler H. and Wolf B., 1989, A\&A 219, 151
\bibitem[]{}
 de Winter D. and van den Ancker M.E., 1997, A\&AS 121, 275
\bibitem[]{}
 Wolf B., Stahl O., 1985, A\&A 148, 412
\bibitem[]{}
 Zsoldos E., 1993, A\&A 268, 149
\bibitem[]{}
 Zuckermann B. and Dyck H.M., 1986, ApJ 311, 345


\end{thebibliography}
\end{document}